\documentclass[]{aastex631}

\usepackage{multirow}

\usepackage[T1]{fontenc}  
\usepackage[utf8]{inputenc} 

\shorttitle{Absolute Physical Parameters of Contact Binaries}
\shortauthors{Xia et al.}
\graphicspath{{./}{figures/}}

\begin{document}

\title{Minute-cadence Observations of the LAMOST Fields with the TMTS: VI. Absolute Physical Parameters of Contact Binaries}

\author{Qiqi Xia}
\affiliation{Physics Department and Tsinghua Center for Astrophysics, Tsinghua University, Beijing, 100084, People's Republic of China}

\author{Xiaofeng Wang}
\altaffiliation{E-mail: wang\_xf@mail.tsinghua.edu.cn}
\affiliation{Physics Department and Tsinghua Center for Astrophysics, Tsinghua University, Beijing, 100084, People's Republic of China}

\author{Kai Li}
\affiliation{Shandong Key Laboratory of Optical Astronomy and Solar-Terrestrial Environment, School of Space Science and Physics, Institute of Space Sciences, Shandong University, Weihai, Shandong 264209, China}

\author{Xiang Gao}
\affiliation{Shandong Key Laboratory of Optical Astronomy and Solar-Terrestrial Environment, School of Space Science and Physics, Institute of Space Sciences, Shandong University, Weihai, Shandong 264209, China}

\author{Fangzhou Guo}
\affiliation{Physics Department and Tsinghua Center for Astrophysics, Tsinghua University, Beijing, 100084, People's Republic of China}

\author{Jie Lin}
\affiliation{Department of Astronomy, University of Science and Technology of China, Hefei 230026, People's Republic of China}
\affiliation{School of Astronomy and Space Science, University of Science and Technology of China, Hefei 230026, People's Republic of China}

\author{Cheng Liu}
\affiliation{Beijing Planetarium, Beijing Academy of Sciences and Technology, Beijing, 100044, People’s Republic of China}

\author{Jun Mo}
\affiliation{Physics Department and Tsinghua Center for Astrophysics, Tsinghua University, Beijing, 100084, People's Republic of China}

\author{Haowei Peng}
\affiliation{Physics Department and Tsinghua Center for Astrophysics, Tsinghua University, Beijing, 100084, People's Republic of China}

\author{Qichun Liu}
\affiliation{Physics Department and Tsinghua Center for Astrophysics, Tsinghua University, Beijing, 100084, People's Republic of China}

\author{Gaobo Xi}
\affiliation{Physics Department and Tsinghua Center for Astrophysics, Tsinghua University, Beijing, 100084, People's Republic of China}

\author{Shengyu Yan}
\affiliation{Physics Department and Tsinghua Center for Astrophysics, Tsinghua University, Beijing, 100084, People's Republic of China}

\author{Xiaojun Jiang}
\affiliation{National Astronomical Observatories, Chinese Academy of Sciences, Beijing 100101, China}

\author{Jicheng Zhang}
\affiliation{School of Physics and Astronomy, Beijing Normal University, Beijing 100875, China.}

\author{Cui-Ying Song}
\affiliation{Physics Department and Tsinghua Center for Astrophysics, Tsinghua University, Beijing, 100084, People's Republic of China}

\author{Jianrong Shi}
\affiliation{National Astronomical Observatories, Chinese Academy of Sciences, Beijing 100101, China}

\author{Xiaoran Ma}
\affiliation{Physics Department and Tsinghua Center for Astrophysics, Tsinghua University, Beijing, 100084, People's Republic of China}

\author{Danfeng Xiang}
\affiliation{Physics Department and Tsinghua Center for Astrophysics, Tsinghua University, Beijing, 100084, People's Republic of China}
\affiliation{Beijing Planetarium, Beijing Academy of Sciences and Technology, Beijing, 100044, People’s Republic of China}

\author{Wenxiong Li}
\affiliation{National Astronomical Observatories, Chinese Academy of Sciences, Beijing 100101, China}



\begin{abstract}

With the development of wide-field surveys, a large amount of data on short-period W UMa contact binaries have been obtained. Continuous and uninterrupted light curves as well as high-resolution spectroscopic data are crucial in determining the absolute physical parameters.
Targets with both TMTS light curves and LAMOST medium-resolution spectra were selected. 
The absolute physical parameters
were inferred with the W-D code for ten systems, all of them are W-type shallow or medium contact binaries.
The O'Connell effect observed in the light curves can be explained by adding a spot on the primary or secondary component in the models.
According to O-C analysis, the orbital periods exhibit a long-term increasing or decreasing trend, amongst which J0132, J1300, and J1402 show periodic variations that may be attributed to the presence of a third body or magnetic activity cycles. 
Spectral subtraction analysis revealed that the equivalent width of H$\alpha$ indicates strong magnetic activity in J0047, J0305, J0638, and J1402.
Among the 10 selected binary systems, except for J0132 and J0913, the more massive components are found to be main-sequence stars while the less massive components have evolved off the main sequence. 
In J0132, both components are in the main sequence, whereas both components of J0913 lie above the terminal-age main sequence. Based on the relationship between orbital angular momentum and total mass for these two systems, as well as their low fill-out factors, it is possible that these two systems are newly formed contact binaries, having recently evolved from the detached configuration.

\end{abstract}



\section{Introduction}
W Ursae Majoris (W UMa) type contact binaries are a class of eclipsing binary stars where both components share a common envelope and are in contact with each other \citep{Lucy+1968b+ApJ+LC,Lucy+1968a+ApJ+structure}. These systems are characterized by their short orbital periods, typically less than a day, and nearly identical spectral types (late type) for two components. 
\cite{Binnendijk+VA+1970} classified W UMa type contact binaries into two types: W-type and A-type.
In W-type systems, the more massive component is cooler than the less massive one, while in A-type systems, the more massive component is hotter \citep{Yildiz+Dogan+2013+MN+AML,Li+etal+2021+AJ+173RV}. The continuous exchange of mass and energy between the two stars leads to complex evolutionary processes, for example, the thermal relaxation oscillation theory \citep{Flannery+1976+ApJ+TRO,Lucy+1976+ApJ+TRO,Robertson+Eggleton+1977+MNRAS+TRO,Lucy+Wilson+1979+ApJ+TRO,Qian+2001+MNRAS+TRO}, the O'Connell effect \citep{OConnell+1951+PRCO}, and long-term increase or decrease in the orbital period \citep{Li+etal+2015+AJ+masstrans,Li+2018+NewA+masstrans,Lee+Park+2018+PASP+masstrans}.

Radial velocity (RV) measurements play a critical role in the precise determination of the absolute physical parameters of the contact binaries \citep[e.g.,][]{Hrivnak+1988+ApJ+RV+I+1,Hrivnak+1989+ApJ+RV+II+1,Rucinski+2000+AJ+RV+III+10}, which requires multiple  short exposure-time spectra with high signal-to-noise ratios (SNR). 
Precise parameters for masses, radii, and luminosities were determined by combining radial velocities with light curves \citep[e.g.,][]{Lu+etal+2007+AJ+RV+LC+1,Alvarez+etal+2015+PASP+MWPav,Liu+etal+2023+AJ+SixRV}. These factors are crucial for explaining the formation and evolution of contact binaries as well as for the mechanisms of mass transfer, angular momentum loss, and the overall stability of the common envelope.
\cite{Latkovic+etal+2021+ApJS+700} compiled 700 contact binaries from individual studies, among which the absolute physical parameters of 159 targets were determined by spectroscopic and photometric observations. \cite{Li+etal+2021+AJ+173RV} compiled the absolute physical parameters of 173 contact binaries with spectroscopic and photometric observations.

In recent years, with the release of large-scale photometric and spectroscopic survey data, e.g., the All Sky Automated Survey \citep[ASAS;][]{Pojmanski+1997+AcA+ASAS,Pojmanski+1998+AcA+ASAS,Pojmanski+2002+AcA+ASAS}, the All-Sky Automated Survey for Supernovae \citep[ASAS-SN;][]{Shappee+etal+2014+ApJ+ASAS-SN,Christy+etal+2023+MNRAS+ASAS-SN}, the Catalina Sky Survey (CSS; \cite{Marsh+etal+MNRAS+2017+CSS}), the Super Wide Angle Search for Planets \citep[SuperWASP;][]{Pallacco+etal+2006+PASP+SuperWASP}, the Transiting Exoplanet Survey Satellite \citep[TESS;][]{Ricker+etal+2010+AAS+TESS,Stassum+etal+AJ+TESS}, the Zwicky Transient Facility, \citep[ZTF;][]{Masci+etal+2019+PASP+ZTF,Bellm+etal+2019+PASP+ZTF}, and the Large Sky Area Multi-Object Fiber Spectroscopic Telescope \citep[LAMOST;][]{Cui+etal+2012+RAA+LAMOST,Zhao+etal+RAA+LAMOST}, a batch of physical parameters of contact binaries have been obtained.
Wilson-Devinney (W-D) program \citep{Wilson+Devinney+1971+ApJ+WD,Wilson+1979+ApJ+WD,Wilson+1990+ApJ+WD,VanHamme+Wilson+2007+ApJ+WD,Wilson+VanHamme+2010+ApJ+WD,Wilson+VanHamme+2014+ApJ+WD} and PHysics Of Eclipsing BinariEs (PHOEBE) Python code \citep{Prsa+2018+book+PHOEBE} are the most commonly used tools to individually solve the orbital parameters of contact binaries \citep[e.g.,][]{Deb+etal+2011+MNRAS+ASAS+WD+EB,Sun+etal+2020+ApJS+CSS+WD+CB,Paki+Poro+2024+arXiv+PHOEBE+TESS+20}. In addition, methods that use machine learning to solve a large number of orbital parameters simultaneously are also being widely developed \citep[e.g.,][]{Ding+etal+2021+PASJ+ML,Ding+etal+2023+MN+TESS+ML,Xiong+etal+2024+ApJS+TESS+ML,Li+etal+ApJS+ASAS-SN+ML}, which has greatly facilitated the study of W UMa binaries.

Continuous, uninterrupted light curves and high-precision radial velocity measurements enable accurate determinations of the absolute physical parameters of these systems, allowing for more detailed investigations of the interactions and evolutionary processes within W UMa systems. The Tsinghua University-Ma Huateng Telescopes for Survey (TMTS) have been continuously monitoring the LAMOST sky areas for the white-light band (TMTS $L$ band covers a wavelength range from 400 to 900 nm) with uninterrupted observation throughout the night \citep{Zhang+etal+2020+PASP+tmts_performance}.
During the first two years of observation, TMTS has detected a series of variable stars with periods shorter than ~7.5 hours \citep{Lin+etal+2022+MNRAS+TMTSI}. A total of 1,100 targets with periods shorter than 2 hours were confirmed \citep{Lin+etal+2023+MNRAS+TMTSII}, which includes a blue large-amplitude pulsator (BLAP) with a pulsation period of only 18.9 min \citep{Lin+etal+2023+NatAs+BLAP}. A theoretical prediction of the hot subdwarf binary of the shortest orbital period was also discovered by TMTS \cite{Lin+etal+2024+NatAs+hotsubdwarf}.
Additionally, using machine learning, 11,638 variable stars were classified, including 5,698 EW-type eclipsing binaries \citep[][hereafter TMTS-V]{Guo+etal+2024+MNRAS+TMTSV}, which are also the data used in this paper.

This paper aims to address the absolute physical parameters derived for the W UMa type contact binaries from the TMTS-V. By analyzing photometric and spectroscopic data, we seek to elucidate the intricate dynamics and evolutionary pathways of W UMa type binaries. The results of this study will contribute to an understanding of the evolution of close binary stars and the physical processes that govern their behaviors.
The data sources and the method are described in Section \ref{sec:data}. Sections \ref{sec:O-C} and \ref{sec:W-D} introduce O-C (observed minus calculated) analysis and independent investigation of photometry and spectroscopy. The evolutionary state and statistical characteristics are discussed in Section \ref{sec:discussion}. Section \ref{sec:summary} provides a brief summary.

\section{Data and method}\label{sec:data}

\subsection{Target Selection}

The Tsinghua University–Ma Huateng Telescopes for Survey (TMTS) is located at the Xinglong Station of the National Astronomical Observatory of Chaina (NAOC), using a multitube telescope system consisting of four 40 cm optical telescopes and a large field of view (FoV) of approximately 18 deg$^2$ \citep{Zhang+etal+2020+PASP+tmts_performance}.
Luminous filter (L-band hereafter) was conducted a wide wavelength range from 390nm to about 900nm\citep{Zhang+etal+2020+PASP+tmts_performance}, similar to $Gaia$ G band \citep[330-1050nm;][]{Gaia+etal+2018+AAP+Gaia}.
During the first two-year survey, TMTS has monitored 449 LAMOST/TMTS plates by the end of 2022. Machine learning classification was performed on the periodic variable sources obtained from uninterrupted light curves, confirming 5698 W UMa type contact binaries, namely EW contact binaries \citep{Guo+etal+2024+MNRAS+TMTSV}.

The Large Sky Area Multi-Object Fiber Spectroscopic Telescope (LAMOST) is a 4 meter Schmidt telescope with a FOV of 5 square degrees, and is equipped with 4,000 fibers \citep{Cui+etal+2012+RAA+LAMOST,Luo+etal+RAA+2015+LAMOST,Zhao+etal+RAA+LAMOST}.
LAMOST medium resolution survey (MRS) Data Release (DR) 10\footnote{https://www.lamost.org/dr10/v1.0/} were used as the source of radial velocities (RVs) for the W UMa contact binaries with a resolution of 7,500. For each observed object, two spectra are obtained within a single exposure, which include a blue (B) side spectrum with a wavelength range of 4950 \AA to 5350 \AA, and a red (R) side spectrum with a wavelength range of 6300 \AA ~to 6800 \AA. The blue arm contains more absorption lines than the red arm, allowing us to measure the RVs with a precision up to 1 km s$^{-1}$ for most stars \citep{Liu+etal+2019+RAA+LAMOST-MRS}.

\begin{figure}[ht!]
\centering
\includegraphics[width=0.99\textwidth]{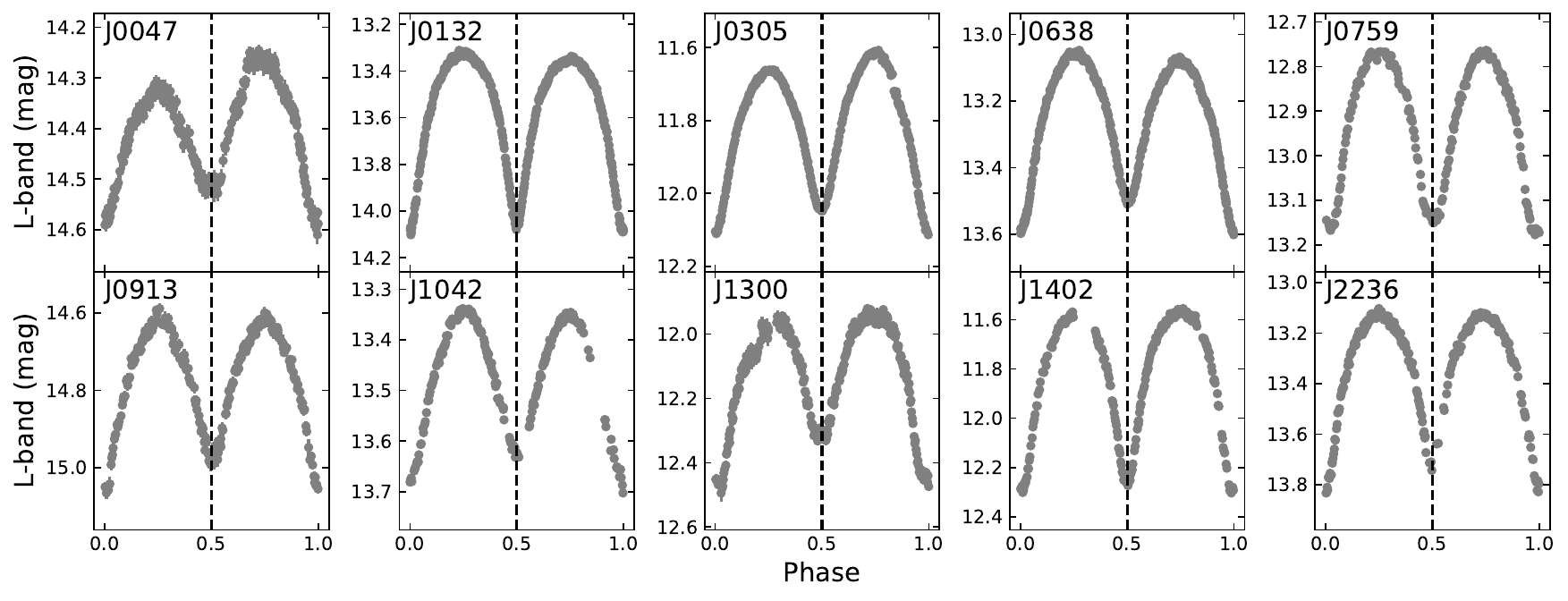}
\caption{Phase-folded light curves of ten targets from TMTS. The L-band is the TMTS Luminous filter.}
\label{fig1:TMTS_LC}
\end{figure}

Firstly, we cross-matched the EW catalog of TMTS with the LAMOST MRS catalog and filtered according to the following criteria:
\begin{itemize}
    \item[1]  Select targets within an angular radius of 3\arcsec.
    \item[2] SNRs of B-band spectra greater than 10.
    \item[3] Number of consecutive exposures greater than or equal to 3.
\end{itemize}
After the above selection process, we obtained 57 targets that met the criteria.
Then, using the period and times of minimum of the TMTS light curves of each target, we calculated the phase corresponding to the observed times for each target. We chose targets with at least three phases between 0.2-0.3 or 0.7-0.8.
After that, we visually inspected their light curves (LCs) again, excluding LCs with insufficient SNR and those targets that might be ellipsoidal variables. Finally ten targets were identified from the LAMOST MRS spectra for subsequent RV calculations, which are listed in Table \ref{tab:observation}. Figure \ref{fig1:TMTS_LC} shows the phase-folded light curves of the ten W UMa contact binaries from the TMTS.

\begin{deluxetable*}{lrcccccccccccc}
\tabletypesize{\scriptsize}
\tablecaption{Photometric observation information of the ten targets\label{tab:observation}}
\tablewidth{0pt}
\tablehead{
\colhead{Source I.D.} & \colhead{Name} & \colhead{R.A.} & \colhead{Dec.} &
\colhead{Obs. Date} & \colhead{$P_{\rm TMTS}$} & \colhead{$P_{\rm VSX}$} &
\colhead{$L_0$}  &  \colhead{$G_{\rm abs}$}   &  \colhead{$B_P-R_P$}  &
\colhead{$N_{\rm LRS}$} &  \colhead{$N_{\rm MRS}$}  &  \colhead{VSX type}  \\
\colhead{}   &  \colhead{}  & \colhead{degree}  &  \colhead{degree}  & \colhead{} &
\colhead{day} & \colhead{day} & \colhead{mag} & \colhead{mag} & \colhead{mag} &
\colhead{} & \colhead{} & \colhead{} & \colhead{} 
}
\startdata
TMTS J00474579+3931052& J0047 &  11.941  &  39.518 &Oct. 6, 2022 &  0.26099 &  0.26108 &14.401 &  5.468 & 1.107  & 2  &  3      &   EW$^{a}$   \\
TMTS J01322049+5512196& J0132 &  23.085  &  55.205 &Oct. 29, 2020 &  0.40898 &  0.40094 &13.577 &  3.550 & 0.680  & 0  &  15     &   EW/KW$^{b}$\\
TMTS J03050520+2934439& J0305 &  46.272  &  29.579 &Dec. 16, 2020 &  0.24118 &  0.24698 &11.804 &  5.922 & 1.181  & 3  &  6      &   EW   \\
TMTS J06380824+4412135& J0638 &  99.534  &  44.204 &Dec. 31, 2021 &  0.34706 &  0.35439 &13.273 &  4.452 & 0.899  & 1  &  22     &   EW   \\
TMTS J07592990+4019452& J0759 &  119.875 &  40.329 &Jan. 15, 2020 &  0.28611 &  0.29595 &12.922 &  4.357 & 0.788  & 1  &  7      &   EW   \\
TMTS J09135321+4354212& J0913 &  138.472 &  43.906 &Jan. 18, 2020 &  0.28207 &  0.27868 &14.763 &  5.347 & 1.098  & 0  &  87     &   EW   \\
TMTS J10421306+3849120& J1042 &  160.554 &  38.820 &Jan. 19, 2020 &  0.32310 &  0.31604 &13.459 &  4.239 & 0.758  & 2  &  100    &   EW   \\
TMTS J13001158+3023102& J1300 &  195.048 &  30.386 &Apr. 7, 2021 &  0.29509 &  0.30199 &12.142 &  4.840 & 0.947  & 2  &  3      &   EW   \\
TMTS J14020545+3402396& J1402 &  210.523 &  34.044 &Apr. 14, 2021 &  0.25532 &  0.26077 &11.835 &  6.103 & 1.318  & 2  &  3      &   EW   \\
TMTS J22364231+3041479& J2236 &  339.176 &  30.697 &Sept. 7, 2022 &  0.33873 &  0.33574 &13.329 &  4.227 & 0.846  & 1  &  3      &   EW   \\
\enddata
\tablecomments{Source I.D.: TMTS catalogue identifier; Name: identifier of the target in this study; P$_{TMTS}$: orbital period from TMTS; P$_{VSX}$: orbital period from VSX; $L_0$: median magnitude from TMTS; $G_{abs}$: absolute magnitude from Gaia; $B_P-R_P$: color index from Gaia; $N_{LRS}$: exposure times of LAMOST LR spectra; $N_{MRS}$: exposrue times of LAMOST MR spectra; VSX type: VSX variability type.
\\
$^{a}$EW represents the W UMa type eclipsing variables from VSX.\\
$^{b}$KW represents contact systems of the W UMa type, with ellipsoidal components of F0-K spectral type.}
\end{deluxetable*}

\subsection{Target Information}

\begin{figure*}[htb]
    \includegraphics[width=0.99\textwidth]{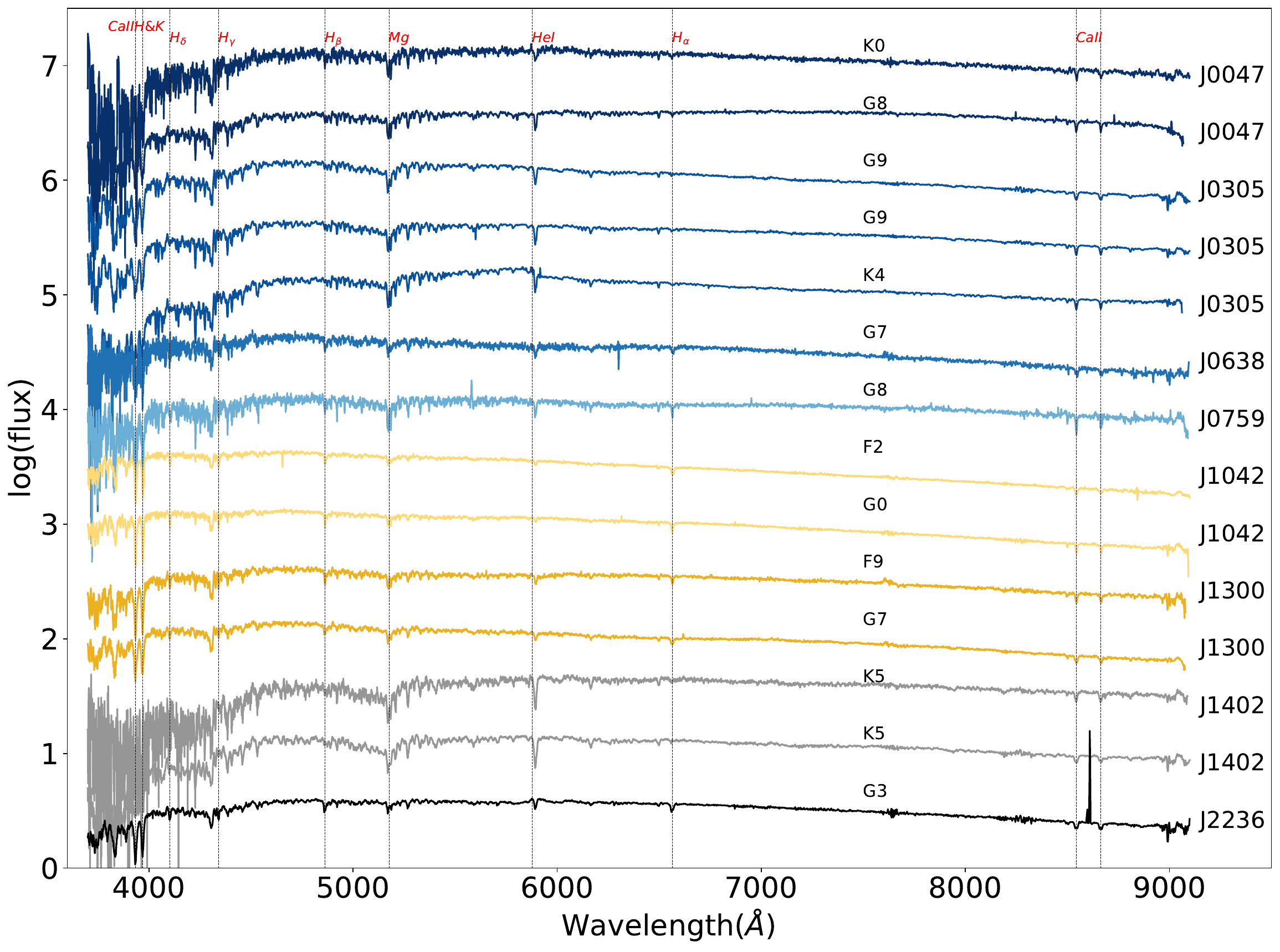}
    \caption{{The LAMOST low-resolution spectra of eight targets. The spectral types and positions of some characteristic spectral lines are labeled in the plot. Each spectrum is accompanied by the corresponding name of the target on the right side.}}
    \label{fig:LAMOST_spec}
\end{figure*}

We also cross-matched these targets with VSX \citep{Watson+etal+2006+VSX}), Gaia \citep{Gaia+etal+2016+AA+Gaia,Gaia+etal+2018+AAP+Gaia}, and the LAMOST low-resolution (LR) spectra. The information obtained is listed in Table \ref{tab:observation} and Table \ref{tab:LAMOST_LRS}. Figure \ref{fig:LAMOST_spec} shows the LR spectra, which are typical FGK-type spectra and characterized by apparent Balmer absorption lines.

Table \ref{tab:observation} provides information on these targets, including their TMTS catalog identifier (Source ID), Name (in our work), right ascension in decimal degrees (R.A.), Declination in decimal degrees (Dec.), observation date (Obs. Date), orbital period from TMTS ($P_{\rm TMTS}$), orbital period from VSX ($P_{\rm VSX}$), median magnitude from TMTS ($L_0$), absolute magnitude derived from Gaia DR2 database ($G_{\rm abs}$), color index derived from Gaia DR2 database ($B_P-R_P$), exposure times of LAMOST LR spectra ($N_{\rm LRS}$), exposure times of LAMOST MR spectra ($N_{\rm MRS}$), and VSX variability type (VSX type). Table \ref{tab:LAMOST_LRS} lists the spectral information of the ten targets from LAMOST DR10, including Name, Obs. Date, spectral type (Spec.type), effective temperature (T$_{eff}$), surface gravity (log$g$), metallicity (Fe/H), Heliocentric radial velocity (RV), and SNR of the $g$ band (SNR$_g$).

All of these targets have been identified as contact binaries or eclipsing binary candidates by the Asteroid Terrestrial-impact Last Alert System (ATLAS) \citep{Heinze+etal+2018+AJ+ATLAS} or {\em Gaia} Data Release 3 \citep{Panchal+Joshi+2021+J0305}, and their orbital periods have been determined, respectively. The periods, which were used for the phase-folded and orbital period analysis, are listed in Table \ref{tab:T0&P}. Among the ten targets, J0132, J0305, J1300 and J1402 have undergone the analysis of multiband photometric solution and studies of period. Detailed explanations of these targets are provided in the following.

(i) J0132 (V471 Cas) was discovered photographically by \cite{Hoffmeister+1966+AN+J0132}, and classified as a W UMa type binary by \cite{Gessner+Meinunger+1973+VSS+J0132}. The photometric solution was first applied to J0132 by \cite{Liu+Tan+1991+APSS+J0132}, in which the mass ratio was determined as 0.5947($\pm$0.0149) and the over-contact factor is 0.19. \cite{Kjurkchieva+etal+2019+J0132} reanalyzed the new light curves and obtained the mass ratio and fill-out factor with 0.635 and 0.078. Through the analysis of orbital period, they revised the period of J0132 to 0.400937 days and confirmed that J0132 may exist a third body with a sinusoidal period of 12.8 yr.

(ii) J0305 (NSVS 6599082) was first classified as a W UMa type binary by \cite{Hoffman+etal+2009+AJ+J0305}. \cite{Panchal+Joshi+2021+J0305} presented the photometric and spectroscopic analysis, which indicates that J0305 shows a long-term increase in orbital period, with $dp/dt$ = 1.78($\pm$1.52)$\times$10$^{-6}$ d yr$^{-1}$. The mass ratio and fill-out factor are 0.31($\pm$0.01) and 0.105, respectively, the equivalent width (EW) of H$\alpha$ is measured as 1.031$\pm$0.018 \AA.

(iii) J1300 (MM Com) was discovered as a contact binary with a period of 0.30199999 days from Robotic Optical Transient Search Experiment (ROTSE) all-sky surveys \citep{Akerlof+etal+2000+AJ+ROTSE}. \cite{Kjurkchieva+etal+2018+RAA+1300} obtained light curves in the $g'$ and $i'$ bands and derived the photometric solution. J0913 is a W-type contact binary, with a mass ratio, orbital inclination, and fill-out factor of 4.66($\pm$0.02), 80.6\degr($\pm$0.03), and 0.2388, respectively.
Then, \cite{Yang+etal+2023+J1300} analyzed new photometric and spectroscopic data for J1300, determining a mass ratio of 4.747($\pm$0.005), an orbital inclination of 79.76\degr($\pm0.16$), and a fill-out factor of 0.32($\pm$0.04). The study of O-C suggests that J0913 may be in a triple system with a period of 20.02($\pm$0.43) yr.

(iv) J1402 (EI CVn) was discovered as an eclipsing binary system by ROTSE with a period of 0.260775 days \citep{Akerlof+etal+2000+AJ+ROTSE}. Then it was carried out as a photometric orbital solution by \cite{Yang+2011+J1402}, which was found to be a W-type contact binary with a mass ratio of 0.461($\pm$0.003) and a fill-out factor of 0.21($\pm$0.07). The $O-C$ analysis indicates that the orbital period of J1402 is decreasing at a long-term rate of $dp/dt$ = -3.11($\pm$0.03)$\times$10$^{-7}$ d yr$^{-1}$. \cite{Alton+2021+AcA+J1402} obtained the new CCD photometric data of J1402 and performed a W-D analysis, resulting in a mass ratio of 0.443($\pm$0.001) and a fill-out factor of 0.15. The study of $O-C$ suggests a long-term period decrease rate of $dp/dt$ = -1.35($\pm$0.01)$\times$10$^{-7}$ d yr$^{-1}$, with a periodic modulation of 10.14$\pm$ 1.13 days.

The other six targets, J0047, J0638, J0759, J0913, J1042, and J2236, have not been systematically analyzed since their discovery.

\begin{table*}
\centering
\caption{LAMOST LR spectroscopic observation log of the ten targets.}
\label{tab:LAMOST_LRS}
\begin{tabular}{llcccccccccccccccc}
\hline\hline
N& Name & Obs. Date & Spec. type & $T_{{\rm eff}}$  & log g & Fe/H & $RV$ & $SNR_{g}$ & EW(H$\alpha$) \\
 &            &           &            & K       &  dex    &  dex      & km/s     &    & \AA     \\
\hline
1& J0047& Dec. 12, 2011   & K0&   5117.3($\pm 93.8$)  & 4.21($\pm 0.16$)  &  -0.242($\pm 0.101$)  & 12.6($\pm 6.8$)  & 14.9 &  0.772($\pm 0.091$)   \\
 &                       & Dec. 9, 2016    & G8&   5107.9($\pm 22.4$)  & 4.38($\pm 0.03$)  &  -0.158($\pm 0.019$)  &  5.8($\pm 3.0$)  & 51.1 &  0.917($\pm 0.033$)   \\
2& J0132&  -             & - &   -                   & -                 &  -                    &  -                &  -    &  -                     \\
3& J0305& Nov. 14, 2014  & G9&   4915.8($\pm 18.0$)  & 4.45($\pm 0.02$)  &  -0.425($\pm 0.013$)  & -24.1($\pm 3.0$)  & 150.1 & 0.916($\pm 0.038$)    \\
 &                       & Nov. 19, 2014  & G9&   4838.1($\pm 29.9$)  & 4.36($\pm 0.04$)  &  -0.412($\pm 0.024$)  & -23.3($\pm 4.2$)  & 93.8  & 1.086($\pm 0.054$)    \\
 &                       & Jan. 3, 2015   & K4&   4719.8($\pm 29.0$)  & 4.41($\pm 0.04$)  &  -0.496($\pm 0.022$)  & -21.8($\pm 4.6$)  & 133.7 & 0.951($\pm 0.045$)    \\
4& J0638& Mar. 9, 2012   & G7&   5518.4($\pm 90.6$)  & 4.38($\pm 0.15$)  &  -0.164($\pm 0.098$)  &  26.5($\pm 6.8$)  & 15.7  & 0.854($\pm 0.100$)     \\
5& J0759& Dec. 12, 2014  & G8&   5143.7($\pm 80.7$)  & 4.69($\pm 0.13$)  &  -0.475($\pm 0.087$)  &  -0.1($\pm 6.0$)  & 15.4  & 0.085($\pm 0.035$)    \\
6& J0913&  -             & - &   -                   & -                 &  -                    &  -                &  -    &                       \\
7& J1042& Apr. 10, 2013  & G0&   5551.9($\pm 25.3$)  & 3.90($\pm 0.04$)  &  -0.99 ($\pm 0.021$)  &  -6.2($\pm 3.4$)  & 67.6  & 0.340($\pm 0.032$)    \\
 &                       & Mar. 3, 2015   & F2&   5751.5($\pm 17.6$)  & 4.08($\pm 0.03$)  &  -0.834($\pm 0.015$)  &  -7.1($\pm 2.4$)  & 47.0  & 0.357($\pm 0.044$)    \\
8& J1300& Feb. 7, 2015   & G7&   5455.0($\pm 25.6$)  & 4.13($\pm 0.04$)  &  -0.237($\pm 0.023$)  & -18.3($\pm 3.5$)  & 41.5  & 0.357($\pm 0.044$)    \\
 &                       & May 18, 2016   & F9&   5548.1($\pm 30.4$)  & 4.17($\pm 0.04$)  &  -0.272($\pm 0.029$)  &   4.0($\pm 3.9$)  & 35.7  & 0.501($\pm 0.064$)    \\
9&J1402& Mar. 25, 2014  & K5&   4504.5($\pm 34.3$)  & 4.19($\pm 0.05$)  &   0.098($\pm 0.029$)  &  -41.0($\pm 4.6$)  & 59.6  & 0.823($\pm 0.029$)   \\
  &                      & Feb. 26, 2018  & K5&   4582.6($\pm 64.6$)  & 4.33($\pm 0.11$)  &   0.176($\pm 0.069$)  & -32.2($\pm 4.9$)  & 16.5  & 0.722($\pm 0.146$)    \\
10&J2236& Sept. 19, 2016 & G3&   5754.1($\pm 27.1$)  & 4.23($\pm 0.04$)  &  -0.193($\pm 0.021$)  & -27.2($\pm 4.1$)  & 114.6 & 0.508($\pm 0.024$)    \\
\hline
\end{tabular}
\end{table*}

\subsection{RV Measurements}

In order to obtain the RVs of these ten targets, we first used the Python package {\it laspec} \citep{Zhang+etal+2020+ApJS+laspec,Zhang+etal+2021+ApJS+RVZP} to process the LAMOST MR spectra and measured the cross-correlation functions (CCF) for each component.
In this process, \textrm{PHOENIX}
\citep{Hauschildt+1993+JQSRT+PHOENIX,Hauschildt+Baron+2006+AAP+PHONIX,Baron+Hauschildt+2007+AAP+PHOENIX} was used to construct template spectra, whose resolution was degraded to match that of the LAMOST MR spectra.
Simultaneously, since RV zero points (RVZPs) of MR spectra may vary with time and different fibers, we used the method of \cite{Zhang+etal+2021+ApJS+RVZP} to calibrate RVZPs.
Then, to determine the positions of the two peaks of the CCF, we used the Python tool {\it GaussPy} to fit the RVs of primary and secondary components, which can implement the Autonomous Gaussian Decomposition (AGD) algorithm \citep{Lindner+etal+2015+AJ+GaussPy}. The RVs obtained for all targets are listed in Table \ref{tab:RV}.

\begin{table*}
\begin{center}
\caption{Radial velocities of the ten targets.}
\label{tab:RV}
\begin{tabular}{lcrrrrlcrrrr}
\hline\hline
JD (Bary.) &  Phase  &RV$_1$&  Errors  & RV$_2$ & Errors &  JD (Bary.) &  Phase  &RV$_1$&  Errors  & RV$_2$ & Errors   \\
2400000+   &         & km s$^{-1}$& km s$^{-1}$ & km s$^{-1}$ &km s$^{-1}$  & 2400000+   &       & km s$^{-1}$& km s$^{-1}$ & km s$^{-1}$ &km s$^{-1}$ \\
\hline
        \multicolumn{12}{c}{J0047} \\
        59532.09939  & 0.70922  & 244.07  & 1.18  & -25.47  & 0.88  & 59532.12987  & 0.76762  & 250.87  & 1.80  & -34.70  & 1.10  \\
        59532.11463  & 0.82597  & 242.28  & 1.31  & -23.66  & 0.94  & ~ & ~ & ~ & ~ & ~ & ~ \\
        \multicolumn{12}{c}{J0132} \\
        58450.02562  & 0.30095  & -236.45  & 1.52  & 96.91  & 1.10  & 59130.25377  & 0.86746  & 128.40  & 0.91  & -145.35  & 0.85  \\
        58450.04197  & 0.34171  & -227.93  & 1.27  & 95.11  & 1.05  & 59130.23744  & 0.82673  & 143.63  & 0.73  & -169.31  & 0.76  \\
        58450.05823  & 0.38227  & -207.58  & 1.05  & 77.66  & 1.04  & 59130.27005  & 0.90807  & 86.75  & 1.74  & -131.80  & 1.53  \\
        \multicolumn{12}{c}{J0305} \\
        58410.17413  & 0.21550  & -206.12  & 1.35  & 104.86  & 1.04  & 58410.20672  & 0.34747  & -180.44  & 1.37  & 75.57  & 1.21  \\
        58410.19044  & 0.28153  & -204.74  & 1.32  & 99.07  & 1.06  & ~ & ~ & ~ & ~ & ~ & ~ \\
        \multicolumn{12}{c}{J0638} \\
        59544.21399  & 0.02171  & 28.63  & 0.68  & 319.60  & 2.35  & 59562.25380  & 0.92498  & 124.04  & 3.42  & -66.83  & 3.33  \\
        59544.22920  & 0.06462  & -43.74  & 5.14  & 122.90  & 5.37  & 59562.26904  & 0.96798  & 135.00  & 2.51  & 17.86  & 4.04  \\
				...  &  ...  &  ...  &  ...  &  ...  &  ...  &  ...  &  ...  &  ...  &  ...  &  ...  &  ...  \\
\hline
\end{tabular}
\end{center}
\tablecomments{
RV$_1$ and RV$_2$ represent the RVs of the primary and secondary stars, respectively.
The errors were determined through the CCF procedure for each RV data.\\
(This table is available in its entirety in machine-readable form.)}
\end{table*}

\section{O-C Analysis of Times of Minimum} \label{sec:O-C}

Since the discovery of these ten targets, orbital period analysis has been conducted only for J0132, J0305, J1300, and J1402, while the other targets remain unanalyzed.
Combining the results from some photometric surveys such as ASAS, ASAS-SN, Brno Regional Network of Observers project (BRNO)\footnote{http://var2.astro.cz/EN/brno/index.php}, CSS, SuperWASP, TESS, and ZTF, we collected as many times of minimum as possible for these ten objects. For survey data spanning long periods, we conducted phase-shifting on them, ensuring that light curves have enough data points within one period.
The O-C (observation minus calculation) time span ranges from approximately 5000 to 8500 days, with the folded periods covering data from 8 to 50 days. By dividing the minimum and maximum folded periods of each target by their respective time intervals, we find that the folded periods account for 0.1\% to 1\% of the entire O-C time interval. The errors resulting from phase shift are acceptable.
Times of minimum were calculated using the K-W method\citep{Kwee+1956+bain+KW}, which is listed in Table \ref{tab:minima}.
The obtained times of minimum are 209, 195, 253, 136, 38, 22, 175, 243, 231, and 213 for J0047, J0132, J0305, J0638, J0759, J0913, J1042, J1300, J1402, and J2236, respectively.

\begin{figure*}[!ht]
\centering
\includegraphics[width=0.97\textwidth]{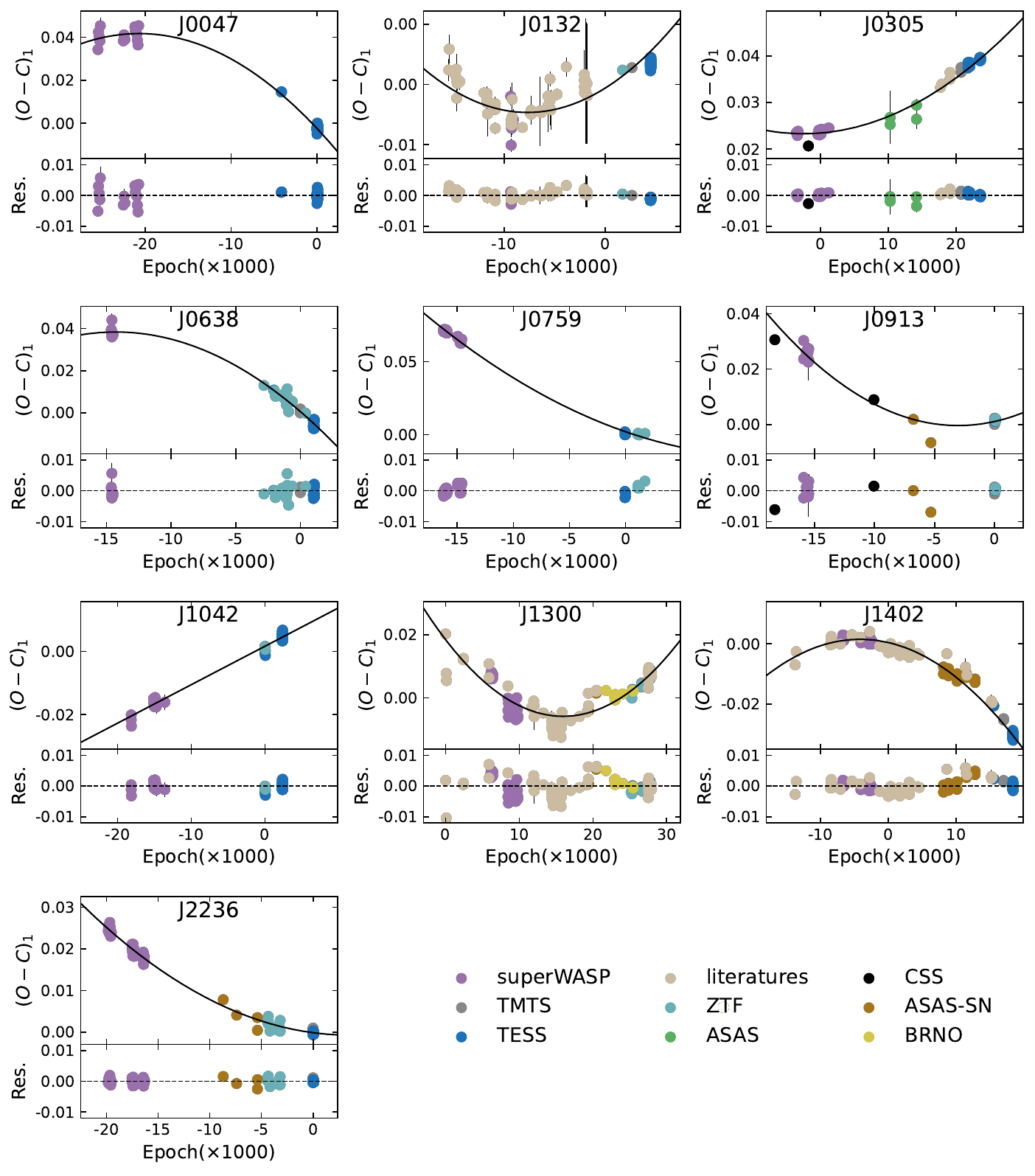}
\caption{The O-C diagram of the 10 targets presented in this paper. The top panel shows the curve $(O-C)_1$ determined by equation \ref{equ:EpochFormula}. The residuals, which remove the quadratic correction term from the $(O-C)_1$ curve, are plotted in the lower panel. The units of the $(O-C)_1$ and Residual are in days. 
The errors not given in Table \ref{tab:minima} are set to 0.001.}
\label{fig:TMTS_oc}
\end{figure*}

Then, we calculated their n$_c$ and O-C using the following formula,
\begin{eqnarray}
    {\rm T = T_0 + P \times n_c,}
\label{equ:EpochFormula}
\end{eqnarray}
with T$_0$ and period provided in Table \ref{tab:T0&P}. T is the computed moment of the observation. {\rm $n_c$} represents the number of cycles.
Then O-C was obtained by subtracting the observed times of the minima from the times of the minima calculated using the above formula.
Figure \ref{fig:TMTS_oc} shows the relationship between n$_c$ and O-C for each target.
All targets exhibit a long-term trend of increasing or decreasing periods. Therefore, we fitted their O-C with a quadratic polynomial. Polynomial coefficients were determined using the least squares method. The corrected epoch, the corrected period and the period change rate ($dp/dt$) are shown in Table \ref{tab:T0&P}. The sign of $dp/dt$ indicates whether the period is increasing (positive) or decreasing (negative) with time.

\begin{figure*}
\centering
\includegraphics[width=0.98\textwidth]{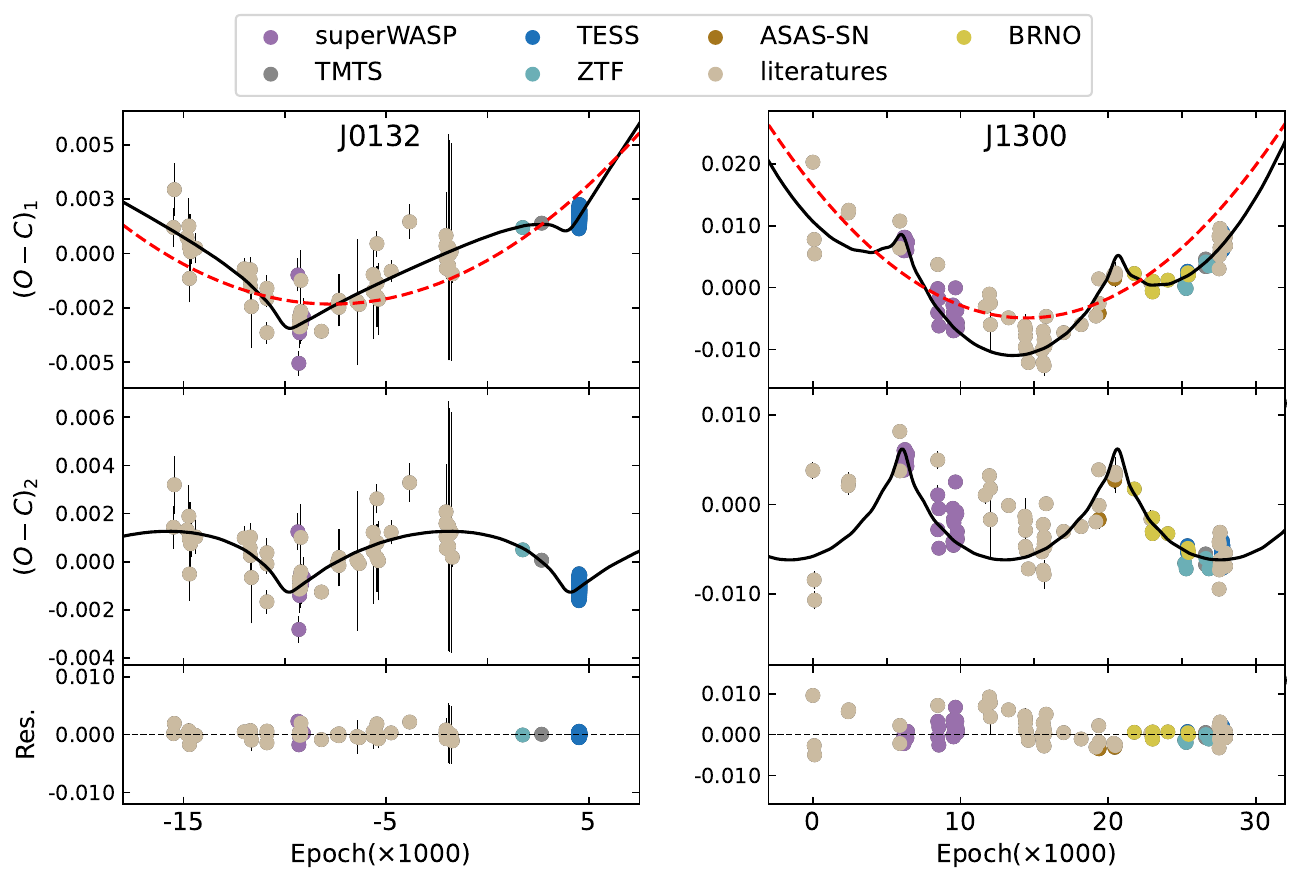}\\
\includegraphics[width=0.49\textwidth]{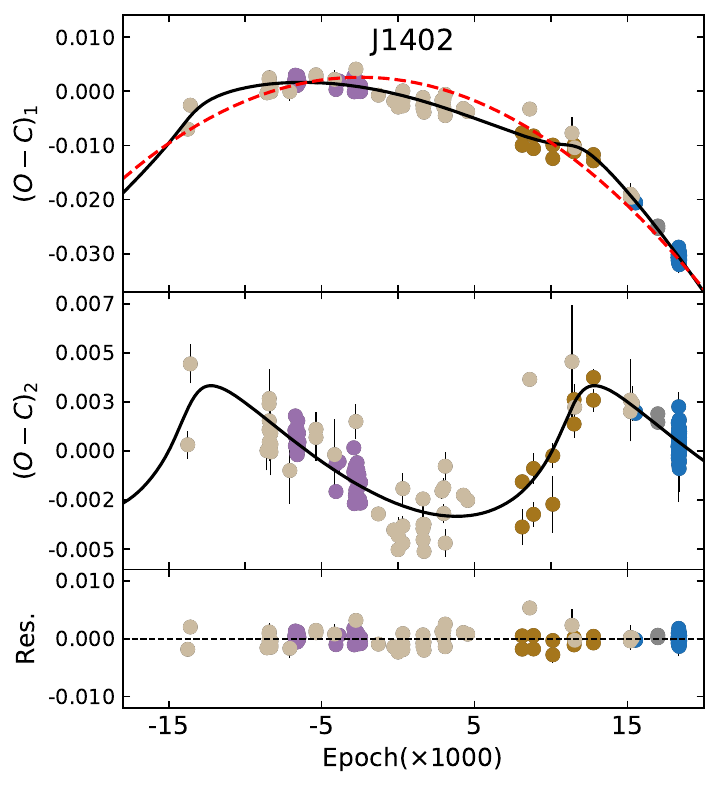}\\
\caption{The O-C diagram of J0132, J1300 and J1402 with periodic variations. The top panel shows the $(O - C)_1$ curve determined by Equation \ref{equ:EpochFormula} as in the top panel of Figure \ref{fig:TMTS_oc}. The $(O -C)_2$, which remove the quadratic correction term from the $(O - C)_1$ curve, are plotted in the middle panel. The residuals from the full ephemeris of Equation \ref{equ:ltte} are displayed in the lower panel. 
The units of $(O-C)_1$, $(O-C)_2$, and Residual are in days. The different colors of the symbols represent different data, as explained in the diagram.}
\label{fig:TMTS_oc_cyc}
\end{figure*}

From the residuals in the bottom panel of each subplot in Figure \ref{fig:TMTS_oc}, it can be seen that J0132, J1300 and J1402 exhibit significant periodic variations. This suggests that these systems may have other physical mechanisms, such as magnetic activity cycles or light time travel effect. Then, we fitted their trends of periodic variation using the following formula from \cite{Irwin+1952+ApJ+ltte}.
\begin{eqnarray}
(O - C)_1 = T_0+\Delta T_0+(P_0+\Delta P_0)E+{\frac{\beta}{2}}E^2 +A[(1-e^2)\frac{\sin(\nu+\omega)}{(1+e\cos\nu)}] \\
=T_0+\Delta T_0+(P_0+\Delta P_0)E+{\frac{\beta}{2}}E^2 
+A[\sqrt{(1-e^2 )}\sin E^*\cos\omega+\cos E^*\sin \omega  - e\sin\omega].\nonumber
\label{equ:ltte}
\end{eqnarray}
T$_0$ and P$_0$ represent the initial epoch and initial period, respectively, same as in Equation \ref{equ:EpochFormula}. $\Delta$ T$_0$ and $\Delta$ P$_0$ are used to modify the initial epoch and the period, respectively. $\beta$ is the long-term change in period, $A$ is the semiamplitude of the cyclic modulation given in days, $e$ is the eccentricity of the supposed third body, $\nu$ is the true anomaly, $\omega$ is the argument of the periastron in the plane of the orbit, and $E^*$ is the eccentric anomaly \citep{Irwin+1952+ApJ+ltte}. According to the fitting, the corresponding fitting curves are shown in Figure \ref{fig:TMTS_oc_cyc}. The fitting parameters for these three targets are listed in Table \ref{tab:oc_cyc}.
The residual panel of each subfigure in Figure \ref{fig:TMTS_oc_cyc} shows smooth variations for J0132 and J1402, except for J1300. The cyclic variation of the O-C for J0132 and J1402 may be due to the light-time effect caused by the presence of a third body. J1300 may be part of a quadruple system. The comprehensive analysis of the orbital period variation will be discussed in Section \ref{sec:discussion}.

\begin{table*}
\begin{center}
\caption{Times of minimum for the ten targets.}
\label{tab:minima}
    \begin{tabular}{cccccccccccc}
    \hline
    \noalign{\smallskip}
        BJD & Error & Method & Ref. & BJD & Error & Method & Ref. & BJD & Error & Method & Ref. \\
        2400000+ & (days) & ~ & ~
        & 2400000+ & (days) & ~ & ~ & 2400000+ & (days) & ~ & ~ \\
        \noalign{\smallskip}
        \hline
        \multicolumn{12}{c}{J0047}  \\
        53201.80826  & 0.00123  & CCD & (1) & 59859.96397  & 0.00045  & CCD & (2) & 59872.75616  & 0.00036  & CCD & (2) \\
        53219.70044  & 0.00304  & CCD & (1) & 59860.09476  & 0.00045  & CCD & (2) & 59872.88773  & 0.00039  & CCD & (2) \\
        53237.84330  & 0.00086  & CCD & (1) & 59860.22457  & 0.00037  & CCD & (2) & 59873.01750  & 0.00051  & CCD & (2) \\
        53258.59713  & 0.00272  & CCD & (1) & 59860.35601  & 0.00035  & CCD & (2) & 59873.14927  & 0.00034  & CCD & (2) \\
        53272.83309  & 0.00372  & CCD & (1) & 59860.48686  & 0.00042  & CCD & (2) & 59873.27857  & 0.00038  & CCD & (2) \\
                \hline
    \end{tabular}
    \end{center}
\textbf{References.}(1) superWASP; (2) TESS; (3) TMTS; (4) ZTF; (5) \cite{Hubscher+2005+J0132min}; (6) BBSAG124;
(7) \cite{Dvorak+2005+J0132min}; (8) \cite{Hubscher+etal+2005+J0132min};
(9) \cite{Hubscher+etal+2006+J0132min}; (10) \cite{Nelson+Robsert+2008+J0132min};
(11) \cite{Hubscher+etal+2008+J0132min}; (12) \cite{Diethelm+2009+J0132min};
(13) \cite{Diethelm+2010+J0132min}; (14) \cite{Hubscher+etal+2010+J0132J1402min};
(15) \cite{Diethelm+2011+J01322min}; (16) \cite{Hubscher+Lehmann+2012+J0132min};
(17) \cite{Diethelm+2012+J032min}; (18) JAAVSO; (19) \cite{Lampens+etal+2017+J0132min};
(20) \cite{Hubscher+2014+J0132min}; (21) \cite{Jurysek+etal+2017+J0132J1300J1402min};
(22) \cite{Hubscher+2017+J0132J1300min}; (23) ASAS; (24) CSS; (25) \cite{Panchal+Joshi+2021+J0305};
(26) ASAS-SN; (27) ROTSE; (28) BRNO; (29) The o-c gateway;
(30) \cite{Lewandowski+etal+2007+J1300min}; (31) \cite{Nelson+2007+J1300min};
(32) \cite{Nelson+2009+J1300min}; (33)\cite{Diethelm+2009+J1300J1402min}; (34) \cite{Hubscher+etal+2010+J1300min};
(35) \cite{Diethelm+2010+J1300min}; (36) \cite{Diethelm+2011+J1300J1402min};
(37) \cite{Hubscher+etal+2012+J1300min}; (38) \cite{Honkova+etal+2013+J1300J1402min};
(39) \cite{Diethelm+2012+J1300J1402min}; (40) \cite{Hubscher+etal+2013+J1300min};
(41) \cite{Nelson+2013+J1300min}; (42) \cite{Honkova+etal+2014+J1300min};
(43) \cite{Honkova+etal+2015+J1300min}; (44) \cite{Yang+etal+2023+J1300};
(45) \cite{Blattler+Diethelm+2003+J1402min}; (46) \cite{Diethelm+2005+J1402min};
(47) \cite{Diethelm+2006+J1402min}; (48) \cite{Diethelm+2007+J1402min}; (49) \cite{Nelson+2009+J1402min};
(50) \cite{Yang+2011+J1402}; (51) \cite{Hubscher+Monninger+2011+J1402min};
(52) \cite{Demircan+etal+2011+J1402min}; (53) \cite{Pagel+2018+J1402min}; (54) \cite{Lehky+etal+2021+J1402min}; \\
(This table is available in its entirety in machine-readable form.)
\end{table*}

\begin{table*}
\begin{center}
\tiny
\caption{Ephemerides and period of the ten targets.
\label{tab:T0&P}
}
    \begin{tabular}{lccccccrrr}
    \hline\hline
        Name & T${_0}$(BJD) & Epoch Ref. & Period & Period Ref. & Corrected Epoch & Corrected Period & ${\rm dp/dt}$                 & ${\rm dM_1/dt} $ \\
             &              &            & days   &             &                 & days             & $\times$ 10$^{-8}$ d yr$^{-1}$ & $\times$ 10$^{-7}$${\rm M_{\odot}}$ yr$^{-1}$   \\
        \hline
        J0047 & 2459859.052923  & (1) & 0.261077  & (2) & 2459859.050409($\pm$ 0.000017) & 0.261076($\pm$ 0.000001)  & -28.84($\pm$ 0.01)  &  0.86($\pm$ 0.01) \\
        J0132 & 2458080.255808  & (3) & 0.400937  & (3) & 2458080.255492($\pm$ 0.000398) & 0.400938($\pm$ 0.000001) & 6.22($\pm$ 0.76)  & -1.44($\pm$ 0.18) \\
        J0305 & 2454085.436000  & (4) & 0.246983  & (4) & 2454085.459429($\pm$ 0.000001) & 0.246983($\pm$ 0.000001) & 7.00($\pm$ 0.01)  & -0.35($\pm$ 0.01) \\
        J0638 & 2459580.005375  & (1) & 0.354390  & (5) & 2459580.005990($\pm$ 0.000036) & 0.354385($\pm$ 0.000001) & -38.03($\pm$ 0.07)  & 7.86($\pm$ 0.01) \\
        J0759 & 2458864.321507  & (1) & 0.295950  & (5) & 2458864.323469($\pm$ 0.000011) & 0.295947($\pm$ 0.000001) & 25.49($\pm$ 0.02)  & -0.60($\pm$ 0.01) \\
        J0913 & 2458867.206314  & (1) & 0.278684  & (2) & 2458867.207365($\pm$ 0.000037) & 0.278685($\pm$ 0.000001) & 41.79($\pm$ 0.12)  & -25.85($\pm$ 0.07) \\
        J1042 & 2458868.309159  & (1) & 0.316037  & (2) & 2458868.310805($\pm$ 0.000063) & 0.316038($\pm$ 0.000001) & -  & - \\
        J1300 & 2451277.840483  & (6) & 0.301990  & (6) & 2451277.856993($\pm$ 0.001430) & 0.301987($\pm$ 0.000001) & 24.73($\pm$ 1.20)  & -0.54($\pm$ 0.03)\\
        J1402 & 2454891.208272  & (7) & 0.260767  & (7) & 2454891.210399($\pm$ 0.000284) & 0.260767($\pm$ 0.000001) & -21.74($\pm$ 0.97)  & 1.49($\pm$ 0.07)\\
        J2236 & 2459830.048809  & (1) & 0.335739  & (2) & 2459830.048690($\pm$ 0.000004) & 0.335739($\pm$ 0.000001) & 10.14($\pm$ 0.01)  & -1.17($\pm$ 0.01)\\
        \hline
    \end{tabular}
    \end{center}
    \textbf{References.}(1) TMTS; (2) \cite{Heinze+etal+2018+AJ+ATLAS}; (3) \cite{Kjurkchieva+etal+2019+J0132}; (4) \cite{Panchal+Joshi+2021+J0305};
    (5) \cite{Sun+etal+2020+ApJS}; (6) \cite{Yang+etal+2023+J1300}; (7) \cite{Yang+2011+J1402}.
\end{table*}

\begin{table*}
\begin{center}
    \caption{The fitted periodic variation parameters of O-C for J0132, J1300 and J1402.}
    \label{tab:oc_cyc}
    \begin{tabular*}{18cm}{@{\extracolsep{\fill}}cccc}
    \hline\hline
        Parameters             & J0132                  & J1300                  & J1402                  \\
        \hline
        A(day)                 & 0.00134 $\pm$ 0.00068  & 0.00628 $\pm$ 0.00044  & 0.00382 $\pm$ 0.00040  \\
        $e$                    & 0.752   $\pm$ 0.376    & 0.857   $\pm$ 0.118    & 0.619   $\pm$ 0.117    \\
        $\omega (deg)$         & 244.4   $\pm$ 41.0     & 99.4    $\pm$ 8.6      & 37.8    $\pm$ 9.6      \\
        P$_3(yr)$              & 15.3    $\pm$ 0.5      & 12.1    $\pm$ 0.2      & 17.9    $\pm$ 0.5      \\
        T$_3$                  & 2427079 $\pm$ 81664    & 2444564 $\pm$ 47358    & 2445689 $\pm$ 63787    \\
                               &                        &                        &                        \\
        $a_{12}^{'}\sin i^{'}$ & 0.232   $\pm$ 0.066    & 1.088   $\pm$ 0.076    & 0.662   $\pm$ 0.069    \\
        $f(m)$($M_\odot$)      & 0.00005 $\pm$ 0.00005  & 0.00880 $\pm$ 0.00185  & 0.00090 $\pm$ 0.00028  \\
        $M_3$($M_\odot$)       & 0.066   $\pm$ 0.032    & 0.312   $\pm$ 0.041    & 0.134   $\pm$ 0.025    \\
        $a_3$($R_\odot$)       & 9.35    $\pm$ 5.23     & 6.26    $\pm$ 0.93     & 8.77    $\pm$ 1.87     \\
        Spec. type              & -                      & M3                     & M5.5                   \\
        $l_3$($L_\odot$)       & -                      & 0.016                  & 0.002                  \\
        $l_3/l$(\%)            & -                      & 1.663                  & 0.512                  \\
        $\Delta Q_1$(g cm$^2$) & 1.28$\times 10^{49}$   & 1.17$\times 10^{49}$ & 3.23$\times 10^{48}$     \\
        $\Delta Q_2$(g cm$^2$)  & 1.90$\times 10^{49}$   & 5.49$\times 10^{49}$ & 5.22$\times 10^{48}$     \\
        \hline
    \end{tabular*}
    \end{center}
\end{table*}

\section{The Investigation of Photometry and Spectroscopy} \label{sec:W-D}

\subsection{W-D Analysis}

We used the 2013 version of the W-D method to analyze the TMTS light curves and the RVs of these ten systems to obtain their orbital and absolute parameters. The initial epoch and period used to convert BJD to phase of each target are the corrected epoch and period in Table \ref{tab:T0&P}. During the W-D analysis, we analyzed every light curve with RV for each target simultaneously.

For targets with LAMOST LR spectra, we used the temperature or the average temperature of multiple spectra provided by LAMOST as the effective temperature of the primary star ($T_1$, the hotter component) during the W-D process. 
For targets without LAMOST LR spectra, we examined their LAMOST MR spectra. Since the MR spectra of J0132 did not yield a temperature and the temperature uncertainties for J0913 exceeded 50 K, the temperatures provided by Gaia were adopted as $T_1$ for these two targets.
The specific values $T_1$ are listed in Table \ref{tab:wd_solution}. Then we fixed the effective temperature of the primary ($T_1$) and adjusted that for the secondary ($T_2$, the cooler component). The gravity-darkening and bolometric albedo coefficients were set to $g_{1,2}$ = 0.32 and $A_{1,2}$ = 0.5 following \cite{Lucy+ZAP+1967+65+89+WD+Gravity} and \cite{Rucinski+1969+AcA+WD+Gravity}. The bolometric limb-darkening and bandpass limb-darkening coefficients were internally computed, which were obtained from \cite{vanHamme+1993+AJ+WD+LimbDarkening}, and the limb-darkening law is the square root law. The weights of the light curves and the radial velocities were set according to the observation errors, using the reciprocal of the square of the error. During the modeling, the fixed parameters were as follows: the efffective temperatures of the primary star $T_1$, the orbital period $P$ and the orbital eccentricity $e$. 
The orbital eccentricity $e$ was fixed to zero, consistent with the assumption that contact binaries typically have circular orbits due to strong tidal interactions.
The adjustable parameters were as follows: the orbital semimajor axis $a$, the systemic radial velocity $V_{\gamma}$, the orbital inclination $i$, the mass ratio $q=M_2/M_1$, the effective temperature of the secondary star $T_2$, the monochromatic luminosity of the primary star $L_1$ and the dimensionless potential of the two components $\Omega_1=\Omega_2$.
The fill-out factor was calculated by $f = (\Omega - \Omega_{in})/(\Omega_{out} - \Omega_{in})$, where $\Omega$, $\Omega_{in}$ and $\Omega_{out}$ represent potentials for the common photospere, the inner and outer contact surfaces, respectively.
When $f$ = 0, the two components just fill the inner contact surface and begin to contact, when $f$ = 1, the two components have already filled the outer contact surface.
After running the automatic iteration, the convergence criterion was established as the condition in which the standard error of the adjustable parameters is greater than their respective correction values, and the difference between the input mean residual and the predicted mean residual does not exceed 1\%.

From their TMTS light curves, it can be seen that the two maxima at phases 0.25 and 0.75 exhibit some difference, a phenomenon known as the O'Connell effect, which is typically attributed to the magnetic activity of the components \citep[e.g.,][]{Guo+etal+2020+RAA+OConnell,Li+etal+2021+ApJ+OConnell,Papageorgiou+etal+2023+AJ+OConnell,Ceki+etal+2024+MNRAS+OConnell}.
Meanwhile, Doppler-boosting also contributes to the observed flux variations\citep{Loeb+Gaudi+ApJL,Zucker+etal+2007+ApJ+Beaming}.
Therefore, a dark spot or a hot spot model was employed during the W-D analysis to obtain a better fit first.
All targets used models with a dark spot added to either the primary or the secondary star. The latitude were both fixed at 90\degr, because the most reliable parameter of a spot is the longitude of the spot, which can be used to study the evolution of the spot and stellar activity cycles \citep{Eker+1999+TJPh+spot,Berdyugina+2005+LRSP+spot}.
Subsequently, we calculated the observed magnitude variations, $\Delta$m ($m_{0.25}$-$m_{0.75}$), in phases 0.25 and 0.75 from the light curves derived from the W-D fitting and listed in Table \ref{tab:wd_solution}. For each target, considering the temperatures ($T_{eff}$) of the components and the observational band ($\lambda$$\approx$650nm), the beaming factors were determined\citep{Zucker+etal+2007+ApJ+Beaming}. Using the radial velocities of each component at phases 0.25 and 0.75, we calculated the beaming-magnitude variations for each target. All of these values are listed in Table \ref{tab:wd_solution}. The observed magnitude variations are on the order of percent of a magnitude, while the beaming-magnitude variations are on the order of thousandths. Therefore, Doppler boosting is unlikely to explain the O'Connell effect. The spot model was adopted to account for the O'Connell effect.

\begin{figure*}
\includegraphics[width=0.95\textwidth]{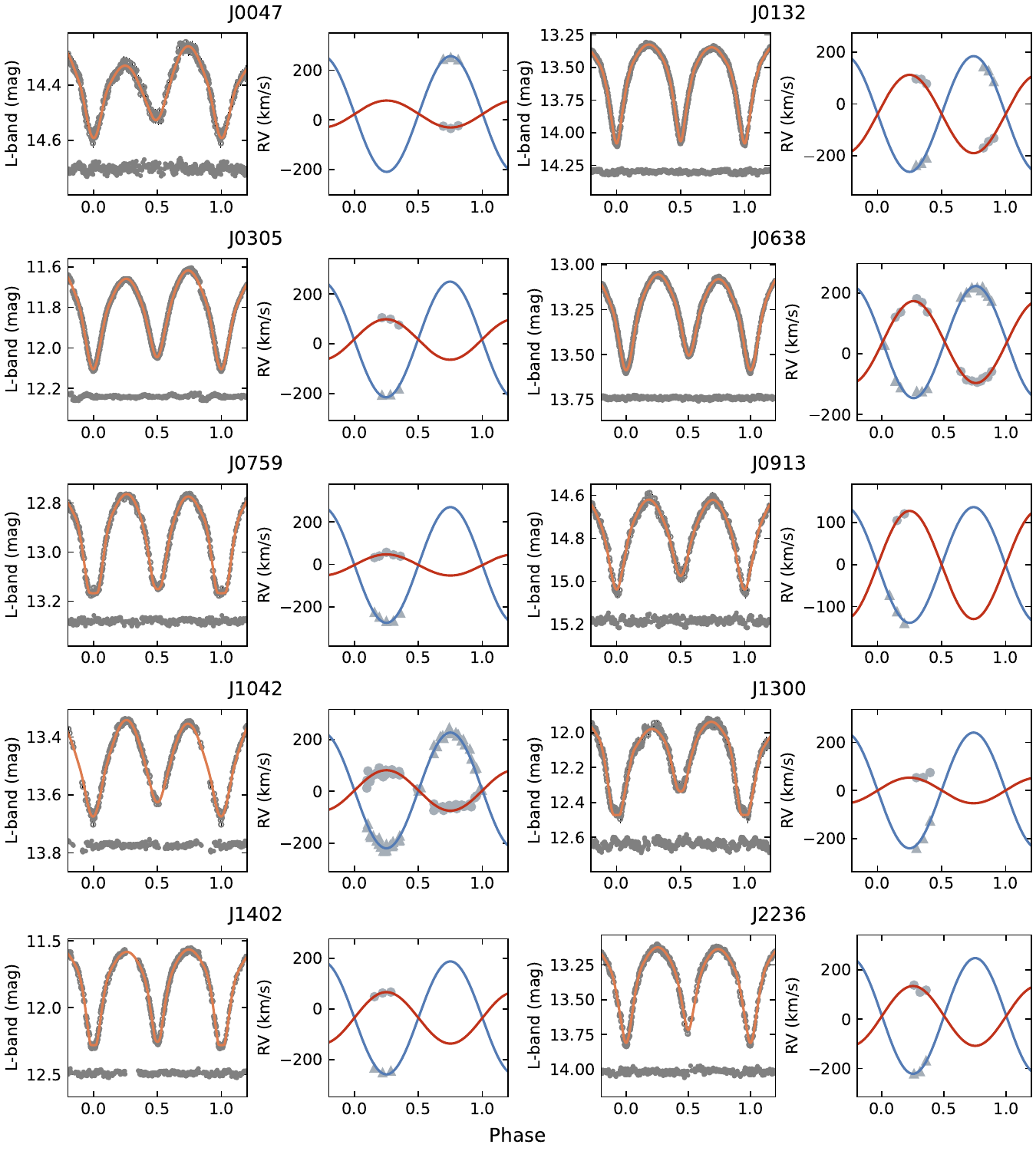}
\caption{Left panel: Theoretical light curves (orange solid line) fitted by W-D code compared to all available TMTS observations (grey dots) for the ten targets. One dark spot is added to the primary or secondary star for each target. Right panel: RVs and fitted curves of W-D code. The grey filled triangles and circles represent the observed primary (hotter) and secondary (cooler) component, respectively. The blue and red solid lines represent the best-fit RV curves.}
\label{fig:TMTS_wd}
\end{figure*}

\begin{figure*}
\centering
\includegraphics[width=0.99\textwidth]{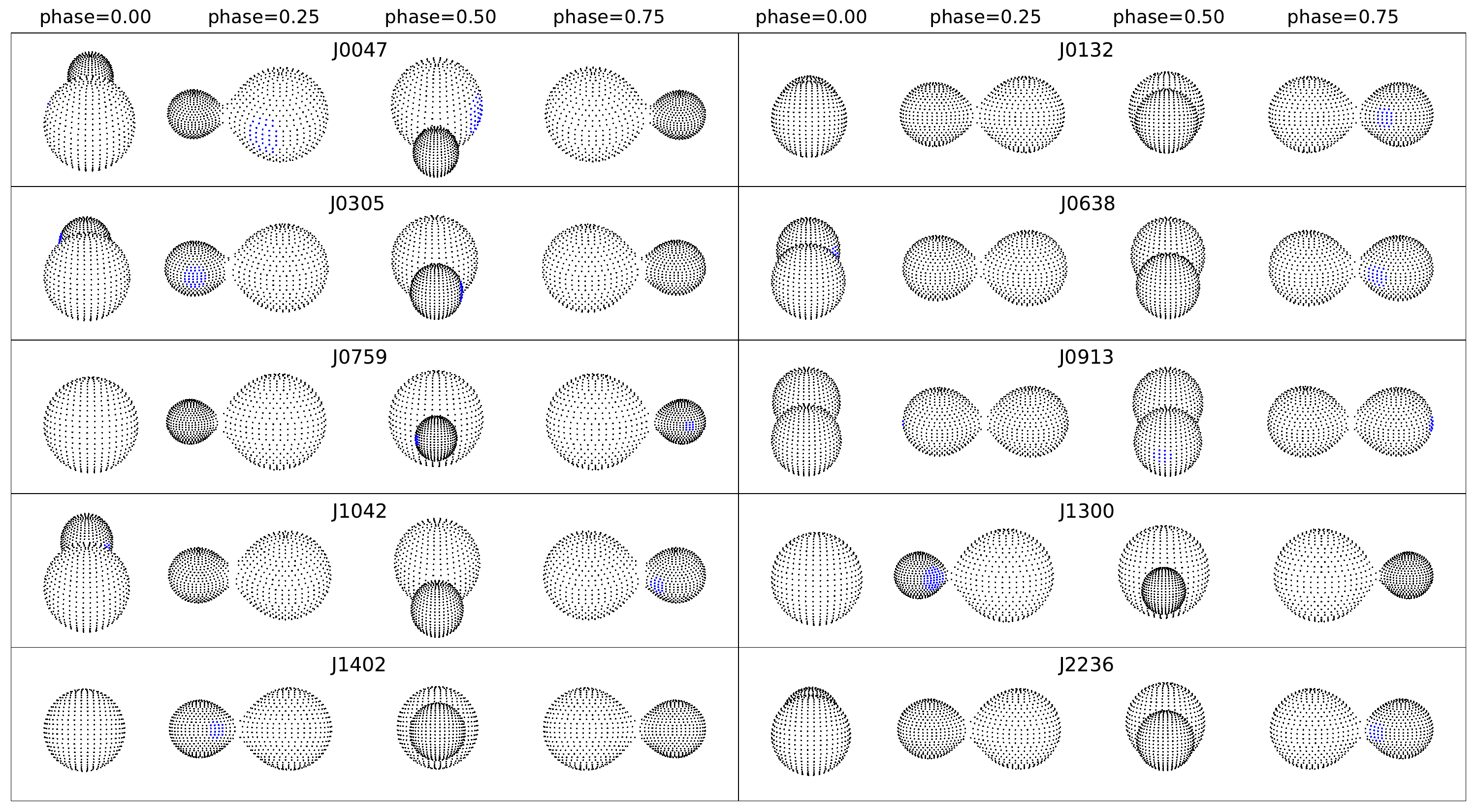}
\caption{Geometrical structures of ten targets at phase 0, 0.25, 0.5, and 0.75. The areas marked with blue cross symbols represent the added dark spots on the components.}
\label{fig:TMTS_structure}
\end{figure*}

The orbital and spot parameters determined by the light curves and the radial velocities are all tabulated in Table \ref{tab:wd_solution}. The uncertainties of the adjustable parameters are only internal uncertainties of the final step of the light curve analysis determined by W-D code. 
To account for the uncertainty of $T_1$ derived from spectroscopy to $T_2$, we conducted the following discussion. Based on photometric data, the W-D code can precisely determine $T_2/T_1$. Therefore, assuming $T_1$ is fixed, the uncertainty of $T_2$ derived from W-D can be used to determine the uncertainty of $T_2/T_1$. Subsequently, combining the uncertainties of $T_2/T_1$ and $T_1$, the accurate uncertainty of $T_2$ can be obtained. The final uncertainties of $T_1$ and $T_2$ are also listed in Table \ref{tab:wd_solution}.
Theoretical light curves and radial velocities for the ten targets are shown in Figure \ref{fig:TMTS_wd}. The smooth residuals of the light curves indicate that the model with the added spot resulted in a good fit for all ten targets. Figure \ref{fig:TMTS_structure} shows the geometric structures of each target, we can clearly see the location and angular radius of the spots. Table \ref{tab:wd_solution} also provides absolute physical parameters from the W-D process, with detailed descriptions presented in Section \ref{sec:discussion}.

\begin{longrotatetable}
\begin{deluxetable*}{ccccccccccc}
\tablecaption{Photometric solutions and absolute parameters of the ten targets \label{tab:wd_solution}}
\tablewidth{700pt}
\tablehead{
\colhead{Parameter} & \colhead{J0047} & \colhead{J0132} & \colhead{J0305} & \colhead{J0638}  &  \colhead{J0759}  & 
                      \colhead{J0913} & \colhead{J1042} & \colhead{J1300} & \colhead{J1402}  &  \colhead{J2236} 
} 
\startdata
      T$_1(K)$                 &   5113            &  6244              &  4822             &  5518                &   5144              &  5523              &  5652              &  5502              &  4544              &  5754               \\
      T$_2(K)$                 &   4610 $\pm$19    &  6165  $\pm$6      &  4641  $\pm$7     &  5263   $\pm$6       &   4947   $\pm$9     &  5142   $\pm$22    &  5313   $\pm$23    &  4970   $\pm$15    &  4408   $\pm$23    &  5503   $\pm$16     \\
      T$_1$-$\sigma$(K)    &   58              &     158            &           26      &             26       &           81        &            24      &             21     &             28     &              49    &            27       \\
      T$_2$-$\sigma$(K)     &   56              &             156    &         26          &           25           &             78        &           31        &           31        &        30           &            48        &            30        \\
      $q(M_2/M_1)$             &   4.29 $\pm$0.05  &  1.48  $\pm$0.02   &  2.85  $\pm$0.01  &  1.37   $\pm$0.01    &   5.47   $\pm$0.04  &  1.07   $\pm$0.01  &  2.86   $\pm$0.05  &  4.51   $\pm$0.04  &  2.19   $\pm$0.05  &  1.93   $\pm$0.07   \\
      $i(deg)$                 &   61.5 $\pm$0.2   &  83.5  $\pm$0.1    &  70.9  $\pm$0.1   &  71.6   $\pm$0.1     &   78.2   $\pm$0.3   &  66.9   $\pm$0.2   &  62.2   $\pm$0.3   &  78.7   $\pm$0.7   &  87.6   $\pm$0.3   &  79.5   $\pm$0.3    \\
      V$_{\gamma}$(km/s)       &   23.5 $\pm$2.1   &  -38.7 $\pm$3.1    &  17.4  $\pm$3.6   &  38.4   $\pm$1.3     &   -2.8   $\pm$2.2   &  -1.1   $\pm$1.5   &  3.5    $\pm$1.4   &  1.0    $\pm$4.6   &  -35.5  $\pm$1.4   &  13.2   $\pm$6.0    \\
      $\Omega_{in}$            &   8.28            &  4.49              &  6.42             &  4.33                &   9.74              &  3.85              &  6.43              &  8.56              &  5.52              &  5.16               \\
      $\Omega_{out}$           &   7.65            &  3.92              &  5.80             &  3.76                &   9.10              &  3.30              &  5.81              &  7.92              &  4.92              &  4.56               \\
      $\Omega_1=\Omega_2$      &   8.13 $\pm$0.01  &  4.42  $\pm$0.03   &  6.26  $\pm$0.05  &  4.26   $\pm$0.01    &   9.75   $\pm$0.04  &  3.83   $\pm$0.02  &  6.26   $\pm$0.07  &  8.52   $\pm$0.05  &  5.44   $\pm$0.07  &  5.08   $\pm$0.10   \\
      $L_{1L}/L_L$             &   0.319$\pm$0.003 &  0.422 $\pm$0.001  &  0.328 $\pm$0.002 &  0.485  $\pm$0.001   &   0.210  $\pm$0.001 &  0.567  $\pm$0.002 &  0.346  $\pm$0.004 &  0.294  $\pm$0.002 &  0.370  $\pm$0.004 &  0.402  $\pm$0.005  \\
      $r_1$                    &   0.273$\pm$0.000 &  0.354 $\pm$0.000  &  0.307 $\pm$0.000 &  0.360  $\pm$0.000   &   0.250  $\pm$0.000 &  0.377  $\pm$0.000 &  0.308  $\pm$0.000 &  0.259  $\pm$0.000 &  0.320  $\pm$0.000 &  0.330  $\pm$0.001  \\
      $r_2$                    &   0.520$\pm$0.001 &  0.425 $\pm$0.002  &  0.490 $\pm$0.002 &  0.416  $\pm$0.001   &   0.537  $\pm$0.001 &  0.387  $\pm$0.001 &  0.487  $\pm$0.003 &  0.516  $\pm$0.002 &  0.457  $\pm$0.003 &  0.448  $\pm$0.006  \\
      $f$                      &   24.1 $\pm$2.3   &  12.8  $\pm$4.4    &  26.4  $\pm$8.1   &  14.5   $\pm$2.3     &   14.5   $\pm$5.9   &  2.6    $\pm$3.0   &  25.7   $\pm$10.7  &  6.2    $\pm$8.6   &  13.4   $\pm$10.7  &  12.6   $\pm$16.7   \\
      spot                     &   Star 2          &  Star 1            &  Star 1           &  Star 1              &   Star 1            &  Star 1            &  Star 1            &  Star 1            &  Star 1            &  Star 1             \\
      Latitude(\degr)       &   90.0            &  90.0              &  90.0             &  90.0                &   90.0              &  90.0              &  90.0              &  90.0              &  90.0              &  90.0               \\
      Longitude(\degr)      &   70.2  $\pm$1.7  &  60.0  $\pm$3.2    &  268.9 $\pm$2.9   &  48.5   $\pm$2.1     &   110.9  $\pm$8.6   &  170.8  $\pm$5.5   &  48.3   $\pm$6.7   &  309.9  $\pm$4.2   &  303.9  $\pm$6.7   &  35.0   $\pm$6.4    \\
      Angular radius(\degr) &   19.0  $\pm$0.3  &  14.1  $\pm$0.3    &  19.7  $\pm$0.3   &  14.8   $\pm$0.3     &   11.9   $\pm$1.0   &  9.3    $\pm$3.4   &  12.9   $\pm$0.6   &  24.2   $\pm$0.6   &  13.6   $\pm$0.6   &  12.6   $\pm$0.8    \\
      T-factor                 &   0.66  $\pm$0.04 &  0.72  $\pm$0.02   &  0.70  $\pm$0.02  &  0.70   $\pm$0.02    &   0.70   $\pm$0.07  &  0.80   $\pm$0.15  &  0.70   $\pm$0.05  &  0.70   $\pm$0.03  &  0.80   $\pm$0.05  &  0.78   $\pm$0.04   \\
      $\Sigma (O-C)^2$         &   6.0$\times$ 10$^{-9}$   &  8.8$\times$ 10$^{-9}$    &  4.0$\times$ 10$^{-8}$  &  8.9$\times$ 10$^{-9}$    &   1.7$\times$ 10$^{-8}$   &  5.1$\times$ 10$^{-9}$  &  8.4$\times$ 10$^{-9}$  &  9.6$\times$ 10$^{-8}$  &  7.1$\times$ 10$^{-8}$ &  1.6$\times$ 10$^{-8}$ \\
      $\Delta$m(observed)     & 0.069 & -0.022 & 0.040 & -0.025 & -0.010 & -0.001 & -0.013 & 0.046 & 0.018 & -0.013  \\ 
      b-factor$_1$      & 4.3 & 3.6 & 4.6 & 4.0 & 4.3 & 4.0 & 3.9 & 4.0 & 4.9 & 3.9 \\ 
      b-factor$_2$             & 4.8 & 3.6 & 4.8 & 4.2 & 4.5 & 4.3 & 4.2 & 4.5 & 5.0 & 4.0 \\
      $\Delta$m(Doppler-boosting) & -0.005 & -0.002 & -0.005 & -0.001 & -0.007 & 0.000 & -0.004 & -0.005 & -0.004 & -0.003 \\    
      \hline
      \multicolumn{11}{c}{Absolute parameters}\\
      \hline
      $a(R_{\odot})$           &   1.69  $\pm$0.03  &  2.98  $\pm$0.02   &  1.62  $\pm$0.04  &  2.35   $\pm$0.02    &   1.93   $\pm$0.02  &  1.60   $\pm$0.01  &  2.21   $\pm$0.02  &  1.79   $\pm$0.07  &  1.68   $\pm$0.02  &  2.40   $\pm$0.06   \\
      $M_1(M_{\odot})$         &   0.179 $\pm$0.012 &  0.897 $\pm$0.022  &  0.244 $\pm$0.019 &  0.590  $\pm$0.020   &   0.170  $\pm$0.007 &  0.285  $\pm$0.006 &  0.338  $\pm$0.017 &  0.153  $\pm$0.018 &  0.291  $\pm$0.017 &  0.561  $\pm$0.059  \\
      $M_2(M_{\odot})$         &   0.768 $\pm$0.060 &  1.325 $\pm$0.047  &  0.695 $\pm$0.064 &  0.807  $\pm$0.033   &   0.926  $\pm$0.047 &  0.422  $\pm$0.015 &  0.967  $\pm$0.065 &  0.692  $\pm$0.089 &  0.638  $\pm$0.065 &  1.084  $\pm$0.155  \\
      $R_1(R_{\odot})$         &   0.464 $\pm$0.009 &  1.065 $\pm$0.007  &  0.502 $\pm$0.012 &  0.855  $\pm$0.009   &   0.484  $\pm$0.006 &  0.568  $\pm$0.003 &  0.661  $\pm$0.008 &  0.468  $\pm$0.018 &  0.541  $\pm$0.008 &  0.799  $\pm$0.023  \\
      $R_2(R_{\odot})$         &   0.880 $\pm$0.018 &  1.269 $\pm$0.012  &  0.794 $\pm$0.022 &  0.984  $\pm$0.012   &   1.034  $\pm$0.016 &  0.678  $\pm$0.006 &  1.046  $\pm$0.018 &  0.926  $\pm$0.035 &  0.770  $\pm$0.018 &  1.075  $\pm$0.042  \\
      $L_1(L_{\odot})$         &   0.132 $\pm$0.005 &  1.556 $\pm$0.022  &  0.123 $\pm$0.006 &  0.611  $\pm$0.013   &   0.148  $\pm$0.004 &  0.270  $\pm$0.003 &  0.402  $\pm$0.010 &  0.181  $\pm$0.014 &  0.113  $\pm$0.010 &  0.631  $\pm$0.036  \\
      $L_2(L_{\odot})$         &   0.316 $\pm$0.013 &  2.098 $\pm$0.041  &  0.264 $\pm$0.015 &  0.670  $\pm$0.016   &   0.578  $\pm$0.018 &  0.290  $\pm$0.006 &  0.787  $\pm$0.027 &  0.472  $\pm$0.036 &  0.202  $\pm$0.027 &  0.955  $\pm$0.075  \\
      log $J_{orb}$                &   51.045$\pm$0.018          &           51.921$\pm$0.003  &   51.129$\pm$0.020         &            51.572$\pm$0.005   &           51.010$\pm$0.010   &            51.053$\pm$0.007 &         51.403$\pm$0.010    &   50.970$\pm$0.034          &    51.179$\pm$0.017         &       51.647$\pm$0.016      \\
\enddata
\end{deluxetable*}
\end{longrotatetable}

\subsection{Spectroscopic Investigation}

When the photosphere or chromosphere of a star exhibits magnetic activity, the atmosphere of the star is heated non-thermally by the magnetic field.
The spectral subtraction technique is usually used to investigate chromospheric activity \citep{Barden+1985+ApJ+SpecSub}. The activity is quantified by the equivalent width (EW) of the emission line, such as the Balmer series (H$\alpha$, H$\beta$, H$\gamma$) and the Ca II infrared triplet (IRT), which serve as useful indicators of chromospheric activity in many contact binaries \citep{Pavlenko+etal+2018+AA+SpecSub,Li+etal+2022+AJ+SpecSub,Liu+etal+2023+MN+SpecSub,Li+etal+2024+MN+SpecSub}.

\begin{figure*}
\centering
\includegraphics[width=0.99\textwidth]{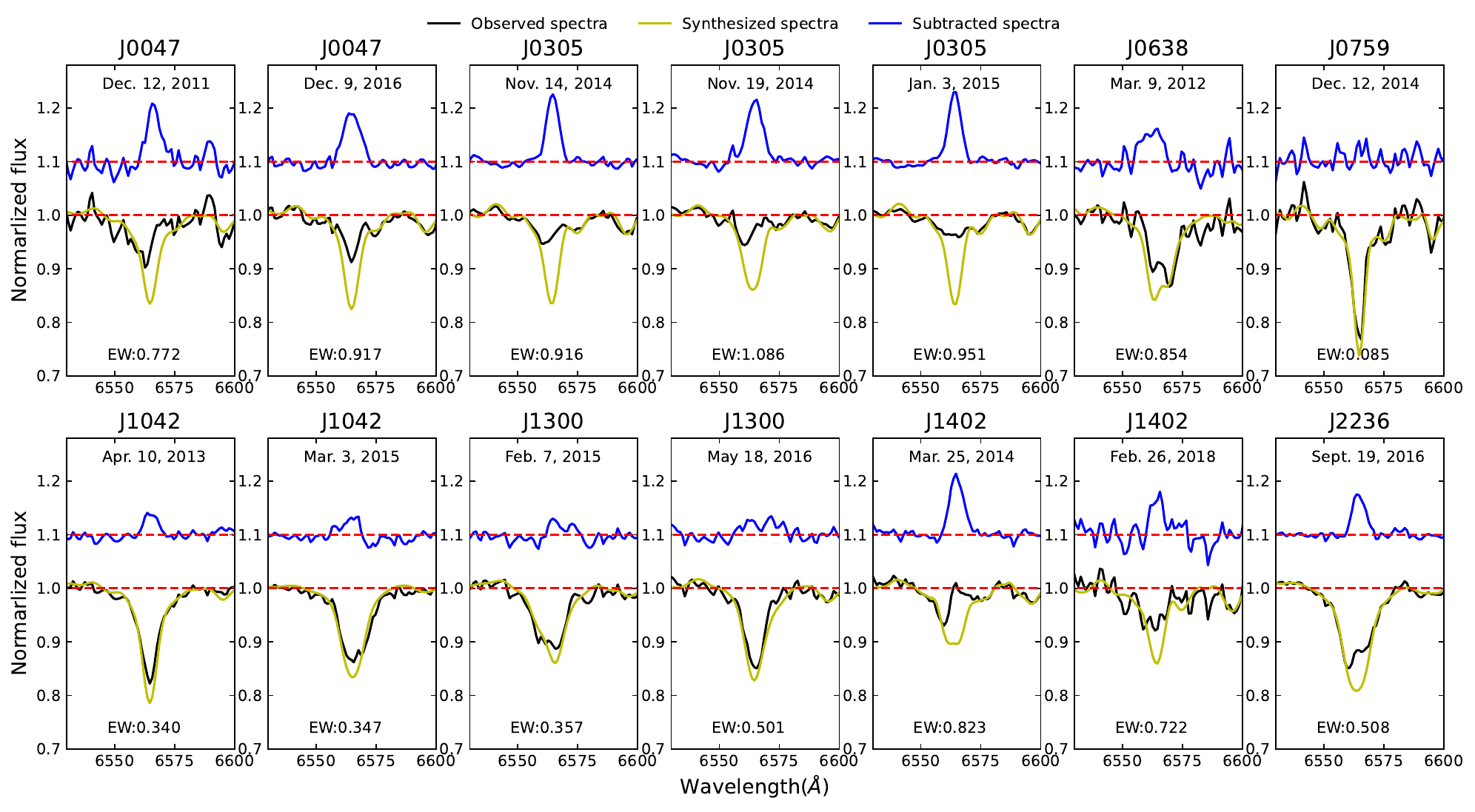}\\
\caption{The H$\alpha$ region of LAMOST LR spectra (black line), synthetic spectra (yellow line) and subtracted spectra (blue line) for the eight targets. The EWs of H$\alpha$ are shown in each panel for each spectra, with units in \AA.}
\label{fig:subspec}
\end{figure*}

We analyzed the LAMOST LR spectra for these ten targets to assess their chromospheric activities. Excluding J0132 and J0913, there are a total of 14 spectra for 8 targets. The analysis process follows three fundamental steps.
Firstly, based on the temperatures of the two components obtained from the W-D solution for each target, we selected template spectra for the primary and secondary stars from the radial velocity standard star catalog of \cite{Huang+etal+2018+AJ+Standard}. The temperatures of the inactive template spectra were chosen to closely match the components of binary, with a difference not exceeding 200 K.
Secondly, we downloaded these template spectra from LAMOST DR10, removing the cosmic rays and normalizing the spectra.
Thirdly, the Fortran code STARMOD was used to construct the synthetic spectra of the binaries, taking into account the radial velocity, the rotationally broadening, and the relative weight of the two components.
The values for the radial velocity and rotationally broadening were referenced from the LAMOST LR spectra, while the weights of the two components were derived from the W-D solution. These parameters were also treated as free during the STARMOD process.
Finally, the subtracted spectra between the LAMOST observed spectra and the synthetic spectra were obtained and displayed in Figure \ref{fig:subspec}, showing only the H$\alpha$ region.

The blue lines in Figure \ref{fig:subspec} represent the subtracted spectra, which show the varying intensities of the emission lines. We used the Python package {\it PySpecKit} to calculate the EWs of the H$\alpha$ emission lines, which are listed in Table \ref{tab:LAMOST_LRS}. In previous studies, stars with EW of H$\alpha$ greater than 0.75\AA ~were considered magnetically active \citep{West+etal+2011+AJ+magnetic,West+etal+2015+ApJ+magnetic}. Therefore, we infer that J0047, J0305, J0638, and J1402 exhibit strong magnetic activities.

\section{Discussion} \label{sec:discussion}

\subsection{Orbital and Absolute Physical Parameters}
According to the W-D code DC and LC program, the orbital and absolute physical parameters of the ten targets were determined, including the mass ratio $q$, inclination $i$, fill-out factor $f$, semi-major axis $a$, masses of the two components ($M_1$, $M_2$), radii of the two components ($R_1$, $R_2$). The luminosities of two components ($L_1$, $L_2$) are determined by the Stefan-Boltzmann law. All orbital and absolute physical parameters are tabulated in Table \ref{tab:wd_solution}. The RV data for these ten targets have been obtained for the first time. Note that four of them have been systematically analyzed in previous studies, with their physical parameters being estimated. Each of these four targets will be discussed below.

(i) For J0132, \cite{Liu+Tan+1991+APSS+J0132} reported a mass ratio of 0.5947 (M$_2$/M$_1$), and \cite{Kjurkchieva+etal+2019+J0132} gave a value of 0.635 (M$_2$/M$_1$). In this work, the mass ratio was determined as 0.676 (M$_1$/M$_2$). 

(ii) For J0305, \cite{Panchal+Joshi+2021+J0305} reported a mass ratio of 0.31 (M$_2$/M$_1$), and our analysis gives a value of 0.35 (M$_1$/M$_2$).

(iii) For J1300, the mass ratios previously determined by \cite{Kjurkchieva+etal+2018+RAA+1300} and \cite{Yang+etal+2023+J1300} were 4.66 (M$_2$/M$_1$) and 4.747 (M$_2$/M$_1$), respectively, which are approximately the same as the value of 4.51 (M$_2$/M$_1$) obtained in this work. 

(iv) For J1402, \cite{Yang+2011+J1402} and \cite{Alton+2021+AcA+J1402} reported mass ratios of 0.461 (M$_1$/M$_2$) and 0.443 (M$_1$/M$_2$), respectively. The mass ratio of 0.457 (M$_1$/M$_2$) determined in this study is also consistent with their results.

Therefore, the mass ratios obtained for these four targets in our study are generally consistent with those reported in previous works. For J0132 and J0305, the temperatures of the two components may have been assigned to different components due to their small temperature differences.

\subsection{The Variation of the Orbital Period}
With the LC minima collected from superWASP, TESS, ZTF, ASAS, CSS, ASAS-SN, BRNO, TMTS, and other literature, we analyzed orbital period variations for the targets. The final analysis revealed that all targets provided the corrected initial epochs and orbital periods. Except for J1042, the periods of the other nine targets exhibit a long-term increasing or decreasing trend, with three targets also showing cyclic period variations.

The long-term period decrease or increase can probably be explained by the mass transfer between two components. Assuming the conservation of mass and angular momentum, we used the following equation to calculate the rate of mass transfer,
\begin{eqnarray}
\frac{\dot{P}}{P} = -3\dot{M_1}(\frac{1}{M_1}-\frac{1}{M_2}).
\label{equ:mass_transfer}
\end{eqnarray}
By combining the period change $\dot{P}$ ($dp/dt$) of each target and the masses of the two components provided in Tables \ref{tab:T0&P} and \ref{tab:wd_solution}, we calculated the {\rm $\dot{M_1}$} ($dM_1/dt$, rate of mass transfer) for each target, which are listed in Table \ref{tab:T0&P}. The positive value indicates that the primary star M$_1$ is receiving mass, while the negative value indicates that the primary star M$_1$ is losing mass.

A long-term decreasing period is caused by mass transfer from the more massive star to the less massive star. This explanation is applied to J0047, J0638, and J1402.
Therefore, we calculated the timescale of mass transfer ($\tau_{mt} \sim \frac{M_{1,2}}{\dot{M_{1,2}}}$) and the thermal timescale ($\tau_{tt} \sim \frac{GM^{2}}{RL}$) for these three targets, and listed in Table \ref{tab:timescale}.
\begin{table}[h!]
\centering
\caption{The mass transfer timescale and thermal timescale for J0047, J0638, and J1402.}
\label{tab:timescale}
\begin{tabular}{ccr}
\hline
 \multirow{2}{*}{Name} & \multirow{2}{*}{$\frac{M_{1,2}}{\dot{M_{1,2}}}$ $\times 10^7$ yr} & \multirow{2}{*}{$\frac{GM^{2}}{RL}$ $\times 10^7$ yr}   \\
 &  &  \\
\hline
 J0047    &  0.21  &  5.20 \\
 J0638    &  0.10  &  9.85  \\
 J1402    &  0.43  &  26.18  \\
\hline
\end{tabular}
\end{table}

The two timescales for these three targets differ significantly, with the mass transfer timescale being 4.00\%, 1.04\%, and 1.64\% of the thermal timescale, respectively. Therefore, mass transfer cannot explain the long-term decrease in period for J0047, J0638, and J1402. The long-term period decrease in these targets is probably due to angular momentum loss (AML). Furthermore, the possibility that O-C of J0047 and J0638 is part of a cyclic variation cannot be ruled out.

For the other seven targets, the long-term increasing trend of the period may be due to mass transfer from the less massive star to the more massive one. Due to the conservation of angular momentum,  the distance between the two components increases as their mass transfer, leading to a decrease in the degree of contact. The systems will evolve from the current contact state to a semi-detached or detached state. According to stellar evolution theory, the more massive star will fill its Roche lobe first. As a result, the mass will be transferred from the more massive star to the less massive one. The mass and energy are then exchanged between the two components, leading to the evolution towards a contact state again. This is called as the thermal relaxation oscillation model (TRO) of contact binaries \citep{Lucy+1976+ApJ+TRO,Flannery+1976+ApJ+TRO,Robertson+Eggleton+1977+MNRAS+TRO}. Therefore, long-term monitoring of these targets is necessary in the future.

The cyclic variation of the $O-C$  parameter can be explained by the light travel-time effect (LTTE) due to the presence of a companion star or by a magnetic activity cycle. The existence of a third body is often used to explain the cyclic variation of orbital period, such as BK Vul \citep{Adalali+Soydugan+2024+NewA+LTTE+BKVul}; CW Aqr \citep{Vijaya+Sriram+2023+RAA+CWAqr}; CSS\_J154915.7+375506 \citep{Wu+etal+2024+MN+CSSJ1549}.
If J0132, J1300, and J1402 are in triple systems, we can apply the function below to describe them:
\begin{equation}
f(m)= {4\pi\over GP^2_{3}}\times(a_{12}^{'}\sin i^{'})^3 = {(M_{3}\sin i^{'})^3\over (M_1+M_2+M_{3})^2},
\end{equation}
where $G$ and $P_3$ are the gravitational constant and the period of the $(O - C)_2$ oscillation, respectively. The amplitude of the oscillation is $A = \frac{a^{'}_{12}{\rm sin} i^{'}}{c}\sqrt{1-e^{'2}{\rm cos^2}\omega^{'}}$, where $a^{'}_{12}$ is the distance between binary and barycenter of the triple body system 
, $c$ is the speed of light and $i^{'}$ is the inclination of the orbit of the third component. $M_1$ and $M_2$ represent the mass of the two components, and $M_3$ is the mass of the third body. We determined $f(m)$ of the additional component for the three targets. The orbital distance between the third body and the central binary can be estimated as $a_3 = (M_1 + M_2) \times a_{12} / M_3$. If the orbital inclination of the third body $i^{'}$ is the same as the binaries ($i^{'}$ = $i$) for the three targets, respectively, the mass and the distance of the tertiary companion are calculated. All parameters are listed in Table \ref{tab:oc_cyc}. Therefore, if the third body is a main sequence star, the spectral type and the luminosity were determined according to the relation of \cite{Cox+2000+AAQ}. Finally, the $M_3$ of J0132 is too small for it to be a main sequence star and may be a brown dwarf. The spectral type and the proportion of the third body to the total luminosity for J1300 and J1402 are determined and listed in Table \ref{tab:oc_cyc}.

Another possible mechanism is magnetic activity. The Applegate mechanism \citep{Applegate+1992+ApJ} involves magnetic activity that produces a change in the variation of the quadrupole moment and, finally, the change in the orbital period. This can be described by using the following equation\citep{Lanza+2002+NA+Applegate},
\begin{eqnarray}
{\Delta{P}\over P}=-9{\Delta Q\over Ma^2},
\label{eq:Applegate}
\end{eqnarray}
where $\Delta P$ is the period of cyclic variation, $P$ is the orbital period, M is the mass for each component of the binary system, and a is the semi-major axis of binary system. The quadrupole momenta of the required variation of both components ($\Delta Q_1$ and $\Delta Q_2$) were determined and tabulated in Table \ref{tab:oc_cyc}. The typical value is usually $10^{51}$ to $10^{52}$ g cm$^2$ for close binaries \citep{Lanza+1999+A&A+Apllegate}, and $\Delta Q$ = $10^{49}$ g cm$^2$ for cataclysmic variables \citep{Lanza+1999+A&A+Apllegate}. Although the quadrupole momenta of many binaires are not in agreement with the typical value, such as DZ Psc~\citep{Yang+etal+2013+AJ+DZPsc}, V1101 Her~\citep{Pi+etal+2017+AJ+V1101Her}, V0474 Cam~\citep{Guo+etal+2018+PASP+V0474Cam}.
They are all similar with the value of $10^{49}$ g cm$^2$. Thus, we cannot exclude the possibility of magnetic activity, especially for J1402. The light curve of J1402 exhibits the O'Connell effect and its spectra show strong magnetic activity.

Therefore, the periodic change for J1300 and J1402 may be caused by the existence of a dim third body, but the Applegate mechanism cannot be excluded because both show evidence of magnetic activity. The cyclic variation of $O - C$ for J0132 may be due to magnetic activity. More observational data are needed in the future to confirm this.

\subsection{Evolutionary State}
\begin{figure*}
\centering
\includegraphics[width=0.99\textwidth]{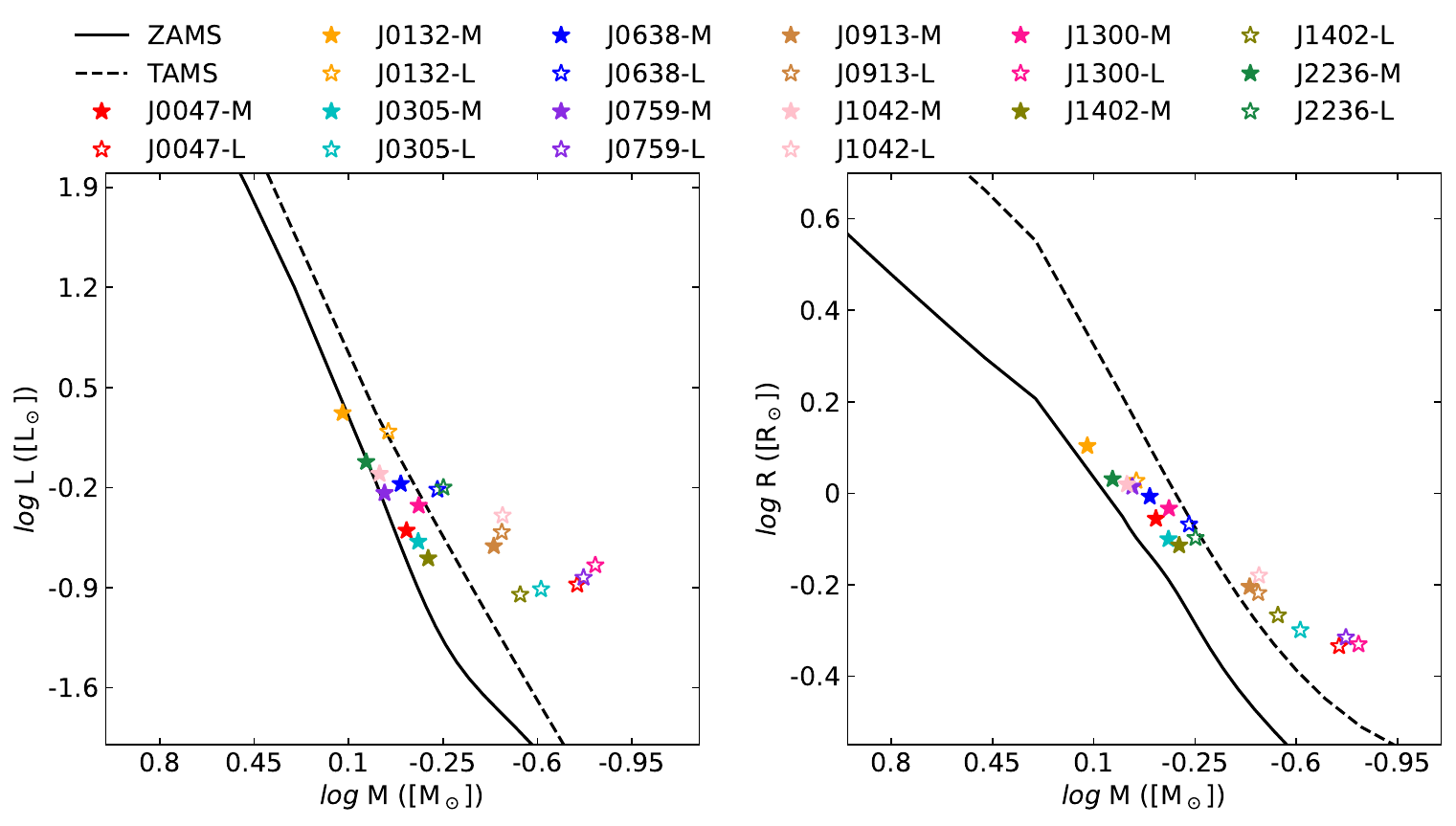}
\caption{Mass-luminosity diagram (left panel) and mass-radius diagram (right panel). The solid and dotted lines represent the ZAMS and TAMS lines constructed using the binary star evolution code provided by~\citet{Hurley+etal+2002+MNRAS+BSE}. The solid and open stars for different colors represent the more massive star (M) and less massive star (L) of the ten targets, respectively.}
\label{fig:M-L-R}
\end{figure*}

With the absolute physical parameters, the positions of binaries on the mass-luminosity (M-L) and mass-radius (M-R) diagrams can be determined. In Figure \ref{fig:M-L-R}, the solid and dotted lines show the zero age main sequence (ZAMS) and the terminal age main sequence (TAMS), constructed with the help of the binary star evolution code (BSE) \citep{Hurley+etal+2002+MNRAS+BSE}.

The more massive stars of all binary sample except J0132 and J0913 are located between the ZAMS line and the TAMS line, while the less massive stars are located above the TAMS line. This indicates that the currently more massive star is a main sequence star, and the currently less massive star is over-luminous and oversized relative to main sequence counterparts with the same mass. This excess in luminosity and radius may be explicable by developments in the early evolution of the system. For J0132, both components are located in the main sequence, while for J0913, both components are above the TAMS. This suggests that these two targets may have evolved more slowly or more rapidly compared to the other eight targets.

\begin{figure*}
\centering
\includegraphics[width=0.65\textwidth]{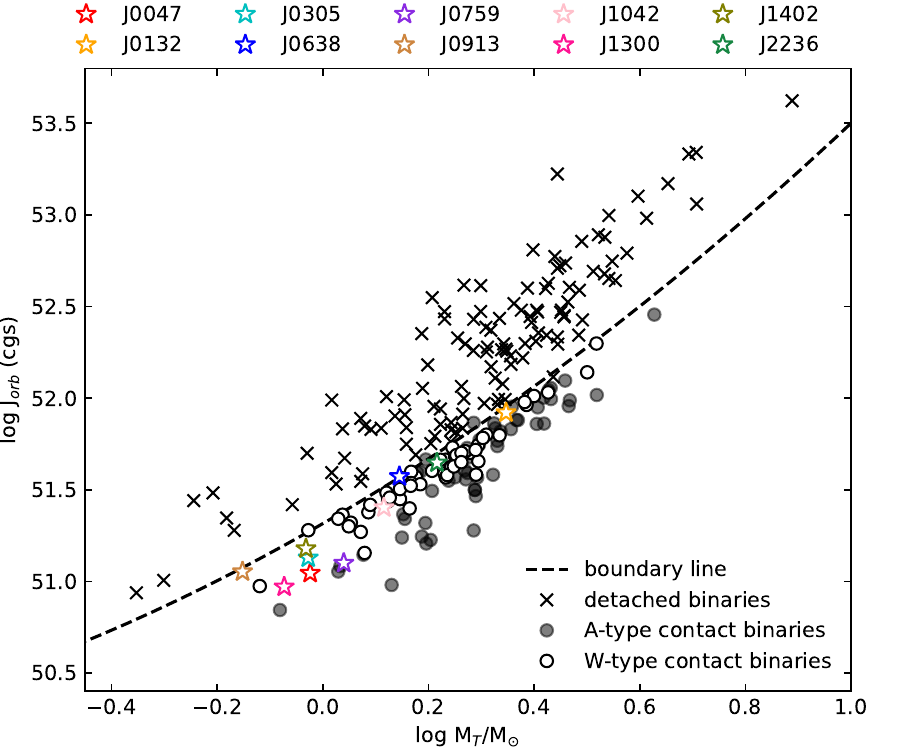}\\
\caption{The relation between orbital angular momentum and total mass for detached and contact binaries. The detached binaries \citep{Eker+etal+2006+MN+J-M} and the contact binary~\citep{Yakut+Eggleton+2005+ApJ+CloseBinary,Eker+etal+2006+MN+J-M} are separated by the boundary line~\citep{Eker+etal+2006+MN+J-M}. The black crosses represent the detached binaries, the black solid and open circles refer to the A-type and W-type contact binaries, respectively. Different colored stars represent different targets.}
\label{fig:MJ}
\end{figure*}

Following \cite{Christopoulou+etal+2013+AJ+MTJorb}, we used the equation below to describe the relationship between mass and angular momentum $J_{orb}$ of contact binaries:
\begin{eqnarray}
J_{{\rm orb}} = 1.24 \times 10^{52} \times M^{3/5}_T \times P^{1/3} \times q \times (1+q)^{-2},
\label{eq:MTJorb}
\end{eqnarray}
where $J_{{\rm orb}}$, $M_{\rm T}$, $P$, and $q$ represent the angular momentum, total mass of the binary system, orbital period, and the mass ratio, respectively. Combining the values of these parameters, the angular momentum of these ten targets was determined and listed in Table \ref{tab:wd_solution}.
The relationship of $M_{\rm T}$ and $J_{\rm orb}$ for detached binaries and overcontact binaries is shown in Figure \ref{fig:MJ}, where the boundary line separates detached binaries and overcontact binaries. Note that the data for the detached binaries in the sample were collected from \cite{Eker+etal+2006+MN+J-M}, while the data for overcontact binaries were collected from \cite{Eker+etal+2006+MN+J-M} and \cite{Yakut+Eggleton+2005+ApJ+CloseBinary}.
The positions of ten targets are marked in Figure \ref{fig:MJ}. It can be seen that all targets are below the boundary line, and J0132 and J0913 are very close to it. Many researchers have proposed that the W UMa binaries may be formed from short period detached binaries by AML \citep[e.g.][]{Stepien+2006+AcA+AML,Stepien+2011+AcA+AML,Yildiz+Dogan+2013+MN+AML,Yildiz+2014+MNRAS+AML,Qian+etal+2017+RAA+AML}.
Considering their positions on the M-L and M-R diagrams in Figure \ref{fig:M-L-R}, along with the overall trends and their low fill-out factors, we infer that these two targets may be newly formed contact binaries.

\section{Summary} \label{sec:summary}

Through the TMTS photometric observations and the LAMOST MR specroscopic observations of the ten targets, physical parameters were obtained from the W-D analysis. The unequal two maxima of the light curves for every target indicate that the O'Connell effect is significant. We adopted the photometric solution with the spot of all targets as the final result. The ten targets are determined to be all W-type contact binaries, in which J0047, J0305, and J1042 are the median contact binaries and the others are shallow contact binaries. Based on all available minima, we analyzed the period change for the ten targets and obtained a long-term increasing or decreasing component. J0132, J1300, and J1402 exhibit cyclic variations for the orbital period, which may be due to their presence in a triple system. Based on spectroscopic analysis, J0047, J0305, J0638, and J1402 exhibit strong magnetic activity. 
The evolutionary states of J0132 and J0913 differ from the other eight targets. Considering their low fill-out factors and relatively high angular momenta, these two targets may be newly formed contact binaries resulting from AML.

In the forthcoming papers, we will construct training and test datasets, employing machine learning methods to publish a catalog of physical parameters for a large sample of contact binaries.
For the interesting targets obtained, phenomena such as low-mass ratios and short-period cutoffs will be discussed and studied in detail. For contact binaries in future TMTS observations, we will continue to obtain their radial velocity data from LAMOST or other surveys as much as possible, enriching and expanding the entire database of absolute physical parameters for contact binaries.

\begin{acknowledgments}
We acknowledge the support of the staff of the Xinglong Observatory of National Astronomical Observatories of China (NAOC) during the installation, commissioning and operation of the Tsinghua University-Ma Huateng Telescopes for Survey (TMTS) system.
This work is supported by the National Science Foundation of China (NSFC grants 12033003 and 12288102), the Ma Huateng Foundation, the Scholar Program of Beijing Academy of Science and Technology (DZ:BS202002), the Tencent Xplorer Prize, Beijing Natural Science Foundation (No. 1242016), and the Talents Program (24CE-YS-08) of Beijing Academy of Science and Technology. J. L. is supported by the National Natural Science Foundation of China (NSFC; Grant Numbers 12403038), the Fundamental Research Funds for the Central Universities (Grant Numbers WK2030000089), and the Cyrus Chun Ying Tang Foundations.

Guoshoujing Telescope (the Large Sky Area Multi-Object Fiber Spectroscopic Telescope LAMOST) is a National Major Scientific Project built by the Chinese Academy of Sciences. Funding for the project has been provided by the National Development and Reform Commission. LAMOST is operated and managed by the National Astronomical Observatories, Chinese Academy of Sciences.

This work has made use of data from the European Space Agency (ESA) mission {\it Gaia} (\url{https://www.cosmos.esa.int/gaia}), processed by the {\it Gaia} Data Processing and Analysis Consortium (DPAC, \url{https://www.cosmos.esa.int/web/gaia/dpac/consortium}). Funding for the DPAC has been provided by national institutions, in particular the institutions participating in the {\it Gaia} Multilateral Agreement.

This research has made use of the International Variable Star Index \citep[VSX,][]{Watson+etal+2006+VSX} data base, operated at the American Association of Variable Star Observers (AAVSO), Cambridge, Massachusetts, USA. Some of the results in this paper have been derived using the HEALPIX \citep{Gorski+etal+2005+ApJ+VSX} package.

We acknowledge the use of data from the SuperWASP project, TESS (Transiting Exoplanet Survey Satellite), ZTF (Zwicky Transient Facility), ASAS (All Sky Automated Survey), CSS (Catalina Sky Survey), ASAS-SN (All-Sky Automated Survey for Supernovae), and Brno Regional Network of Observers (B.R.N.O.). These resources and efforts have been invaluable to this study and we sincerely thank the teams and organizations for their publicly available data and continuous efforts.
\end{acknowledgments}


\bibliography{references}{}

\begin{thebibliography}{}
\expandafter\ifx\csname natexlab\endcsname\relax\def\natexlab#1{#1}\fi
\providecommand{\url}[1]{\href{#1}{#1}}
\providecommand{\dodoi}[1]{doi:~\href{http://doi.org/#1}{\nolinkurl{#1}}}
\providecommand{\doeprint}[1]{\href{http://ascl.net/#1}{\nolinkurl{http://ascl.net/#1}}}
\providecommand{\doarXiv}[1]{\href{https://arxiv.org/abs/#1}{\nolinkurl{https://arxiv.org/abs/#1}}}

\bibitem[{{Adalal{\i}} \& {Soydugan}(2024)}]{Adalali+Soydugan+2024+NewA+LTTE+BKVul}
{Adalal{\i}}, S., \& {Soydugan}, E. 2024, \na, 112, 102270, \dodoi{10.1016/j.newast.2024.102270}

\bibitem[{{Akerlof} {et~al.}(2000){Akerlof}, {Amrose}, {Balsano}, {Bloch}, {Casperson}, {Fletcher}, {Gisler}, {Hills}, {Kehoe}, {Lee}, {Marshall}, {McKay}, {Pawl}, {Schaefer}, {Szymanski}, \& {Wren}}]{Akerlof+etal+2000+AJ+ROTSE}
{Akerlof}, C., {Amrose}, S., {Balsano}, R., {et~al.} 2000, \aj, 119, 1901, \dodoi{10.1086/301321}

\bibitem[{{Alton} \& {St{\k{e}}pie{\'n}}(2021)}]{Alton+2021+AcA+J1402}
{Alton}, K.~B., \& {St{\k{e}}pie{\'n}}, K. 2021, \actaa, 71, 123, \dodoi{10.32023/0001-5237/71.2.4}

\bibitem[{{Alvarez} {et~al.}(2015){Alvarez}, {Sowell}, {Williamon}, \& {Lapasset}}]{Alvarez+etal+2015+PASP+MWPav}
{Alvarez}, G.~E., {Sowell}, J.~R., {Williamon}, R.~M., \& {Lapasset}, E. 2015, \pasp, 127, 742, \dodoi{10.1086/682388}

\bibitem[{{Applegate}(1992)}]{Applegate+1992+ApJ}
{Applegate}, J.~H. 1992, \apj, 385, 621, \dodoi{10.1086/170967}

\bibitem[{{Barden}(1985)}]{Barden+1985+ApJ+SpecSub}
{Barden}, S.~C. 1985, \apj, 295, 162, \dodoi{10.1086/163361}

\bibitem[{{Baron} \& {Hauschildt}(2007)}]{Baron+Hauschildt+2007+AAP+PHOENIX}
{Baron}, E., \& {Hauschildt}, P.~H. 2007, \aap, 468, 255, \dodoi{10.1051/0004-6361:20066755}

\bibitem[{{Bellm} {et~al.}(2019){Bellm}, {Kulkarni}, {Graham}, {Dekany}, {Smith}, {Riddle}, {Masci}, {Helou}, {Prince}, {Adams}, {Barbarino}, {Barlow}, {Bauer}, {Beck}, {Belicki}, {Biswas}, {Blagorodnova}, {Bodewits}, {Bolin}, {Brinnel}, {Brooke}, {Bue}, {Bulla}, {Burruss}, {Cenko}, {Chang}, {Connolly}, {Coughlin}, {Cromer}, {Cunningham}, {De}, {Delacroix}, {Desai}, {Duev}, {Eadie}, {Farnham}, {Feeney}, {Feindt}, {Flynn}, {Franckowiak}, {Frederick}, {Fremling}, {Gal-Yam}, {Gezari}, {Giomi}, {Goldstein}, {Golkhou}, {Goobar}, {Groom}, {Hacopians}, {Hale}, {Henning}, {Ho}, {Hover}, {Howell}, {Hung}, {Huppenkothen}, {Imel}, {Ip}, {Ivezi{\'c}}, {Jackson}, {Jones}, {Juric}, {Kasliwal}, {Kaspi}, {Kaye}, {Kelley}, {Kowalski}, {Kramer}, {Kupfer}, {Landry}, {Laher}, {Lee}, {Lin}, {Lin}, {Lunnan}, {Giomi}, {Mahabal}, {Mao}, {Miller}, {Monkewitz}, {Murphy}, {Ngeow}, {Nordin}, {Nugent}, {Ofek}, {Patterson}, {Penprase}, {Porter}, {Rauch}, {Rebbapragada}, {Reiley}, {Rigault}, {Rodriguez}, {van Roestel}, {Rusholme}, {van
  Santen}, {Schulze}, {Shupe}, {Singer}, {Soumagnac}, {Stein}, {Surace}, {Sollerman}, {Szkody}, {Taddia}, {Terek}, {Van Sistine}, {van Velzen}, {Vestrand}, {Walters}, {Ward}, {Ye}, {Yu}, {Yan}, \& {Zolkower}}]{Bellm+etal+2019+PASP+ZTF}
{Bellm}, E.~C., {Kulkarni}, S.~R., {Graham}, M.~J., {et~al.} 2019, \pasp, 131, 018002, \dodoi{10.1088/1538-3873/aaecbe}

\bibitem[{{Berdyugina}(2005)}]{Berdyugina+2005+LRSP+spot}
{Berdyugina}, S.~V. 2005, Living Reviews in Solar Physics, 2, 8, \dodoi{10.12942/lrsp-2005-8}

\bibitem[{{Binnendijk}(1970)}]{Binnendijk+VA+1970}
{Binnendijk}, L. 1970, Vistas in Astronomy, 12, 217, \dodoi{10.1016/0083-6656(70)90041-3}

\bibitem[{{Blattler} \& {Diethelm}(2003)}]{Blattler+Diethelm+2003+J1402min}
{Blattler}, E., \& {Diethelm}, R. 2003, Information Bulletin on Variable Stars, 5403, 1

\bibitem[{{Christopoulou} \& {Papageorgiou}(2013)}]{Christopoulou+etal+2013+AJ+MTJorb}
{Christopoulou}, P.~E., \& {Papageorgiou}, A. 2013, \aj, 146, 157, \dodoi{10.1088/0004-6256/146/6/157}

\bibitem[{{Christy} {et~al.}(2023){Christy}, {Jayasinghe}, {Stanek}, {Kochanek}, {Thompson}, {Shappee}, {Holoien}, {Prieto}, {Dong}, \& {Giles}}]{Christy+etal+2023+MNRAS+ASAS-SN}
{Christy}, C.~T., {Jayasinghe}, T., {Stanek}, K.~Z., {et~al.} 2023, \mnras, 519, 5271, \dodoi{10.1093/mnras/stac3801}

\bibitem[{{Cox}(2000)}]{Cox+2000+AAQ}
{Cox}, A.~N. 2000, {Allen's astrophysical quantities}

\bibitem[{{Cui} {et~al.}(2012){Cui}, {Zhao}, {Chu}, {Li}, {Li}, {Zhang}, {Su}, {Yao}, {Wang}, {Xing}, {Li}, {Zhu}, {Wang}, {Gu}, {Luo}, {Xu}, {Zhang}, {Liu}, {Zhang}, {Yang}, {Cao}, {Chen}, {Chen}, {Chen}, {Chen}, {Chu}, {Feng}, {Gong}, {Hou}, {Hu}, {Hu}, {Hu}, {Jia}, {Jiang}, {Jiang}, {Jiang}, {Jin}, {Li}, {Li}, {Li}, {Liu}, {Liu}, {Lu}, {Mao}, {Men}, {Qi}, {Qi}, {Shi}, {Tang}, {Tao}, {Wang}, {Wang}, {Wang}, {Wang}, {Wang}, {Wang}, {Wang}, {Wang}, {Wang}, {Wang}, {Wang}, {Wang}, {Xu}, {Xu}, {Yang}, {Yu}, {Yuan}, {Yuan}, {Zhai}, {Zhang}, {Zhang}, {Zhang}, {Zhao}, {Zhou}, {Zhou}, {Zhu}, \& {Zou}}]{Cui+etal+2012+RAA+LAMOST}
{Cui}, X.-Q., {Zhao}, Y.-H., {Chu}, Y.-Q., {et~al.} 2012, Research in Astronomy and Astrophysics, 12, 1197, \dodoi{10.1088/1674-4527/12/9/003}

\bibitem[{{Deb} \& {Singh}(2011)}]{Deb+etal+2011+MNRAS+ASAS+WD+EB}
{Deb}, S., \& {Singh}, H.~P. 2011, \mnras, 412, 1787, \dodoi{10.1111/j.1365-2966.2010.18016.x}

\bibitem[{{Demircan} {et~al.}(2011){Demircan}, {Gurol}, {Gokay}, {Terzioglu}, {Saral}, {Gursoytrak}, {Okan}, {Demirhan}, {Coker}, \& {Derman}}]{Demircan+etal+2011+J1402min}
{Demircan}, Y., {Gurol}, B., {Gokay}, G., {et~al.} 2011, Information Bulletin on Variable Stars, 5965, 1

\bibitem[{{Diethelm}(2005)}]{Diethelm+2005+J1402min}
{Diethelm}, R. 2005, Information Bulletin on Variable Stars, 5653, 1

\bibitem[{{Diethelm}(2006)}]{Diethelm+2006+J1402min}
---. 2006, Information Bulletin on Variable Stars, 5713, 1

\bibitem[{{Diethelm}(2007)}]{Diethelm+2007+J1402min}
---. 2007, Information Bulletin on Variable Stars, 5781, 1

\bibitem[{{Diethelm}(2009{\natexlab{a}})}]{Diethelm+2009+J0132min}
---. 2009{\natexlab{a}}, Information Bulletin on Variable Stars, 5871, 1

\bibitem[{{Diethelm}(2009{\natexlab{b}})}]{Diethelm+2009+J1300J1402min}
---. 2009{\natexlab{b}}, Information Bulletin on Variable Stars, 5894, 1

\bibitem[{{Diethelm}(2010{\natexlab{a}})}]{Diethelm+2010+J0132min}
---. 2010{\natexlab{a}}, Information Bulletin on Variable Stars, 5920, 1

\bibitem[{{Diethelm}(2010{\natexlab{b}})}]{Diethelm+2010+J1300min}
---. 2010{\natexlab{b}}, Information Bulletin on Variable Stars, 5945, 1

\bibitem[{{Diethelm}(2011{\natexlab{a}})}]{Diethelm+2011+J01322min}
---. 2011{\natexlab{a}}, Information Bulletin on Variable Stars, 5960, 1

\bibitem[{{Diethelm}(2011{\natexlab{b}})}]{Diethelm+2011+J1300J1402min}
---. 2011{\natexlab{b}}, Information Bulletin on Variable Stars, 5992, 1

\bibitem[{{Diethelm}(2012{\natexlab{a}})}]{Diethelm+2012+J032min}
---. 2012{\natexlab{a}}, Information Bulletin on Variable Stars, 6011, 1

\bibitem[{{Diethelm}(2012{\natexlab{b}})}]{Diethelm+2012+J1300J1402min}
---. 2012{\natexlab{b}}, Information Bulletin on Variable Stars, 6029, 1

\bibitem[{{Ding} {et~al.}(2023){Ding}, {Ji}, {Li}, {Xiong}, {Cheng}, \& {Wang}}]{Ding+etal+2023+MN+TESS+ML}
{Ding}, X., {Ji}, K., {Li}, X., {et~al.} 2023, \mnras, 525, 4596, \dodoi{10.1093/mnras/stad2565}

\bibitem[{{Ding} {et~al.}(2021){Ding}, {Ji}, \& {Li}}]{Ding+etal+2021+PASJ+ML}
{Ding}, X., {Ji}, K.-F., \& {Li}, X.-Z. 2021, \pasj, 73, 786, \dodoi{10.1093/pasj/psab042}

\bibitem[{{Dvorak}(2005)}]{Dvorak+2005+J0132min}
{Dvorak}, S.~W. 2005, Information Bulletin on Variable Stars, 5603, 1

\bibitem[{{Eker}(1999)}]{Eker+1999+TJPh+spot}
{Eker}, Z. 1999, Turkish Journal of Physics, 23, 357

\bibitem[{{Eker} {et~al.}(2006){Eker}, {Demircan}, {Bilir}, \& {Karata{\c{s}}}}]{Eker+etal+2006+MN+J-M}
{Eker}, Z., {Demircan}, O., {Bilir}, S., \& {Karata{\c{s}}}, Y. 2006, \mnras, 373, 1483, \dodoi{10.1111/j.1365-2966.2006.11073.x}

\bibitem[{{Flannery}(1976)}]{Flannery+1976+ApJ+TRO}
{Flannery}, B.~P. 1976, \apj, 205, 217, \dodoi{10.1086/154266}

\bibitem[{{Gaia Collaboration} {et~al.}(2016){Gaia Collaboration}, {Prusti}, {de Bruijne}, {Brown}, {Vallenari}, {Babusiaux}, {Bailer-Jones}, {Bastian}, {Biermann}, {Evans}, {Eyer}, {Jansen}, {Jordi}, {Klioner}, {Lammers}, {Lindegren}, {Luri}, {Mignard}, {Milligan}, {Panem}, {Poinsignon}, {Pourbaix}, {Randich}, {Sarri}, {Sartoretti}, {Siddiqui}, {Soubiran}, {Valette}, {van Leeuwen}, {Walton}, {Aerts}, {Arenou}, {Cropper}, {Drimmel}, {H{\o}g}, {Katz}, {Lattanzi}, {O'Mullane}, {Grebel}, {Holland}, {Huc}, {Passot}, {Bramante}, {Cacciari}, {Casta{\~n}eda}, {Chaoul}, {Cheek}, {De Angeli}, {Fabricius}, {Guerra}, {Hern{\'a}ndez}, {Jean-Antoine-Piccolo}, {Masana}, {Messineo}, {Mowlavi}, {Nienartowicz}, {Ord{\'o}{\~n}ez-Blanco}, {Panuzzo}, {Portell}, {Richards}, {Riello}, {Seabroke}, {Tanga}, {Th{\'e}venin}, {Torra}, {Els}, {Gracia-Abril}, {Comoretto}, {Garcia-Reinaldos}, {Lock}, {Mercier}, {Altmann}, {Andrae}, {Astraatmadja}, {Bellas-Velidis}, {Benson}, {Berthier}, {Blomme}, {Busso}, {Carry}, {Cellino}, {Clementini},
  {Cowell}, {Creevey}, {Cuypers}, {Davidson}, {De Ridder}, {de Torres}, {Delchambre}, {Dell'Oro}, {Ducourant}, {Fr{\'e}mat}, {Garc{\'\i}a-Torres}, {Gosset}, {Halbwachs}, {Hambly}, {Harrison}, {Hauser}, {Hestroffer}, {Hodgkin}, {Huckle}, {Hutton}, {Jasniewicz}, {Jordan}, {Kontizas}, {Korn}, {Lanzafame}, {Manteiga}, {Moitinho}, {Muinonen}, {Osinde}, {Pancino}, {Pauwels}, {Petit}, {Recio-Blanco}, {Robin}, {Sarro}, {Siopis}, {Smith}, {Smith}, {Sozzetti}, {Thuillot}, {van Reeven}, {Viala}, {Abbas}, {Abreu Aramburu}, {Accart}, {Aguado}, {Allan}, {Allasia}, {Altavilla}, {{\'A}lvarez}, {Alves}, {Anderson}, {Andrei}, {Anglada Varela}, {Antiche}, {Antoja}, {Ant{\'o}n}, {Arcay}, {Atzei}, {Ayache}, {Bach}, {Baker}, {Balaguer-N{\'u}{\~n}ez}, {Barache}, {Barata}, {Barbier}, {Barblan}, {Baroni}, {Barrado y Navascu{\'e}s}, {Barros}, {Barstow}, {Becciani}, {Bellazzini}, {Bellei}, {Bello Garc{\'\i}a}, {Belokurov}, {Bendjoya}, {Berihuete}, {Bianchi}, {Bienaym{\'e}}, {Billebaud}, {Blagorodnova}, {Blanco-Cuaresma}, {Boch},
  {Bombrun}, {Borrachero}, {Bouquillon}, {Bourda}, {Bouy}, {Bragaglia}, {Breddels}, {Brouillet}, {Br{\"u}semeister}, {Bucciarelli}, {Budnik}, {Burgess}, {Burgon}, {Burlacu}, {Busonero}, {Buzzi}, {Caffau}, {Cambras}, {Campbell}, {Cancelliere}, {Cantat-Gaudin}, {Carlucci}, {Carrasco}, {Castellani}, {Charlot}, {Charnas}, {Charvet}, {Chassat}, {Chiavassa}, {Clotet}, {Cocozza}, {Collins}, {Collins}, {Costigan}, {Crifo}, {Cross}, {Crosta}, {Crowley}, {Dafonte}, {Damerdji}, {Dapergolas}, {David}, {David}, {De Cat}, {de Felice}, {de Laverny}, {De Luise}, {De March}, {de Martino}, {de Souza}, {Debosscher}, {del Pozo}, {Delbo}, {Delgado}, {Delgado}, {di Marco}, {Di Matteo}, {Diakite}, {Distefano}, {Dolding}, {Dos Anjos}, {Drazinos}, {Dur{\'a}n}, {Dzigan}, {Ecale}, {Edvardsson}, {Enke}, {Erdmann}, {Escolar}, {Espina}, {Evans}, {Eynard Bontemps}, {Fabre}, {Fabrizio}, {Faigler}, {Falc{\~a}o}, {Farr{\`a}s Casas}, {Faye}, {Federici}, {Fedorets}, {Fern{\'a}ndez-Hern{\'a}ndez}, {Fernique}, {Fienga}, {Figueras}, {Filippi},
  {Findeisen}, {Fonti}, {Fouesneau}, {Fraile}, {Fraser}, {Fuchs}, {Furnell}, {Gai}, {Galleti}, {Galluccio}, {Garabato}, {Garc{\'\i}a-Sedano}, {Gar{\'e}}, {Garofalo}, {Garralda}, {Gavras}, {Gerssen}, {Geyer}, {Gilmore}, {Girona}, {Giuffrida}, {Gomes}, {Gonz{\'a}lez-Marcos}, {Gonz{\'a}lez-N{\'u}{\~n}ez}, {Gonz{\'a}lez-Vidal}, {Granvik}, {Guerrier}, {Guillout}, {Guiraud}, {G{\'u}rpide}, {Guti{\'e}rrez-S{\'a}nchez}, {Guy}, {Haigron}, {Hatzidimitriou}, {Haywood}, {Heiter}, {Helmi}, {Hobbs}, {Hofmann}, {Holl}, {Holland}, {Hunt}, {Hypki}, {Icardi}, {Irwin}, {Jevardat de Fombelle}, {Jofr{\'e}}, {Jonker}, {Jorissen}, {Julbe}, {Karampelas}, {Kochoska}, {Kohley}, {Kolenberg}, {Kontizas}, {Koposov}, {Kordopatis}, {Koubsky}, {Kowalczyk}, {Krone-Martins}, {Kudryashova}, {Kull}, {Bachchan}, {Lacoste-Seris}, {Lanza}, {Lavigne}, {Le Poncin-Lafitte}, {Lebreton}, {Lebzelter}, {Leccia}, {Leclerc}, {Lecoeur-Taibi}, {Lemaitre}, {Lenhardt}, {Leroux}, {Liao}, {Licata}, {Lindstr{\o}m}, {Lister}, {Livanou}, {Lobel}, {L{\"o}ffler},
  {L{\'o}pez}, {Lopez-Lozano}, {Lorenz}, {Loureiro}, {MacDonald}, {Magalh{\~a}es Fernandes}, {Managau}, {Mann}, {Mantelet}, {Marchal}, {Marchant}, {Marconi}, {Marie}, {Marinoni}, {Marrese}, {Marschalk{\'o}}, {Marshall}, {Mart{\'\i}n-Fleitas}, {Martino}, {Mary}, {Matijevi{\v{c}}}, {Mazeh}, {McMillan}, {Messina}, {Mestre}, {Michalik}, {Millar}, {Miranda}, {Molina}, {Molinaro}, {Molinaro}, {Moln{\'a}r}, {Moniez}, {Montegriffo}, {Monteiro}, {Mor}, {Mora}, {Morbidelli}, {Morel}, {Morgenthaler}, {Morley}, {Morris}, {Mulone}, {Muraveva}, {Musella}, {Narbonne}, {Nelemans}, {Nicastro}, {Noval}, {Ord{\'e}novic}, {Ordieres-Mer{\'e}}, {Osborne}, {Pagani}, {Pagano}, {Pailler}, {Palacin}, {Palaversa}, {Parsons}, {Paulsen}, {Pecoraro}, {Pedrosa}, {Pentik{\"a}inen}, {Pereira}, {Pichon}, {Piersimoni}, {Pineau}, {Plachy}, {Plum}, {Poujoulet}, {Pr{\v{s}}a}, {Pulone}, {Ragaini}, {Rago}, {Rambaux}, {Ramos-Lerate}, {Ranalli}, {Rauw}, {Read}, {Regibo}, {Renk}, {Reyl{\'e}}, {Ribeiro}, {Rimoldini}, {Ripepi}, {Riva}, {Rixon},
  {Roelens}, {Romero-G{\'o}mez}, {Rowell}, {Royer}, {Rudolph}, {Ruiz-Dern}, {Sadowski}, {Sagrist{\`a} Sell{\'e}s}, {Sahlmann}, {Salgado}, {Salguero}, {Sarasso}, {Savietto}, {Schnorhk}, {Schultheis}, {Sciacca}, {Segol}, {Segovia}, {Segransan}, {Serpell}, {Shih}, {Smareglia}, {Smart}, {Smith}, {Solano}, {Solitro}, {Sordo}, {Soria Nieto}, {Souchay}, {Spagna}, {Spoto}, {Stampa}, {Steele}, {Steidelm{\"u}ller}, {Stephenson}, {Stoev}, {Suess}, {S{\"u}veges}, {Surdej}, {Szabados}, {Szegedi-Elek}, {Tapiador}, {Taris}, {Tauran}, {Taylor}, {Teixeira}, {Terrett}, {Tingley}, {Trager}, {Turon}, {Ulla}, {Utrilla}, {Valentini}, {van Elteren}, {Van Hemelryck}, {van Leeuwen}, {Varadi}, {Vecchiato}, {Veljanoski}, {Via}, {Vicente}, {Vogt}, {Voss}, {Votruba}, {Voutsinas}, {Walmsley}, {Weiler}, {Weingrill}, {Werner}, {Wevers}, {Whitehead}, {Wyrzykowski}, {Yoldas}, {{\v{Z}}erjal}, {Zucker}, {Zurbach}, {Zwitter}, {Alecu}, {Allen}, {Allende Prieto}, {Amorim}, {Anglada-Escud{\'e}}, {Arsenijevic}, {Azaz}, {Balm}, {Beck}, {Bernstein},
  {Bigot}, {Bijaoui}, {Blasco}, {Bonfigli}, {Bono}, {Boudreault}, {Bressan}, {Brown}, {Brunet}, {Bunclark}, {Buonanno}, {Butkevich}, {Carret}, {Carrion}, {Chemin}, {Ch{\'e}reau}, {Corcione}, {Darmigny}, {de Boer}, {de Teodoro}, {de Zeeuw}, {Delle Luche}, {Domingues}, {Dubath}, {Fodor}, {Fr{\'e}zouls}, {Fries}, {Fustes}, {Fyfe}, {Gallardo}, {Gallegos}, {Gardiol}, {Gebran}, {Gomboc}, {G{\'o}mez}, {Grux}, {Gueguen}, {Heyrovsky}, {Hoar}, {Iannicola}, {Isasi Parache}, {Janotto}, {Joliet}, {Jonckheere}, {Keil}, {Kim}, {Klagyivik}, {Klar}, {Knude}, {Kochukhov}, {Kolka}, {Kos}, {Kutka}, {Lainey}, {LeBouquin}, {Liu}, {Loreggia}, {Makarov}, {Marseille}, {Martayan}, {Martinez-Rubi}, {Massart}, {Meynadier}, {Mignot}, {Munari}, {Nguyen}, {Nordlander}, {Ocvirk}, {O'Flaherty}, {Olias Sanz}, {Ortiz}, {Osorio}, {Oszkiewicz}, {Ouzounis}, {Palmer}, {Park}, {Pasquato}, {Peltzer}, {Peralta}, {P{\'e}turaud}, {Pieniluoma}, {Pigozzi}, {Poels}, {Prat}, {Prod'homme}, {Raison}, {Rebordao}, {Risquez}, {Rocca-Volmerange}, {Rosen},
  {Ruiz-Fuertes}, {Russo}, {Sembay}, {Serraller Vizcaino}, {Short}, {Siebert}, {Silva}, {Sinachopoulos}, {Slezak}, {Soffel}, {Sosnowska}, {Strai{\v{z}}ys}, {ter Linden}, {Terrell}, {Theil}, {Tiede}, {Troisi}, {Tsalmantza}, {Tur}, {Vaccari}, {Vachier}, {Valles}, {Van Hamme}, {Veltz}, {Virtanen}, {Wallut}, {Wichmann}, {Wilkinson}, {Ziaeepour}, \& {Zschocke}}]{Gaia+etal+2016+AA+Gaia}
{Gaia Collaboration}, {Prusti}, T., {de Bruijne}, J.~H.~J., {et~al.} 2016, \aap, 595, A1, \dodoi{10.1051/0004-6361/201629272}

\bibitem[{{Gaia Collaboration} {et~al.}(2018){Gaia Collaboration}, {Brown}, {Vallenari}, {Prusti}, {de Bruijne}, {Babusiaux}, {Bailer-Jones}, {Biermann}, {Evans}, {Eyer}, {Jansen}, {Jordi}, {Klioner}, {Lammers}, {Lindegren}, {Luri}, {Mignard}, {Panem}, {Pourbaix}, {Randich}, {Sartoretti}, {Siddiqui}, {Soubiran}, {van Leeuwen}, {Walton}, {Arenou}, {Bastian}, {Cropper}, {Drimmel}, {Katz}, {Lattanzi}, {Bakker}, {Cacciari}, {Casta{\~n}eda}, {Chaoul}, {Cheek}, {De Angeli}, {Fabricius}, {Guerra}, {Holl}, {Masana}, {Messineo}, {Mowlavi}, {Nienartowicz}, {Panuzzo}, {Portell}, {Riello}, {Seabroke}, {Tanga}, {Th{\'e}venin}, {Gracia-Abril}, {Comoretto}, {Garcia-Reinaldos}, {Teyssier}, {Altmann}, {Andrae}, {Audard}, {Bellas-Velidis}, {Benson}, {Berthier}, {Blomme}, {Burgess}, {Busso}, {Carry}, {Cellino}, {Clementini}, {Clotet}, {Creevey}, {Davidson}, {De Ridder}, {Delchambre}, {Dell'Oro}, {Ducourant}, {Fern{\'a}ndez-Hern{\'a}ndez}, {Fouesneau}, {Fr{\'e}mat}, {Galluccio}, {Garc{\'\i}a-Torres},
  {Gonz{\'a}lez-N{\'u}{\~n}ez}, {Gonz{\'a}lez-Vidal}, {Gosset}, {Guy}, {Halbwachs}, {Hambly}, {Harrison}, {Hern{\'a}ndez}, {Hestroffer}, {Hodgkin}, {Hutton}, {Jasniewicz}, {Jean-Antoine-Piccolo}, {Jordan}, {Korn}, {Krone-Martins}, {Lanzafame}, {Lebzelter}, {L{\"o}ffler}, {Manteiga}, {Marrese}, {Mart{\'\i}n-Fleitas}, {Moitinho}, {Mora}, {Muinonen}, {Osinde}, {Pancino}, {Pauwels}, {Petit}, {Recio-Blanco}, {Richards}, {Rimoldini}, {Robin}, {Sarro}, {Siopis}, {Smith}, {Sozzetti}, {S{\"u}veges}, {Torra}, {van Reeven}, {Abbas}, {Abreu Aramburu}, {Accart}, {Aerts}, {Altavilla}, {{\'A}lvarez}, {Alvarez}, {Alves}, {Anderson}, {Andrei}, {Anglada Varela}, {Antiche}, {Antoja}, {Arcay}, {Astraatmadja}, {Bach}, {Baker}, {Balaguer-N{\'u}{\~n}ez}, {Balm}, {Barache}, {Barata}, {Barbato}, {Barblan}, {Barklem}, {Barrado}, {Barros}, {Barstow}, {Bartholom{\'e} Mu{\~n}oz}, {Bassilana}, {Becciani}, {Bellazzini}, {Berihuete}, {Bertone}, {Bianchi}, {Bienaym{\'e}}, {Blanco-Cuaresma}, {Boch}, {Boeche}, {Bombrun}, {Borrachero},
  {Bossini}, {Bouquillon}, {Bourda}, {Bragaglia}, {Bramante}, {Breddels}, {Bressan}, {Brouillet}, {Br{\"u}semeister}, {Brugaletta}, {Bucciarelli}, {Burlacu}, {Busonero}, {Butkevich}, {Buzzi}, {Caffau}, {Cancelliere}, {Cannizzaro}, {Cantat-Gaudin}, {Carballo}, {Carlucci}, {Carrasco}, {Casamiquela}, {Castellani}, {Castro-Ginard}, {Charlot}, {Chemin}, {Chiavassa}, {Cocozza}, {Costigan}, {Cowell}, {Crifo}, {Crosta}, {Crowley}, {Cuypers}, {Dafonte}, {Damerdji}, {Dapergolas}, {David}, {David}, {de Laverny}, {De Luise}, {De March}, {de Martino}, {de Souza}, {de Torres}, {Debosscher}, {del Pozo}, {Delbo}, {Delgado}, {Delgado}, {Di Matteo}, {Diakite}, {Diener}, {Distefano}, {Dolding}, {Drazinos}, {Dur{\'a}n}, {Edvardsson}, {Enke}, {Eriksson}, {Esquej}, {Eynard Bontemps}, {Fabre}, {Fabrizio}, {Faigler}, {Falc{\~a}o}, {Farr{\`a}s Casas}, {Federici}, {Fedorets}, {Fernique}, {Figueras}, {Filippi}, {Findeisen}, {Fonti}, {Fraile}, {Fraser}, {Fr{\'e}zouls}, {Gai}, {Galleti}, {Garabato}, {Garc{\'\i}a-Sedano}, {Garofalo},
  {Garralda}, {Gavel}, {Gavras}, {Gerssen}, {Geyer}, {Giacobbe}, {Gilmore}, {Girona}, {Giuffrida}, {Glass}, {Gomes}, {Granvik}, {Gueguen}, {Guerrier}, {Guiraud}, {Guti{\'e}rrez-S{\'a}nchez}, {Haigron}, {Hatzidimitriou}, {Hauser}, {Haywood}, {Heiter}, {Helmi}, {Heu}, {Hilger}, {Hobbs}, {Hofmann}, {Holland}, {Huckle}, {Hypki}, {Icardi}, {Jan{\ss}en}, {Jevardat de Fombelle}, {Jonker}, {Juh{\'a}sz}, {Julbe}, {Karampelas}, {Kewley}, {Klar}, {Kochoska}, {Kohley}, {Kolenberg}, {Kontizas}, {Kontizas}, {Koposov}, {Kordopatis}, {Kostrzewa-Rutkowska}, {Koubsky}, {Lambert}, {Lanza}, {Lasne}, {Lavigne}, {Le Fustec}, {Le Poncin-Lafitte}, {Lebreton}, {Leccia}, {Leclerc}, {Lecoeur-Taibi}, {Lenhardt}, {Leroux}, {Liao}, {Licata}, {Lindstr{\o}m}, {Lister}, {Livanou}, {Lobel}, {L{\'o}pez}, {Managau}, {Mann}, {Mantelet}, {Marchal}, {Marchant}, {Marconi}, {Marinoni}, {Marschalk{\'o}}, {Marshall}, {Martino}, {Marton}, {Mary}, {Massari}, {Matijevi{\v{c}}}, {Mazeh}, {McMillan}, {Messina}, {Michalik}, {Millar}, {Molina}, {Molinaro},
  {Moln{\'a}r}, {Montegriffo}, {Mor}, {Morbidelli}, {Morel}, {Morris}, {Mulone}, {Muraveva}, {Musella}, {Nelemans}, {Nicastro}, {Noval}, {O'Mullane}, {Ord{\'e}novic}, {Ord{\'o}{\~n}ez-Blanco}, {Osborne}, {Pagani}, {Pagano}, {Pailler}, {Palacin}, {Palaversa}, {Panahi}, {Pawlak}, {Piersimoni}, {Pineau}, {Plachy}, {Plum}, {Poggio}, {Poujoulet}, {Pr{\v{s}}a}, {Pulone}, {Racero}, {Ragaini}, {Rambaux}, {Ramos-Lerate}, {Regibo}, {Reyl{\'e}}, {Riclet}, {Ripepi}, {Riva}, {Rivard}, {Rixon}, {Roegiers}, {Roelens}, {Romero-G{\'o}mez}, {Rowell}, {Royer}, {Ruiz-Dern}, {Sadowski}, {Sagrist{\`a} Sell{\'e}s}, {Sahlmann}, {Salgado}, {Salguero}, {Sanna}, {Santana-Ros}, {Sarasso}, {Savietto}, {Schultheis}, {Sciacca}, {Segol}, {Segovia}, {S{\'e}gransan}, {Shih}, {Siltala}, {Silva}, {Smart}, {Smith}, {Solano}, {Solitro}, {Sordo}, {Soria Nieto}, {Souchay}, {Spagna}, {Spoto}, {Stampa}, {Steele}, {Steidelm{\"u}ller}, {Stephenson}, {Stoev}, {Suess}, {Surdej}, {Szabados}, {Szegedi-Elek}, {Tapiador}, {Taris}, {Tauran}, {Taylor},
  {Teixeira}, {Terrett}, {Teyssandier}, {Thuillot}, {Titarenko}, {Torra Clotet}, {Turon}, {Ulla}, {Utrilla}, {Uzzi}, {Vaillant}, {Valentini}, {Valette}, {van Elteren}, {Van Hemelryck}, {van Leeuwen}, {Vaschetto}, {Vecchiato}, {Veljanoski}, {Viala}, {Vicente}, {Vogt}, {von Essen}, {Voss}, {Votruba}, {Voutsinas}, {Walmsley}, {Weiler}, {Wertz}, {Wevers}, {Wyrzykowski}, {Yoldas}, {{\v{Z}}erjal}, {Ziaeepour}, {Zorec}, {Zschocke}, {Zucker}, {Zurbach}, \& {Zwitter}}]{Gaia+etal+2018+AAP+Gaia}
{Gaia Collaboration}, {Brown}, A.~G.~A., {Vallenari}, A., {et~al.} 2018, \aap, 616, A1, \dodoi{10.1051/0004-6361/201833051}

\bibitem[{{Gessner} \& {Meinunger}(1973)}]{Gessner+Meinunger+1973+VSS+J0132}
{Gessner}, H., \& {Meinunger}, I. 1973, Veroeffentlichungen der Sternwarte Sonneberg, 7, 607

\bibitem[{{G{\'o}rski} {et~al.}(2005){G{\'o}rski}, {Hivon}, {Banday}, {Wandelt}, {Hansen}, {Reinecke}, \& {Bartelmann}}]{Gorski+etal+2005+ApJ+VSX}
{G{\'o}rski}, K.~M., {Hivon}, E., {Banday}, A.~J., {et~al.} 2005, \apj, 622, 759, \dodoi{10.1086/427976}

\bibitem[{{Guo} {et~al.}(2018){Guo}, {Li}, {Hu}, \& {Chen}}]{Guo+etal+2018+PASP+V0474Cam}
{Guo}, D.~F., {Li}, K., {Hu}, S.~M., \& {Chen}, X. 2018, \pasp, 130, 064201, \dodoi{10.1088/1538-3873/aaba50}

\bibitem[{{Guo} {et~al.}(2024){Guo}, {Lin}, {Wang}, {Chen}, {Li}, {Chen}, {Xia}, {Mo}, {Xi}, {Zhang}, {Liu}, {Jiang}, {Yan}, {Peng}, {Liu}, {Li}, {Lin}, {Xiang}, {Ma}, \& {Cai}}]{Guo+etal+2024+MNRAS+TMTSV}
{Guo}, F., {Lin}, J., {Wang}, X., {et~al.} 2024, \mnras, 528, 6997, \dodoi{10.1093/mnras/stae404}

\bibitem[{{Guo} {et~al.}(2020){Guo}, {Li}, {Xia}, {Gao}, {Jiang}, \& {Liu}}]{Guo+etal+2020+RAA+OConnell}
{Guo}, Y.-N., {Li}, K., {Xia}, Q.-Q., {et~al.} 2020, Research in Astronomy and Astrophysics, 20, 179, \dodoi{10.1088/1674-4527/20/11/179}

\bibitem[{{Hauschildt}(1993)}]{Hauschildt+1993+JQSRT+PHOENIX}
{Hauschildt}, P.~H. 1993, \jqsrt, 50, 301, \dodoi{10.1016/0022-4073(93)90080-2}

\bibitem[{{Hauschildt} \& {Baron}(2006)}]{Hauschildt+Baron+2006+AAP+PHONIX}
{Hauschildt}, P.~H., \& {Baron}, E. 2006, \aap, 451, 273, \dodoi{10.1051/0004-6361:20053846}

\bibitem[{{Heinze} {et~al.}(2018){Heinze}, {Tonry}, {Denneau}, {Flewelling}, {Stalder}, {Rest}, {Smith}, {Smartt}, \& {Weiland}}]{Heinze+etal+2018+AJ+ATLAS}
{Heinze}, A.~N., {Tonry}, J.~L., {Denneau}, L., {et~al.} 2018, \aj, 156, 241, \dodoi{10.3847/1538-3881/aae47f}

\bibitem[{{Hoffman} {et~al.}(2009){Hoffman}, {Harrison}, \& {McNamara}}]{Hoffman+etal+2009+AJ+J0305}
{Hoffman}, D.~I., {Harrison}, T.~E., \& {McNamara}, B.~J. 2009, \aj, 138, 466, \dodoi{10.1088/0004-6256/138/2/466}

\bibitem[{{Hoffmeister}(1966)}]{Hoffmeister+1966+AN+J0132}
{Hoffmeister}, C. 1966, Astronomische Nachrichten, 289, 139, \dodoi{10.1002/asna.19662890306}

\bibitem[{{Honkov{\'a}} {et~al.}(2014){Honkov{\'a}}, {Jury{\v{s}}ek}, {Lehk{\'y}}, {{\v{S}}melcer}, {Trnka}, {Ma{\v{s}}ek}, {Urban{\'\i}k}, {Auer}, {Vra{\v{s}}t{\'a}k}, {Ku{\v{c}}{\'a}kov{\'a}}, {Ruocco}, {Magris}, {Pol{\'a}k}, {Br{\'a}t}, {Audejean}, {Banfi}, {Moudr{\'a}}, {Lomoz}, {P{\v{r}}ib{\'\i}k}, {D{\v{r}}ev{\v{e}}n{\'y}}, {Scaggiante}, {Koci{\'a}n}, {Caga{\v{s}}}, {Poddan{\'y}}, {Z{\'\i}bar}, {Jacobsen}, {Marek}, {Colazo}, {Zardin}, {Sobotka}, {Starzomski}, {Hlad{\'\i}k}, {Vincenzi}, {Skarka}, {Walter}, {Chapman}, {D{\'\i}az}, {Aceti}, {Singh}, {Kalista}, {Kamenec}, {Zejda}, {Marchi}, {B{\'\i}lek}, {Guzzo}, {Corfini}, {Onderkov{\'a}}, {He{\v{c}}ko}, {Mina}, {V{\'\i}tek}, {Barsa}, {Quinones}, {Taormina}, {Melia}, {Schneiter}, {Scavuzzo}, {Marcionni}, {Ehrenberger}, {Tapia}, {Fasseta}, {Suarez}, {Scaggiante}, {Artusi}, {Garcia}, {Grnja}, {Fi{\v{s}}er}, {Hynek}, {Vil{\'a}{\v{s}}ek}, {Rozehnal}, {Kalisch}, {Lang}, {Gorkov{\'a}}, {Novysedl{\'a}k}, {Salvaggio}, {Smy{\v{c}}ka}, {Spurn{\'y}}, {Wikander},
  {Mravik}, {{\v{S}}ucha{\'n}}, \& {{\v{C}}aloud}}]{Honkova+etal+2014+J1300min}
{Honkov{\'a}}, K., {Jury{\v{s}}ek}, J., {Lehk{\'y}}, M., {et~al.} 2014, Open European Journal on Variable Stars, 165, 1

\bibitem[{{Honkova} {et~al.}(2015){Honkova}, {Jurysek}, {Lehky}, {Smelcer}, {Masek}, {Mazanec}, {Hanzl}, {Urbanik}, {Magris}, {Vrastak}, {Walter}, {Hladik}, {Medulka}, {Bilek}, {Trnka}, {Jacobsen}, {Benacek}, {Kuchtak}, {Audejean}, {Ogmen}, {Zibar}, {Fatka}, {Marchi}, {Poddany}, {Quinones}, {Tapia}, {Scaggiante}, {Zardin}, {Corfini}, {Hajek}, {Lomoz}, {Mravik}, {Grnja}, {Campos}, {Caloud}, {Esseiva}, {Jaks}, {Hornik}, {Filip}, {Uhlar}, {Mina}, {Artola}, {Zalazar}, {Muller}, {Pintr}, \& {Divisova}}]{Honkova+etal+2015+J1300min}
{Honkova}, K., {Jurysek}, J., {Lehky}, M., {et~al.} 2015, Open European Journal on Variable Stars, 168, 1, \dodoi{10.48550/arXiv.1606.00369}

\bibitem[{{Ho{\v{n}}kov{\'a}} {et~al.}(2013){Ho{\v{n}}kov{\'a}}, {Jury{\v{s}}ek}, {Lehk{\'y}}, {{\v{S}}melcer}, {Trnka}, {Ma{\v{s}}ek}, {Urban{\'\i}k}, {Auer}, {Vra{\v{s}}t{\'a}k}, {Ku{\v{c}}{\'a}kov{\'a}}, {Ruocco}, {Magris}, {Pol{\'a}k}, {Br{\'a}t}, {Audejean}, {Banfi}, {Moudr{\'a}}, {Lomoz}, {P{\v{r}}ib{\'\i}k}, {D{\v{r}}ev{\v{e}}n{\'y}}, {Scaggiante}, {Koci{\'a}n}, {Caga{\v{s}}}, {Poddan{\'y}}, {Z{\'\i}bar}, {Jacobsen}, {Marek}, {Colazo}, {Zardin}, {Sobotka}, {Starzomski}, {Hlad{\'\i}k}, {Vincenzi}, {Skarka}, {Walter}, {Chapman}, {D{\'\i}az}, {Aceti}, {Singh}, {Kalista}, {Kamenec}, {Zejda}, {Marchi}, {B{\'\i}lek}, {Guzzo}, {Corfini}, {Onderkov{\'a}}, {He{\v{c}}ko}, {Mina}, {V{\'\i}tek}, {Barsa}, {Quinones}, {Taormina}, {Melia}, {Schneiter}, {Scavuzzo}, {Marcionni}, {Ehrenberger}, {Tapia}, {Fasseta}, {Suarez}, {Scaggiante}, {Artusi}, {Garcia}, {Grnja}, {Fi{\v{s}}er}, {Hynek}, {Vil{\'a}{\v{s}}ek}, {Rozehnal}, {Kalisch}, {Lang}, {Gorkov{\'a}}, {Novysedl{\'a}k}, {Salvaggio}, {Smy{\v{c}}ka}, {Spurn{\'y}},
  {Wikander}, {Mravik}, {{\v{S}}ucha{\v{n}}}, \& {{\v{C}}aloud}}]{Honkova+etal+2013+J1300J1402min}
{Ho{\v{n}}kov{\'a}}, K., {Jury{\v{s}}ek}, J., {Lehk{\'y}}, M., {et~al.} 2013, Open European Journal on Variable Stars, 160, 1

\bibitem[{{Hrivnak}(1988)}]{Hrivnak+1988+ApJ+RV+I+1}
{Hrivnak}, B.~J. 1988, \apj, 335, 319, \dodoi{10.1086/166930}

\bibitem[{{Hrivnak}(1989)}]{Hrivnak+1989+ApJ+RV+II+1}
---. 1989, \apj, 340, 458, \dodoi{10.1086/167408}

\bibitem[{{Huang} {et~al.}(2018){Huang}, {Liu}, {Chen}, {Zhang}, {Yuan}, {Xiang}, {Wang}, \& {Tian}}]{Huang+etal+2018+AJ+Standard}
{Huang}, Y., {Liu}, X.~W., {Chen}, B.~Q., {et~al.} 2018, \aj, 156, 90, \dodoi{10.3847/1538-3881/aacda5}

\bibitem[{{Hubscher}(2005)}]{Hubscher+2005+J0132min}
{Hubscher}, J. 2005, Information Bulletin on Variable Stars, 5643, 1

\bibitem[{{Hubscher}(2014)}]{Hubscher+2014+J0132min}
---. 2014, Information Bulletin on Variable Stars, 6118, 1

\bibitem[{{Hubscher}(2017)}]{Hubscher+2017+J0132J1300min}
---. 2017, Information Bulletin on Variable Stars, 6196, 1, \dodoi{10.22444/IBVS.6196}

\bibitem[{{Hubscher} {et~al.}(2013){Hubscher}, {Braune}, \& {Lehmann}}]{Hubscher+etal+2013+J1300min}
{Hubscher}, J., {Braune}, W., \& {Lehmann}, P.~B. 2013, Information Bulletin on Variable Stars, 6048, 1

\bibitem[{{Hubscher} \& {Lehmann}(2012)}]{Hubscher+Lehmann+2012+J0132min}
{Hubscher}, J., \& {Lehmann}, P.~B. 2012, Information Bulletin on Variable Stars, 6026, 1

\bibitem[{{Hubscher} {et~al.}(2010{\natexlab{a}}){Hubscher}, {Lehmann}, {Monninger}, {Steinbach}, \& {Walter}}]{Hubscher+etal+2010+J0132J1402min}
{Hubscher}, J., {Lehmann}, P.~B., {Monninger}, G., {Steinbach}, H.-M., \& {Walter}, F. 2010{\natexlab{a}}, Information Bulletin on Variable Stars, 5941, 1

\bibitem[{{Hubscher} {et~al.}(2010{\natexlab{b}}){Hubscher}, {Lehmann}, {Monninger}, {Steinbach}, \& {Walter}}]{Hubscher+etal+2010+J1300min}
---. 2010{\natexlab{b}}, Information Bulletin on Variable Stars, 5918, 1

\bibitem[{{Hubscher} {et~al.}(2012){Hubscher}, {Lehmann}, \& {Walter}}]{Hubscher+etal+2012+J1300min}
{Hubscher}, J., {Lehmann}, P.~B., \& {Walter}, F. 2012, Information Bulletin on Variable Stars, 6010, 1

\bibitem[{{Hubscher} \& {Monninger}(2011)}]{Hubscher+Monninger+2011+J1402min}
{Hubscher}, J., \& {Monninger}, G. 2011, Information Bulletin on Variable Stars, 5959, 1

\bibitem[{{Hubscher} {et~al.}(2005){Hubscher}, {Paschke}, \& {Walter}}]{Hubscher+etal+2005+J0132min}
{Hubscher}, J., {Paschke}, A., \& {Walter}, F. 2005, Information Bulletin on Variable Stars, 5657, 1

\bibitem[{{Hubscher} {et~al.}(2006){Hubscher}, {Paschke}, \& {Walter}}]{Hubscher+etal+2006+J0132min}
---. 2006, Information Bulletin on Variable Stars, 5731, 1

\bibitem[{{Hubscher} {et~al.}(2008){Hubscher}, {Steinbach}, \& {Walter}}]{Hubscher+etal+2008+J0132min}
{Hubscher}, J., {Steinbach}, H.-M., \& {Walter}, F. 2008, Information Bulletin on Variable Stars, 5830, 1

\bibitem[{{Hurley} {et~al.}(2002){Hurley}, {Tout}, \& {Pols}}]{Hurley+etal+2002+MNRAS+BSE}
{Hurley}, J.~R., {Tout}, C.~A., \& {Pols}, O.~R. 2002, \mnras, 329, 897, \dodoi{10.1046/j.1365-8711.2002.05038.x}

\bibitem[{{Irwin}(1952)}]{Irwin+1952+ApJ+ltte}
{Irwin}, J.~B. 1952, \apj, 116, 211, \dodoi{10.1086/145604}

\bibitem[{{Jury{\v{s}}ek} {et~al.}(2017){Jury{\v{s}}ek}, {Ho{\v{n}}kov{\'a}}, {{\v{S}}melcer}, {Ma{\v{s}}ek}, {Lehk{\'y}}, {B{\'\i}lek}, {Mazanec}, {Han{\v{z}}l}, {Magris}, {Nos{\'a}l}, {Bragagnolo}, {Medulka}, {Vra{\v{s}}\&tacute}, {{\'a}k}, {Urban{\'\i}k}, {Auer}, {Sergey}, {Jacobsen}, {Alessandroni}, {Andreatta}, {Antonio}, {Artola}, {Audejean}, {Balanzino}, {Banfi}, {Baz{\'a}n}, {Borgonovo}, {Caga{\v{s}}}, {{\v{C}}aloud}, {Campos}, {{\v{C}}apkov{\'a}}, {{\v{C}}ern{\'\i}kov{\'a}}, {{\v{C}}ervinka}, {Chiavassa}, {D{\v{r}}ev{\v{e}}n{\'y}}, {Durantini}, {Ferraro}, {Ferrero}, {Girardini}, {Gudmundsson}, {Guzzo}, {Guevara}, {Hlad{\'\i}k}, {Horn{\'\i}k}, {Jak{\v{s}}}, {Jano{\v{s}}tiak}, {Jel{\'\i}nek}, {Kal{\'a}{\v{s}}ek}, {Kalmbach}, {Kubica}, {Ku{\v{c}}{\'a}kov{\'a}}, {Li{\v{s}}ka}, {Lomoz}, {L{\'o}pez}, {Lovato}, {Morero}, {Mrll{\'a}k}, {Mr{\v{n}}{\'a}k}, {Persha}, {Pignata}, {Pintr}, {Popov}, {Portillo}, {Qui{\~n}ones}, {Rodriguez}, {Ruocco}, {Scaggiante}, {Scavuzzo}, {{\v{S}}ebela}, {{\v{S}}imkovi{\v{c}}},
  {{\v{S}}koln{\'\i}k}, {Skub{\'a}k}, {Smolka}, {{\v{S}}peci{\'a}n}, {{\v{S}}ucha{\v{n}}}, {Tornatore}, {Trnka}, {Tyl{\v{s}}ar}, {Walter}, {Zardin}, {Zejda}, {Z{\'\i}bar}, \& {Zikov{\'a}}}]{Jurysek+etal+2017+J0132J1300J1402min}
{Jury{\v{s}}ek}, J., {Ho{\v{n}}kov{\'a}}, K., {{\v{S}}melcer}, L., {et~al.} 2017, Open European Journal on Variable Stars, 179, 1

\bibitem[{{Kjurkchieva} {et~al.}(2019){Kjurkchieva}, {Popov}, {Eneva}, \& {Petrov}}]{Kjurkchieva+etal+2019+J0132}
{Kjurkchieva}, D.~P., {Popov}, V.~A., {Eneva}, Y., \& {Petrov}, N.~I. 2019, Research in Astronomy and Astrophysics, 19, 014, \dodoi{10.1088/1674-4527/19/1/14}

\bibitem[{{Kjurkchieva} {et~al.}(2018){Kjurkchieva}, {Popov}, {Lyubenova Vasileva}, \& {Petrov}}]{Kjurkchieva+etal+2018+RAA+1300}
{Kjurkchieva}, D.~P., {Popov}, V.~A., {Lyubenova Vasileva}, D., \& {Petrov}, N.~I. 2018, Research in Astronomy and Astrophysics, 18, 046, \dodoi{10.1088/1674-4527/18/4/46}

\bibitem[{{Kwee} \& {van Woerden}(1956)}]{Kwee+1956+bain+KW}
{Kwee}, K.~K., \& {van Woerden}, H. 1956, \bain, 12, 327

\bibitem[{{Lampens} {et~al.}(2017){Lampens}, {Van Cauteren}, {Ayiomamitis}, {Kleidis}, {Panagiotopoulos}, {Vanleenhove}, {Hambsch}, {Hautecler}, {Van Wassenhove}, \& {Vermeylen}}]{Lampens+etal+2017+J0132min}
{Lampens}, P., {Van Cauteren}, P., {Ayiomamitis}, A., {et~al.} 2017, Information Bulletin on Variable Stars, 6230, 1, \dodoi{10.22444/IBVS.6230}

\bibitem[{{Lanza} \& {Rodon{\`o}}(1999)}]{Lanza+1999+A&A+Apllegate}
{Lanza}, A.~F., \& {Rodon{\`o}}, M. 1999, \aap, 349, 887

\bibitem[{{Lanza} \& {Rodon{\`o}}(2002)}]{Lanza+2002+NA+Applegate}
---. 2002, Astronomische Nachrichten, 323, 424, \dodoi{10.1002/1521-3994(200208)323:3/4<424::AID-ASNA424>3.0.CO;2-1}

\bibitem[{{Latkovi{\'c}} {et~al.}(2021){Latkovi{\'c}}, {{\v{C}}eki}, \& {Lazarevi{\'c}}}]{Latkovic+etal+2021+ApJS+700}
{Latkovi{\'c}}, O., {{\v{C}}eki}, A., \& {Lazarevi{\'c}}, S. 2021, \apjs, 254, 10, \dodoi{10.3847/1538-4365/abeb23}

\bibitem[{{Lee} \& {Park}(2018)}]{Lee+Park+2018+PASP+masstrans}
{Lee}, J.~W., \& {Park}, J.-H. 2018, \pasp, 130, 034201, \dodoi{10.1088/1538-3873/aaa390}

\bibitem[{{Lehk{\'y}} {et~al.}(2021){Lehk{\'y}}, {Ho{\v{n}}kov{\'a}}, {{\v{S}}melcer}, {Souza de Joode}, {J{\'\i}lek}, {Ma{\v{s}}ek}, {Urban{\'\i}k}, {Walter}, {Dienstbier}, {Bragagnolo}, {Nos{\'a}{\'l}}, {{\v{C}}ervinka}, {Mazanec}, {Vra{\v{s}}{\v{t}}{\'a}k}, {Lomoz}, {Han{\v{z}}l}, {Sergey}, {Gudmundsson}, {Jacobsen}, {Ehrenberger}, {Hlad{\'\i}k}, {Magris}, {Tyl{\v{s}}ar}, {Persha}, {{\v{S}}koln{\'\i}k}, {Smolka}, {Audejean}, {Trnka}, {Medulka}, {{\v{S}}ucha{\v{n}}}, {Salvaggio}, {Papini}, {Marchini}, {Colaco}, {Vala}, {Starck}, {Qui{\v{n}}ones}, {Auer}, {Melia}, {Ruocco}, {Kal{\'a}{\v{s}}ek}, {J{\'\i}ra}, {Versari}, {Girardici}, {Tornatore}, {Bokov{\'a}}, {Malinak}, {Novotn{\'y}}, {Mokr{\'y}}, {Banfi}, {Castillo}, {Durantini}, {Hr{\'a}dek}, {Kubica}, {Lambersk{\'a}}, {L{\'o}pez}, {Lyachov{\'a}}, {Mr{\v{n}}{\'a}k}, {Pavl{\'\i}kov{\'a}}, {Vilchis}, \& {Za{\v{c}}al}}]{Lehky+etal+2021+J1402min}
{Lehk{\'y}}, M., {Ho{\v{n}}kov{\'a}}, K., {{\v{S}}melcer}, L., {et~al.} 2021, Open European Journal on Variable Stars, 211, 1, \dodoi{10.5817/OEJV2021-0211}

\bibitem[{{Lewandowski} {et~al.}(2007){Lewandowski}, {Niedzielski}, \& {Maciejewski}}]{Lewandowski+etal+2007+J1300min}
{Lewandowski}, M., {Niedzielski}, A., \& {Maciejewski}, G. 2007, Information Bulletin on Variable Stars, 5784, 1

\bibitem[{{Li}(2018)}]{Li+2018+NewA+masstrans}
{Li}, K. 2018, \na, 59, 60, \dodoi{10.1016/j.newast.2017.09.004}

\bibitem[{{Li} {et~al.}(2022){Li}, {Gao}, {Liu}, {Gao}, {Li}, {Chen}, \& {Sun}}]{Li+etal+2022+AJ+SpecSub}
{Li}, K., {Gao}, X., {Liu}, X.-Y., {et~al.} 2022, \aj, 164, 202, \dodoi{10.3847/1538-3881/ac8ff2}

\bibitem[{{Li} {et~al.}(2015){Li}, {Hu}, {Guo}, {Jiang}, {Gao}, {Chen}, \& {Odell}}]{Li+etal+2015+AJ+masstrans}
{Li}, K., {Hu}, S.~M., {Guo}, D.~F., {et~al.} 2015, \aj, 149, 120, \dodoi{10.1088/0004-6256/149/4/120}

\bibitem[{{Li} {et~al.}(2021{\natexlab{a}}){Li}, {Xia}, {Kim}, {Hu}, {Guo}, {Jeong}, {Chen}, \& {Gao}}]{Li+etal+2021+ApJ+OConnell}
{Li}, K., {Xia}, Q.-Q., {Kim}, C.-H., {et~al.} 2021{\natexlab{a}}, \apj, 922, 122, \dodoi{10.3847/1538-4357/ac242f}

\bibitem[{{Li} {et~al.}(2021{\natexlab{b}}){Li}, {Xia}, {Kim}, {Gao}, {Hu}, {Guo}, {Gao}, {Chen}, \& {Guo}}]{Li+etal+2021+AJ+173RV}
---. 2021{\natexlab{b}}, \aj, 162, 13, \dodoi{10.3847/1538-3881/abfc53}

\bibitem[{{Li} {et~al.}(2024{\natexlab{a}}){Li}, {Li}, {Gao}, {Chen}, {Gao}, \& {Sun}}]{Li+etal+2024+MN+SpecSub}
{Li}, L.-Z., {Li}, K., {Gao}, X., {et~al.} 2024{\natexlab{a}}, \mnras, 527, 3982, \dodoi{10.1093/mnras/stad3251}

\bibitem[{{Li} {et~al.}(2024{\natexlab{b}}){Li}, {Zhu}, {Ding}, {Xu}, {Zheng}, {Qiu}, \& {Liu}}]{Li+etal+ApJS+ASAS-SN+ML}
{Li}, X.-Z., {Zhu}, Q.-F., {Ding}, X., {et~al.} 2024{\natexlab{b}}, \apjs, 271, 32, \dodoi{10.3847/1538-4365/ad226a}

\bibitem[{{Lin} {et~al.}(2022){Lin}, {Wang}, {Mo}, {Xi}, {Zhang}, {Jiang}, {Shi}, {Zhang}, {Zhang}, {Wei}, {Ye}, {Wu}, {Yan}, {Chen}, {Li}, {Li}, {Lin}, {Lin}, {Sai}, {Xiang}, \& {Zhang}}]{Lin+etal+2022+MNRAS+TMTSI}
{Lin}, J., {Wang}, X., {Mo}, J., {et~al.} 2022, \mnras, 509, 2362, \dodoi{10.1093/mnras/stab2812}

\bibitem[{{Lin} {et~al.}(2023{\natexlab{a}}){Lin}, {Wang}, {Mo}, {Xi}, {Filippenko}, {Yan}, {Brink}, {Yang}, {Wu}, {N{\'e}meth}, {Li}, {Guo}, {Guo}, {Cai}, {Xiong}, {Zheng}, {Liu}, {Zhang}, {Jiang}, {Chen}, {Xia}, {Peng}, {Chen}, {Li}, {Lin}, {Xiang}, {Ma}, \& {Liu}}]{Lin+etal+2023+MNRAS+TMTSII}
---. 2023{\natexlab{a}}, \mnras, 523, 2172, \dodoi{10.1093/mnras/stad994}

\bibitem[{{Lin} {et~al.}(2023{\natexlab{b}}){Lin}, {Wu}, {Wang}, {N{\'e}meth}, {Xiong}, {Wu}, {Filippenko}, {Cai}, {Brink}, {Yan}, {Zeng}, {Luo}, {Xiang}, {Zhang}, {Zheng}, {Yang}, {Mo}, {Xi}, {Zhang}, {Iskandar}, {Esamdin}, {Jiang}, {Sai}, {Wei}, {Chen}, {Guo}, {Chen}, {Li}, {Lin}, {Lin}, \& {Zhang}}]{Lin+etal+2023+NatAs+BLAP}
{Lin}, J., {Wu}, C., {Wang}, X., {et~al.} 2023{\natexlab{b}}, Nature Astronomy, 7, 223, \dodoi{10.1038/s41550-022-01783-z}

\bibitem[{{Lin} {et~al.}(2024){Lin}, {Wu}, {Xiong}, {Wang}, {N{\'e}meth}, {Han}, {Li}, {Elias-Rosa}, {Salmaso}, {Filippenko}, {Brink}, {Yang}, {Chen}, {Yan}, {Zhang}, {Guo}, {Cai}, {Mo}, {Xi}, {Liu}, {Guo}, {Xia}, {Xiang}, {Li}, {Li}, {Zheng}, {Zhang}, {Liu}, {Guo}, {Chen}, \& {Li}}]{Lin+etal+2024+NatAs+hotsubdwarf}
{Lin}, J., {Wu}, C., {Xiong}, H., {et~al.} 2024, Nature Astronomy, 8, 491, \dodoi{10.1038/s41550-023-02188-2}

\bibitem[{{Lindner} {et~al.}(2015){Lindner}, {Vera-Ciro}, {Murray}, {Stanimirovi{\'c}}, {Babler}, {Heiles}, {Hennebelle}, {Goss}, \& {Dickey}}]{Lindner+etal+2015+AJ+GaussPy}
{Lindner}, R.~R., {Vera-Ciro}, C., {Murray}, C.~E., {et~al.} 2015, \aj, 149, 138, \dodoi{10.1088/0004-6256/149/4/138}

\bibitem[{{Liu} {et~al.}(2019){Liu}, {Fu}, {Zong}, {Shi}, {Luo}, {Zhang}, {Cui}, {Hou}, {Pan}, {Shan}, {Chen}, {Bai}, {Chen}, {Du}, {Hou}, {Liu}, {Tian}, {Wang}, {Wang}, {Wu}, {Wu}, {Yan}, \& {Zuo}}]{Liu+etal+2019+RAA+LAMOST-MRS}
{Liu}, N., {Fu}, J.-N., {Zong}, W., {et~al.} 2019, Research in Astronomy and Astrophysics, 19, 075, \dodoi{10.1088/1674-4527/19/5/75}

\bibitem[{{Liu} {et~al.}(2023{\natexlab{a}}){Liu}, {Qian}, {Liao}, {Huang}, \& {Yuan}}]{Liu+etal+2023+AJ+SixRV}
{Liu}, N.~P., {Qian}, S.~B., {Liao}, W.~P., {Huang}, Y., \& {Yuan}, Z.~L. 2023{\natexlab{a}}, \aj, 165, 259, \dodoi{10.3847/1538-3881/acd04e}

\bibitem[{{Liu} \& {Tan}(1991)}]{Liu+Tan+1991+APSS+J0132}
{Liu}, X., \& {Tan}, H. 1991, \apss, 183, 237

\bibitem[{{Liu} {et~al.}(2023{\natexlab{b}}){Liu}, {Li}, {Michel}, {Gao}, {Gao}, {Liu}, {Yin}, {Wang}, \& {Sun}}]{Liu+etal+2023+MN+SpecSub}
{Liu}, X.-Y., {Li}, K., {Michel}, R., {et~al.} 2023{\natexlab{b}}, \mnras, 519, 5760, \dodoi{10.1093/mnras/stad026}

\bibitem[{{Loeb} \& {Gaudi}(2003)}]{Loeb+Gaudi+ApJL}
{Loeb}, A., \& {Gaudi}, B.~S. 2003, \apjl, 588, L117, \dodoi{10.1086/375551}

\bibitem[{{Lu} {et~al.}(2007){Lu}, {Hrivnak}, \& {Rush}}]{Lu+etal+2007+AJ+RV+LC+1}
{Lu}, W., {Hrivnak}, B.~J., \& {Rush}, B.~W. 2007, \aj, 133, 255, \dodoi{10.1086/509604}

\bibitem[{{Lucy}(1967)}]{Lucy+ZAP+1967+65+89+WD+Gravity}
{Lucy}, L.~B. 1967, \zap, 65, 89

\bibitem[{{Lucy}(1968{\natexlab{a}})}]{Lucy+1968b+ApJ+LC}
---. 1968{\natexlab{a}}, \apj, 153, 877, \dodoi{10.1086/149712}

\bibitem[{{Lucy}(1968{\natexlab{b}})}]{Lucy+1968a+ApJ+structure}
---. 1968{\natexlab{b}}, \apj, 151, 1123, \dodoi{10.1086/149510}

\bibitem[{{Lucy}(1976)}]{Lucy+1976+ApJ+TRO}
---. 1976, \apj, 205, 208, \dodoi{10.1086/154265}

\bibitem[{{Lucy} \& {Wilson}(1979)}]{Lucy+Wilson+1979+ApJ+TRO}
{Lucy}, L.~B., \& {Wilson}, R.~E. 1979, \apj, 231, 502, \dodoi{10.1086/157212}

\bibitem[{{Luo} {et~al.}(2015){Luo}, {Zhao}, {Zhao}, {Deng}, {Liu}, {Jing}, {Wang}, {Zhang}, {Shi}, {Cui}, {Chu}, {Li}, {Bai}, {Wu}, {Cai}, {Cao}, {Cao}, {Carlin}, {Chen}, {Chen}, {Chen}, {Chen}, {Chen}, {Chen}, {Chen}, {Christlieb}, {Chu}, {Cui}, {Dong}, {Du}, {Fan}, {Feng}, {Fu}, {Gao}, {Gong}, {Gu}, {Guo}, {Han}, {He}, {Hou}, {Hou}, {Hou}, {Hu}, {Hu}, {Hu}, {Huo}, {Jia}, {Jiang}, {Jiang}, {Jiang}, {Jin}, {Kong}, {Kong}, {Lei}, {Li}, {Li}, {Li}, {Li}, {Li}, {Li}, {Li}, {Li}, {Li}, {Li}, {Li}, {Li}, {Liang}, {Lin}, {Liu}, {Liu}, {Liu}, {Liu}, {Lu}, {Luo}, {Mao}, {Newberg}, {Ni}, {Qi}, {Qi}, {Shen}, {Shi}, {Song}, {Song}, {Su}, {Su}, {Tang}, {Tao}, {Tian}, {Wang}, {Wang}, {Wang}, {Wang}, {Wang}, {Wang}, {Wang}, {Wang}, {Wang}, {Wang}, {Wang}, {Wang}, {Wang}, {Wang}, {Wang}, {Wang}, {Wang}, {Wang}, {Wang}, {Wang}, {Wei}, {Wei}, {Wu}, {Wu}, {Wu}, {Wu}, {Xing}, {Xu}, {Xu}, {Xu}, {Yan}, {Yang}, {Yang}, {Yang}, {Yang}, {Yao}, {Yu}, {Yuan}, {Yuan}, {Yuan}, {Yuan}, {Zhai}, {Zhang}, {Zhang}, {Zhang}, {Zhang},
  {Zhang}, {Zhang}, {Zhang}, {Zhang}, {Zhao}, {Zhou}, {Zhou}, {Zhu}, {Zhu}, {Zou}, \& {Zuo}}]{Luo+etal+RAA+2015+LAMOST}
{Luo}, A.~L., {Zhao}, Y.-H., {Zhao}, G., {et~al.} 2015, Research in Astronomy and Astrophysics, 15, 1095, \dodoi{10.1088/1674-4527/15/8/002}

\bibitem[{{Marsh} {et~al.}(2017){Marsh}, {Prince}, {Mahabal}, {Bellm}, {Drake}, \& {Djorgovski}}]{Marsh+etal+MNRAS+2017+CSS}
{Marsh}, F.~M., {Prince}, T.~A., {Mahabal}, A.~A., {et~al.} 2017, \mnras, 465, 4678, \dodoi{10.1093/mnras/stw2110}

\bibitem[{{Masci} {et~al.}(2019){Masci}, {Laher}, {Rusholme}, {Shupe}, {Groom}, {Surace}, {Jackson}, {Monkewitz}, {Beck}, {Flynn}, {Terek}, {Landry}, {Hacopians}, {Desai}, {Howell}, {Brooke}, {Imel}, {Wachter}, {Ye}, {Lin}, {Cenko}, {Cunningham}, {Rebbapragada}, {Bue}, {Miller}, {Mahabal}, {Bellm}, {Patterson}, {Juri{\'c}}, {Golkhou}, {Ofek}, {Walters}, {Graham}, {Kasliwal}, {Dekany}, {Kupfer}, {Burdge}, {Cannella}, {Barlow}, {Van Sistine}, {Giomi}, {Fremling}, {Blagorodnova}, {Levitan}, {Riddle}, {Smith}, {Helou}, {Prince}, \& {Kulkarni}}]{Masci+etal+2019+PASP+ZTF}
{Masci}, F.~J., {Laher}, R.~R., {Rusholme}, B., {et~al.} 2019, \pasp, 131, 018003, \dodoi{10.1088/1538-3873/aae8ac}

\bibitem[{{Nelson}(2007)}]{Nelson+2007+J1300min}
{Nelson}, R.~H. 2007, Information Bulletin on Variable Stars, 5760, 1

\bibitem[{{Nelson}(2008)}]{Nelson+Robsert+2008+J0132min}
---. 2008, Information Bulletin on Variable Stars, 5820, 1

\bibitem[{{Nelson}(2009{\natexlab{a}})}]{Nelson+2009+J1300min}
---. 2009{\natexlab{a}}, Information Bulletin on Variable Stars, 5875, 1

\bibitem[{{Nelson}(2009{\natexlab{b}})}]{Nelson+2009+J1402min}
---. 2009{\natexlab{b}}, Information Bulletin on Variable Stars, 5875, 1

\bibitem[{{Nelson}(2013)}]{Nelson+2013+J1300min}
---. 2013, Information Bulletin on Variable Stars, 6050, 1

\bibitem[{{O'Connell}(1951)}]{OConnell+1951+PRCO}
{O'Connell}, D.~J.~K. 1951, Publications of the Riverview College Observatory, 2, 85

\bibitem[{{Pagel}(2018)}]{Pagel+2018+J1402min}
{Pagel}, L. 2018, Information Bulletin on Variable Stars, 6244, 1, \dodoi{10.22444/IBVS.6244}

\bibitem[{{Paki} \& {Poro}(2024)}]{Paki+Poro+2024+arXiv+PHOEBE+TESS+20}
{Paki}, E., \& {Poro}, A. 2024, arXiv e-prints, arXiv:2405.18618, \dodoi{10.48550/arXiv.2405.18618}

\bibitem[{{Panchal} \& {Joshi}(2021)}]{Panchal+Joshi+2021+J0305}
{Panchal}, A., \& {Joshi}, Y.~C. 2021, \aj, 161, 221, \dodoi{10.3847/1538-3881/abea0c}

\bibitem[{{Papageorgiou} {et~al.}(2023){Papageorgiou}, {Christopoulou}, {Ferreira Lopes}, {Lalounta}, {Catelan}, \& {Drake}}]{Papageorgiou+etal+2023+AJ+OConnell}
{Papageorgiou}, A., {Christopoulou}, P.-E., {Ferreira Lopes}, C.~E., {et~al.} 2023, \aj, 165, 80, \dodoi{10.3847/1538-3881/aca65a}

\bibitem[{{Pavlenko} {et~al.}(2018){Pavlenko}, {Evans}, {Banerjee}, {Southworth}, {Shahbandeh}, \& {Davis}}]{Pavlenko+etal+2018+AA+SpecSub}
{Pavlenko}, Y.~V., {Evans}, A., {Banerjee}, D.~P.~K., {et~al.} 2018, \aap, 615, A120, \dodoi{10.1051/0004-6361/201832717}

\bibitem[{{Pi} {et~al.}(2017){Pi}, {Zhang}, {Bi}, {Han}, {Wang}, \& {Lu}}]{Pi+etal+2017+AJ+V1101Her}
{Pi}, Q.-f., {Zhang}, L.-y., {Bi}, S.-l., {et~al.} 2017, \aj, 154, 260, \dodoi{10.3847/1538-3881/aa9438}

\bibitem[{{Pojmanski}(1997)}]{Pojmanski+1997+AcA+ASAS}
{Pojmanski}, G. 1997, \actaa, 47, 467, \dodoi{10.48550/arXiv.astro-ph/9712146}

\bibitem[{{Pojmanski}(1998)}]{Pojmanski+1998+AcA+ASAS}
---. 1998, \actaa, 48, 35, \dodoi{10.48550/arXiv.astro-ph/9802330}

\bibitem[{{Pojmanski}(2002)}]{Pojmanski+2002+AcA+ASAS}
---. 2002, \actaa, 52, 397, \dodoi{10.48550/arXiv.astro-ph/0210283}

\bibitem[{{Pollacco} {et~al.}(2006){Pollacco}, {Skillen}, {Collier Cameron}, {Christian}, {Hellier}, {Irwin}, {Lister}, {Street}, {West}, {Anderson}, {Clarkson}, {Deeg}, {Enoch}, {Evans}, {Fitzsimmons}, {Haswell}, {Hodgkin}, {Horne}, {Kane}, {Keenan}, {Maxted}, {Norton}, {Osborne}, {Parley}, {Ryans}, {Smalley}, {Wheatley}, \& {Wilson}}]{Pallacco+etal+2006+PASP+SuperWASP}
{Pollacco}, D.~L., {Skillen}, I., {Collier Cameron}, A., {et~al.} 2006, \pasp, 118, 1407, \dodoi{10.1086/508556}

\bibitem[{{Pr{\v{s}}a}(2018)}]{Prsa+2018+book+PHOEBE}
{Pr{\v{s}}a}, A. 2018, {Modeling and Analysis of Eclipsing Binary Stars; The theory and design principles of PHOEBE}, \dodoi{10.1088/978-0-7503-1287-5}

\bibitem[{{Qian}(2001)}]{Qian+2001+MNRAS+TRO}
{Qian}, S. 2001, \mnras, 328, 914, \dodoi{10.1046/j.1365-8711.2001.04921.x}

\bibitem[{{Qian} {et~al.}(2017){Qian}, {He}, {Zhang}, {Zhu}, {Shi}, {Zhao}, \& {Zhou}}]{Qian+etal+2017+RAA+AML}
{Qian}, S.-B., {He}, J.-J., {Zhang}, J., {et~al.} 2017, Research in Astronomy and Astrophysics, 17, 087, \dodoi{10.1088/1674-4527/17/8/87}

\bibitem[{{Ricker} {et~al.}(2010){Ricker}, {Latham}, {Vanderspek}, {Ennico}, {Bakos}, {Brown}, {Burgasser}, {Charbonneau}, {Clampin}, {Deming}, {Doty}, {Dunham}, {Elliot}, {Holman}, {Ida}, {Jenkins}, {Jernigan}, {Kawai}, {Laughlin}, {Lissauer}, {Martel}, {Sasselov}, {Schingler}, {Seager}, {Torres}, {Udry}, {Villasenor}, {Winn}, \& {Worden}}]{Ricker+etal+2010+AAS+TESS}
{Ricker}, G.~R., {Latham}, D.~W., {Vanderspek}, R.~K., {et~al.} 2010, in American Astronomical Society Meeting Abstracts, Vol. 215, American Astronomical Society Meeting Abstracts \#215, 450.06

\bibitem[{{Robertson} \& {Eggleton}(1977)}]{Robertson+Eggleton+1977+MNRAS+TRO}
{Robertson}, J.~A., \& {Eggleton}, P.~P. 1977, \mnras, 179, 359, \dodoi{10.1093/mnras/179.3.359}

\bibitem[{{Ruci{\'n}ski}(1969)}]{Rucinski+1969+AcA+WD+Gravity}
{Ruci{\'n}ski}, S.~M. 1969, \actaa, 19, 245

\bibitem[{{Rucinski} {et~al.}(2000){Rucinski}, {Lu}, \& {Mochnacki}}]{Rucinski+2000+AJ+RV+III+10}
{Rucinski}, S.~M., {Lu}, W., \& {Mochnacki}, S.~W. 2000, \aj, 120, 1133, \dodoi{10.1086/301458}

\bibitem[{{Shappee} {et~al.}(2014){Shappee}, {Prieto}, {Grupe}, {Kochanek}, {Stanek}, {De Rosa}, {Mathur}, {Zu}, {Peterson}, {Pogge}, {Komossa}, {Im}, {Jencson}, {Holoien}, {Basu}, {Beacom}, {Szczygie{\l}}, {Brimacombe}, {Adams}, {Campillay}, {Choi}, {Contreras}, {Dietrich}, {Dubberley}, {Elphick}, {Foale}, {Giustini}, {Gonzalez}, {Hawkins}, {Howell}, {Hsiao}, {Koss}, {Leighly}, {Morrell}, {Mudd}, {Mullins}, {Nugent}, {Parrent}, {Phillips}, {Pojmanski}, {Rosing}, {Ross}, {Sand}, {Terndrup}, {Valenti}, {Walker}, \& {Yoon}}]{Shappee+etal+2014+ApJ+ASAS-SN}
{Shappee}, B.~J., {Prieto}, J.~L., {Grupe}, D., {et~al.} 2014, \apj, 788, 48, \dodoi{10.1088/0004-637X/788/1/48}

\bibitem[{{Stassun} {et~al.}(2018){Stassun}, {Oelkers}, {Pepper}, {Paegert}, {De Lee}, {Torres}, {Latham}, {Charpinet}, {Dressing}, {Huber}, {Kane}, {L{\'e}pine}, {Mann}, {Muirhead}, {Rojas-Ayala}, {Silvotti}, {Fleming}, {Levine}, \& {Plavchan}}]{Stassum+etal+AJ+TESS}
{Stassun}, K.~G., {Oelkers}, R.~J., {Pepper}, J., {et~al.} 2018, \aj, 156, 102, \dodoi{10.3847/1538-3881/aad050}

\bibitem[{{Stepien}(2006)}]{Stepien+2006+AcA+AML}
{Stepien}, K. 2006, \actaa, 56, 347, \dodoi{10.48550/arXiv.astro-ph/0701529}

\bibitem[{{St{\k{e}}pie{\'n}}(2011)}]{Stepien+2011+AcA+AML}
{St{\k{e}}pie{\'n}}, K. 2011, \actaa, 61, 139, \dodoi{10.48550/arXiv.1105.2645}

\bibitem[{{Sun} {et~al.}(2020{\natexlab{a}}){Sun}, {Chen}, {Deng}, \& {de Grijs}}]{Sun+etal+2020+ApJS+CSS+WD+CB}
{Sun}, W., {Chen}, X., {Deng}, L., \& {de Grijs}, R. 2020{\natexlab{a}}, \apjs, 247, 50, \dodoi{10.3847/1538-4365/ab7894}

\bibitem[{{Sun} {et~al.}(2020{\natexlab{b}}){Sun}, {Chen}, {Deng}, \& {de Grijs}}]{Sun+etal+2020+ApJS}
---. 2020{\natexlab{b}}, \apjs, 247, 50, \dodoi{10.3847/1538-4365/ab7894}

\bibitem[{{van Hamme}(1993)}]{vanHamme+1993+AJ+WD+LimbDarkening}
{van Hamme}, W. 1993, \aj, 106, 2096, \dodoi{10.1086/116788}

\bibitem[{{Van Hamme} \& {Wilson}(2007)}]{VanHamme+Wilson+2007+ApJ+WD}
{Van Hamme}, W., \& {Wilson}, R.~E. 2007, \apj, 661, 1129, \dodoi{10.1086/517870}

\bibitem[{{{\v{C}}eki} {et~al.}(2024){{\v{C}}eki}, {{\c{S}}enavc{\i}}, {Latkovi{\'c}}, {Uzun{\c{c}}am}, {Yorulmaz}, \& {Bahar}}]{Ceki+etal+2024+MNRAS+OConnell}
{{\v{C}}eki}, A., {{\c{S}}enavc{\i}}, H.~V., {Latkovi{\'c}}, O., {et~al.} 2024, \mnras, 532, 3582, \dodoi{10.1093/mnras/stae1709}

\bibitem[{{Vijaya} \& {Sriram}(2023)}]{Vijaya+Sriram+2023+RAA+CWAqr}
{Vijaya}, A., \& {Sriram}, K. 2023, Research in Astronomy and Astrophysics, 23, 055009, \dodoi{10.1088/1674-4527/acc154}

\bibitem[{{Watson} {et~al.}(2006){Watson}, {Henden}, \& {Price}}]{Watson+etal+2006+VSX}
{Watson}, C.~L., {Henden}, A.~A., \& {Price}, A. 2006, Society for Astronomical Sciences Annual Symposium, 25, 47

\bibitem[{{West} {et~al.}(2015){West}, {Weisenburger}, {Irwin}, {Berta-Thompson}, {Charbonneau}, {Dittmann}, \& {Pineda}}]{West+etal+2015+ApJ+magnetic}
{West}, A.~A., {Weisenburger}, K.~L., {Irwin}, J., {et~al.} 2015, \apj, 812, 3, \dodoi{10.1088/0004-637X/812/1/3}

\bibitem[{{West} {et~al.}(2011){West}, {Morgan}, {Bochanski}, {Andersen}, {Bell}, {Kowalski}, {Davenport}, {Hawley}, {Schmidt}, {Bernat}, {Hilton}, {Muirhead}, {Covey}, {Rojas-Ayala}, {Schlawin}, {Gooding}, {Schluns}, {Dhital}, {Pineda}, \& {Jones}}]{West+etal+2011+AJ+magnetic}
{West}, A.~A., {Morgan}, D.~P., {Bochanski}, J.~J., {et~al.} 2011, \aj, 141, 97, \dodoi{10.1088/0004-6256/141/3/97}

\bibitem[{{Wilson}(1979)}]{Wilson+1979+ApJ+WD}
{Wilson}, R.~E. 1979, \apj, 234, 1054, \dodoi{10.1086/157588}

\bibitem[{{Wilson}(1990)}]{Wilson+1990+ApJ+WD}
---. 1990, \apj, 356, 613, \dodoi{10.1086/168867}

\bibitem[{{Wilson} \& {Devinney}(1971)}]{Wilson+Devinney+1971+ApJ+WD}
{Wilson}, R.~E., \& {Devinney}, E.~J. 1971, \apj, 166, 605, \dodoi{10.1086/150986}

\bibitem[{{Wilson} \& {Van Hamme}(2014)}]{Wilson+VanHamme+2014+ApJ+WD}
{Wilson}, R.~E., \& {Van Hamme}, W. 2014, \apj, 780, 151, \dodoi{10.1088/0004-637X/780/2/151}

\bibitem[{{Wilson} {et~al.}(2010){Wilson}, {Van Hamme}, \& {Terrell}}]{Wilson+VanHamme+2010+ApJ+WD}
{Wilson}, R.~E., {Van Hamme}, W., \& {Terrell}, D. 2010, \apj, 723, 1469, \dodoi{10.1088/0004-637X/723/2/1469}

\bibitem[{{Wu} {et~al.}(2024){Wu}, {Zhu}, {Matekov}, {Li}, {Ehgamberdiev}, {Asfandiyarov}, {Wang}, {Zhang}, \& {Meng}}]{Wu+etal+2024+MN+CSSJ1549}
{Wu}, J.-F., {Zhu}, L.-Y., {Matekov}, A., {et~al.} 2024, \mnras, 529, 3113, \dodoi{10.1093/mnras/stae590}

\bibitem[{{Xiong} {et~al.}(2024){Xiong}, {Ding}, {Li}, {Ge}, {Cheng}, {Ji}, {Han}, \& {Chen}}]{Xiong+etal+2024+ApJS+TESS+ML}
{Xiong}, J., {Ding}, X., {Li}, J., {et~al.} 2024, \apjs, 270, 20, \dodoi{10.3847/1538-4365/ad0ceb}

\bibitem[{{Yakut} \& {Eggleton}(2005)}]{Yakut+Eggleton+2005+ApJ+CloseBinary}
{Yakut}, K., \& {Eggleton}, P.~P. 2005, \apj, 629, 1055, \dodoi{10.1086/431300}

\bibitem[{{Yang} {et~al.}(2023){Yang}, {Michel}, {Yuan}, {Wang}, \& {Tamayo}}]{Yang+etal+2023+J1300}
{Yang}, Y., {Michel}, R., {Yuan}, H., {Wang}, S., \& {Tamayo}, F. 2023, \mnras, 522, 3076, \dodoi{10.1093/mnras/stad1141}

\bibitem[{{Yang}(2011)}]{Yang+2011+J1402}
{Yang}, Y.-G. 2011, Research in Astronomy and Astrophysics, 11, 181, \dodoi{10.1088/1674-4527/11/2/006}

\bibitem[{{Yang} {et~al.}(2013){Yang}, {Qian}, {Zhang}, {Dai}, \& {Soonthornthum}}]{Yang+etal+2013+AJ+DZPsc}
{Yang}, Y.~G., {Qian}, S.~B., {Zhang}, L.~Y., {Dai}, H.~F., \& {Soonthornthum}, B. 2013, \aj, 146, 35, \dodoi{10.1088/0004-6256/146/2/35}

\bibitem[{{Y{\i}ld{\i}z}(2014)}]{Yildiz+2014+MNRAS+AML}
{Y{\i}ld{\i}z}, M. 2014, \mnras, 437, 185, \dodoi{10.1093/mnras/stt1874}

\bibitem[{{Yildiz} \& {Do{\u{g}}an}(2013)}]{Yildiz+Dogan+2013+MN+AML}
{Yildiz}, M., \& {Do{\u{g}}an}, T. 2013, \mnras, 430, 2029, \dodoi{10.1093/mnras/stt028}

\bibitem[{{Zhang} {et~al.}(2020{\natexlab{a}}){Zhang}, {Liu}, \& {Deng}}]{Zhang+etal+2020+ApJS+laspec}
{Zhang}, B., {Liu}, C., \& {Deng}, L.-C. 2020{\natexlab{a}}, \apjs, 246, 9, \dodoi{10.3847/1538-4365/ab55ef}

\bibitem[{{Zhang} {et~al.}(2021){Zhang}, {Li}, {Yang}, {Xiong}, {Fu}, {Liu}, {Tian}, {Li}, {Wang}, {Liang}, {Zhou}, {Zong}, {Yang}, {Liu}, \& {Hou}}]{Zhang+etal+2021+ApJS+RVZP}
{Zhang}, B., {Li}, J., {Yang}, F., {et~al.} 2021, \apjs, 256, 14, \dodoi{10.3847/1538-4365/ac0834}

\bibitem[{{Zhang} {et~al.}(2020{\natexlab{b}}){Zhang}, {Wang}, {Mo}, {Xi}, {Lin}, {Jiang}, {Zhang}, {Li}, {Yan}, {Chen}, {Hu}, {Li}, {Lin}, {Lin}, {Miao}, {Rui}, {Sai}, {Xiang}, \& {Zhang}}]{Zhang+etal+2020+PASP+tmts_performance}
{Zhang}, J.-C., {Wang}, X.-F., {Mo}, J., {et~al.} 2020{\natexlab{b}}, \pasp, 132, 125001, \dodoi{10.1088/1538-3873/abbea2}

\bibitem[{{Zhao} {et~al.}(2012){Zhao}, {Zhao}, {Chu}, {Jing}, \& {Deng}}]{Zhao+etal+RAA+LAMOST}
{Zhao}, G., {Zhao}, Y.-H., {Chu}, Y.-Q., {Jing}, Y.-P., \& {Deng}, L.-C. 2012, Research in Astronomy and Astrophysics, 12, 723, \dodoi{10.1088/1674-4527/12/7/002}

\bibitem[{{Zucker} {et~al.}(2007){Zucker}, {Mazeh}, \& {Alexander}}]{Zucker+etal+2007+ApJ+Beaming}
{Zucker}, S., {Mazeh}, T., \& {Alexander}, T. 2007, \apj, 670, 1326, \dodoi{10.1086/521389}

\end{thebibliography}
\bibliographystyle{aasjournal}



\end{document}